\documentclass[conference]{IEEEtran}

\usepackage[latin1]{inputenc}

\usepackage[english]{babel}

% ...

\usepackage{amsthm}

% ...

\usepackage{cite}

\usepackage[final]{microtype}

\usepackage{changebar}

\renewcommand{\baselinestretch}{0.975}

\usepackage{etoolbox}

\usepackage{fixltx2e}

\usepackage{xargs}

\usepackage{xspace}

\usepackage{xstring}

\usepackage{boolexpr}

\usepackage[cmex10]{mathtools}
\usepackage{amssymb}
\usepackage{amsfonts}
\usepackage{mathrsfs}
\usepackage{latexsym}
\usepackage{textcomp}
\usepackage{pifont}

\newcommand{\argemp}[2]
	{\if&#1&\else#2\fi}

\newcommand{\argdef}[2]
	{\if&#1&#2\else#1\fi}

\newcommand{\argint}[3]
	{\if&#2&\else#1#2#3\fi}

\newcommand{\argext}[3]
	{\if&#1&#3\else#1\if&#3&\else#2#3\fi\fi}

\newcommandx{\argsubsup}[3][2=, 3=]
	{\def\argsubscript{{#2}}\def\argsuperscript{{#3}}#1}

\newcommandx{\argind}[9][2=, 3=, 4=, 5=, 6=, 7=, 8=, 9=]
	{%
	\switch[#1=]%
		\case{0}#2%
		\case{1}#3%
		\case{2}#4%
		\case{3}#5%
		\case{4}#6%
		\case{5}#7%
		\case{6}#8%
		\case{7}#9%
		\otherwise\ensuremath{\clubsuit}%
	\endswitch%
	}

\newcommand{\arga}[1]
	{#1}
\newcommand{\argb}[2]
	{\argext{\arga{#1}}{, \allowbreak}{#2}}
\newcommand{\argc}[3]
	{\argext{\argb{#1}{#2}}{, \allowbreak}{#3}}
\newcommand{\argd}[4]
	{\argext{\argc{#1}{#2}{#3}}{, \allowbreak}{#4}}
\newcommand{\arge}[5]
	{\argext{\argd{#1}{#2}{#3}{#4}}{, \allowbreak}{#5}}
\newcommand{\argf}[6]
	{\argext{\arge{#1}{#2}{#3}{#4}{#5}}{, \allowbreak}{#6}}
\newcommand{\argg}[7]
	{\argext{\argf{#1}{#2}{#3}{#4}{#5}{#6}}{, \allowbreak}{#7}}
\newcommand{\argh}[8]
	{\argext{\argg{#1}{#2}{#3}{#4}{#5}{#6}{#7}}{, \allowbreak}{#8}}
\newcommand{\argi}[9]
	{\argext{\argh{#1}{#2}{#3}{#4}{#5}{#6}{#7}{#8}}{, \allowbreak}{#9}}

\newcommand{\txtfnt}[2][]
	{{%
	\IfStrEq{#1}{}
		{#2}
		{%
		\StrLeft{#1}{2}[\optbgn]%
		\StrGobbleLeft{#1}{2}[\optend]%
		\IfStrEqCase{\optbgn}
			{%
			{Rm}{\rmfamily\txtfnt[\optend]{#2}}%
			{Sf}{\sffamily\txtfnt[\optend]{#2}}%
			{Tt}{\ttfamily\txtfnt[\optend]{#2}}%
			{Up}{\upshape\txtfnt[\optend]{#2}}%
			{It}{\itshape\txtfnt[\optend]{#2}}%
			{Sl}{\slshape\txtfnt[\optend]{#2}}%
			{Sc}{\scshape\txtfnt[\optend]{#2}}%
			{Md}{\mdseries\txtfnt[\optend]{#2}}%
			{Bf}{\bfseries\txtfnt[\optend]{#2}}%
			{Em}{\emph{\txtfnt[\optend]{#2}}}%
			}
			[\ensuremath{\clubsuit}]%
		}%
	}}

\newcommand{\txtsub}[2][]
	{\argemp{#2}{\ensuremath{_{\text{\txtfnt[#1]{#2}}}}}}

\newcommand{\txtsup}[2][]
	{\argemp{#2}{\ensuremath{^{\text{\txtfnt[#1]{#2}}}}}}

\newcommandx{\txt}[4][1=, 3=, 4=]
	{%
	\ensuremath{\text{%
		\txtfnt[#1]{#2}\ensuremath{\txtsub[#1]{#3}\txtsup[#1]{#4}}%
	}}%
	}

\newcommandx{\txtarg}[5][1=, 3=, 4=]
	{{\txt[#1]{#2}[#3][#4]\argint{(}{#5}{)}}}

\newcommand{\txtstyname}{RmScMd}
\newcommand{\txtname}[1][]
	{\txt[\argdef{#1}{\txtstyname}]}
\newcommand{\txtargname}[1][]
	{\txtarg[\argdef{#1}{\txtstyname}]}

\newcommand{\txtstyabr}{Em}
\newcommand{\txtabr}[1][]
	{\txt[\argdef{#1}{\txtstyabr}]}

\newcommandx{\mthfnt}[3][1=, 2=0]
	{{%
	\IfStrEqCase{#1}
		{%
		{}%
			{#3}%
		{Name}%
			{%
			\IfStrEqCase{#2}
				{%
				{0}{\mathcal{#3}}%
				{1}{\mathscr{#3}}%
				{2}{\mathfrak{#3}}%
				{3}{\mathbb{#3}}%
				}
				[\ensuremath{\clubsuit}]%
			}%
		{Set}%
			{%
			\IfStrEqCase{#2}
				{%
				{0}{\mathrm{#3}}%
				{1}{\mathsf{#3}}%
				{2}{\mathbb{#3}}%
				{3}{\mathbf{#3}}%
				}
				[\ensuremath{\clubsuit}]%
			}%
		{Fun}%
			{%
			\IfStrEqCase{#2}
				{%
				{0}{\mathsf{#3}}%
				{1}{\mathrm{#3}}%
				}
				[\ensuremath{\clubsuit}]%
			}%
		{Rel}%
			{%
			\IfStrEqCase{#2}
				{%
				{0}{\mathit{#3}}%
				{1}{\mathtt{#3}}%
				}
				[\ensuremath{\clubsuit}]%
			}%
		{Sym}%
			{%
			\IfStrEqCase{#2}
				{%
				{0}{\mathtt{#3}}%
				{1}{\mathbf{#3}}%
				}
				[\ensuremath{\clubsuit}]%
			}%
		{Elm}%
			{\mathnormal{#3}}
		}
		[\ensuremath{\clubsuit}]%
	}}

\newcommand{\mthsub}[1]
	{\argemp{#1}{\ensuremath{_{\mathnormal{#1}}}}}

\newcommand{\mthsup}[1]
	{\argemp{#1}{\ensuremath{^{\mathnormal{#1}}}}}

\newcommandx{\mth}[5][1=, 2=0, 4=, 5=]
	{{\ensuremath{\mthfnt[#1][#2]{#3}\mthsub{#4}\mthsup{#5}}}}

\newcommandx{\mtharg}[6][1=, 2=0, 4=, 5=]
	{{\mth[#1][#2]{#3}[#4][#5]\ensuremath{\argint{(}{#6}{)}}}}

\newcommand{\mthempty}
	{\mth[][]}

\newcommand{\mthstyname}{0}
\newcommand{\mthname}[1][]
	{\mth[Name][\argdef{#1}{\mthstyname}]}

\newcommand{\mthstyset}{0}
\newcommand{\mthset}[1][]
	{\mth[Set][\argdef{#1}{\mthstyset}]}
\newcommand{\mthargset}[1][]
	{\mtharg[Set][\argdef{#1}{\mthstyset}]}

\newcommand{\mthstyfun}{0}
\newcommand{\mthfun}[1][]
	{\mth[Fun][\argdef{#1}{\mthstyfun}]}
\newcommand{\mthargfun}[1][]
	{\mtharg[Fun][\argdef{#1}{\mthstyfun}]}

\newcommand{\mthstyrel}{0}
\newcommand{\mthrel}[1][]
	{\mth[Rel][\argdef{#1}{\mthstyrel}]}

\newcommand{\mthstysym}{0}
\newcommand{\mthsym}[1][]
	{\mth[Sym][\argdef{#1}{\mthstysym}]}

\newcommand{\mthstyelm}{0}
\newcommand{\mthelm}[1][]
	{\mth[Elm][\argdef{#1}{\mthstyelm}]}

\newcommandx{\AName}[4][1=, 2=, 3=, 4=]{\mthname[#4]{A#3}[#1][#2]}
\newcommandx{\BName}[4][1=, 2=, 3=, 4=]{\mthname[#4]{B#3}[#1][#2]}
\newcommandx{\CName}[4][1=, 2=, 3=, 4=]{\mthname[#4]{C#3}[#1][#2]}
\newcommandx{\DName}[4][1=, 2=, 3=, 4=]{\mthname[#4]{D#3}[#1][#2]}
\newcommandx{\EName}[4][1=, 2=, 3=, 4=]{\mthname[#4]{E#3}[#1][#2]}
\newcommandx{\FName}[4][1=, 2=, 3=, 4=]{\mthname[#4]{F#3}[#1][#2]}
\newcommandx{\GName}[4][1=, 2=, 3=, 4=]{\mthname[#4]{G#3}[#1][#2]}
\newcommandx{\HName}[4][1=, 2=, 3=, 4=]{\mthname[#4]{H#3}[#1][#2]}
\newcommandx{\IName}[4][1=, 2=, 3=, 4=]{\mthname[#4]{I#3}[#1][#2]}
\newcommandx{\JName}[4][1=, 2=, 3=, 4=]{\mthname[#4]{J#3}[#1][#2]}
\newcommandx{\KName}[4][1=, 2=, 3=, 4=]{\mthname[#4]{K#3}[#1][#2]}
\newcommandx{\LName}[4][1=, 2=, 3=, 4=]{\mthname[#4]{L#3}[#1][#2]}
\newcommandx{\MName}[4][1=, 2=, 3=, 4=]{\mthname[#4]{M#3}[#1][#2]}
\newcommandx{\NName}[4][1=, 2=, 3=, 4=]{\mthname[#4]{N#3}[#1][#2]}
\newcommandx{\OName}[4][1=, 2=, 3=, 4=]{\mthname[#4]{O#3}[#1][#2]}
\newcommandx{\PName}[4][1=, 2=, 3=, 4=]{\mthname[#4]{P#3}[#1][#2]}
\newcommandx{\QName}[4][1=, 2=, 3=, 4=]{\mthname[#4]{Q#3}[#1][#2]}
\newcommandx{\RName}[4][1=, 2=, 3=, 4=]{\mthname[#4]{R#3}[#1][#2]}
\newcommandx{\SName}[4][1=, 2=, 3=, 4=]{\mthname[#4]{S#3}[#1][#2]}
\newcommandx{\TName}[4][1=, 2=, 3=, 4=]{\mthname[#4]{T#3}[#1][#2]}
\newcommandx{\UName}[4][1=, 2=, 3=, 4=]{\mthname[#4]{U#3}[#1][#2]}
\newcommandx{\VName}[4][1=, 2=, 3=, 4=]{\mthname[#4]{V#3}[#1][#2]}
\newcommandx{\WName}[4][1=, 2=, 3=, 4=]{\mthname[#4]{W#3}[#1][#2]}
\newcommandx{\XName}[4][1=, 2=, 3=, 4=]{\mthname[#4]{X#3}[#1][#2]}
\newcommandx{\YName}[4][1=, 2=, 3=, 4=]{\mthname[#4]{Y#3}[#1][#2]}
\newcommandx{\ZName}[4][1=, 2=, 3=, 4=]{\mthname[#4]{Z#3}[#1][#2]}

\newcommandx{\ASet}[4][1=, 2=, 3=, 4=]{\mthset[#4]{A#3}[#1][#2]}
\newcommandx{\BSet}[4][1=, 2=, 3=, 4=]{\mthset[#4]{B#3}[#1][#2]}
\newcommandx{\CSet}[4][1=, 2=, 3=, 4=]{\mthset[#4]{C#3}[#1][#2]}
\newcommandx{\DSet}[4][1=, 2=, 3=, 4=]{\mthset[#4]{D#3}[#1][#2]}
\newcommandx{\ESet}[4][1=, 2=, 3=, 4=]{\mthset[#4]{E#3}[#1][#2]}
\newcommandx{\FSet}[4][1=, 2=, 3=, 4=]{\mthset[#4]{F#3}[#1][#2]}
\newcommandx{\GSet}[4][1=, 2=, 3=, 4=]{\mthset[#4]{G#3}[#1][#2]}
\newcommandx{\HSet}[4][1=, 2=, 3=, 4=]{\mthset[#4]{H#3}[#1][#2]}
\newcommandx{\ISet}[4][1=, 2=, 3=, 4=]{\mthset[#4]{I#3}[#1][#2]}
\newcommandx{\JSet}[4][1=, 2=, 3=, 4=]{\mthset[#4]{J#3}[#1][#2]}
\newcommandx{\KSet}[4][1=, 2=, 3=, 4=]{\mthset[#4]{K#3}[#1][#2]}
\newcommandx{\LSet}[4][1=, 2=, 3=, 4=]{\mthset[#4]{L#3}[#1][#2]}
\newcommandx{\MSet}[4][1=, 2=, 3=, 4=]{\mthset[#4]{M#3}[#1][#2]}
\newcommandx{\NSet}[4][1=, 2=, 3=, 4=]{\mthset[#4]{N#3}[#1][#2]}
\newcommandx{\OSet}[4][1=, 2=, 3=, 4=]{\mthset[#4]{O#3}[#1][#2]}
\newcommandx{\PSet}[4][1=, 2=, 3=, 4=]{\mthset[#4]{P#3}[#1][#2]}
\newcommandx{\QSet}[4][1=, 2=, 3=, 4=]{\mthset[#4]{Q#3}[#1][#2]}
\newcommandx{\RSet}[4][1=, 2=, 3=, 4=]{\mthset[#4]{R#3}[#1][#2]}
\newcommandx{\SSet}[4][1=, 2=, 3=, 4=]{\mthset[#4]{S#3}[#1][#2]}
\newcommandx{\TSet}[4][1=, 2=, 3=, 4=]{\mthset[#4]{T#3}[#1][#2]}
\newcommandx{\USet}[4][1=, 2=, 3=, 4=]{\mthset[#4]{U#3}[#1][#2]}
\newcommandx{\VSet}[4][1=, 2=, 3=, 4=]{\mthset[#4]{V#3}[#1][#2]}
\newcommandx{\WSet}[4][1=, 2=, 3=, 4=]{\mthset[#4]{W#3}[#1][#2]}
\newcommandx{\XSet}[4][1=, 2=, 3=, 4=]{\mthset[#4]{X#3}[#1][#2]}
\newcommandx{\YSet}[4][1=, 2=, 3=, 4=]{\mthset[#4]{Y#3}[#1][#2]}
\newcommandx{\ZSet}[4][1=, 2=, 3=, 4=]{\mthset[#4]{Z#3}[#1][#2]}

\newcommandx{\aSet}[4][1=, 2=, 3=, 4=]{\mthset[#4]{a#3}[#1][#2]}
\newcommandx{\bSet}[4][1=, 2=, 3=, 4=]{\mthset[#4]{b#3}[#1][#2]}
\newcommandx{\cSet}[4][1=, 2=, 3=, 4=]{\mthset[#4]{c#3}[#1][#2]}
\newcommandx{\dSet}[4][1=, 2=, 3=, 4=]{\mthset[#4]{d#3}[#1][#2]}
\newcommandx{\eSet}[4][1=, 2=, 3=, 4=]{\mthset[#4]{e#3}[#1][#2]}
\newcommandx{\fSet}[4][1=, 2=, 3=, 4=]{\mthset[#4]{f#3}[#1][#2]}
\newcommandx{\gSet}[4][1=, 2=, 3=, 4=]{\mthset[#4]{g#3}[#1][#2]}
\newcommandx{\hSet}[4][1=, 2=, 3=, 4=]{\mthset[#4]{h#3}[#1][#2]}
\newcommandx{\iSet}[4][1=, 2=, 3=, 4=]{\mthset[#4]{i#3}[#1][#2]}
\newcommandx{\jSet}[4][1=, 2=, 3=, 4=]{\mthset[#4]{j#3}[#1][#2]}
\newcommandx{\kSet}[4][1=, 2=, 3=, 4=]{\mthset[#4]{k#3}[#1][#2]}
\newcommandx{\lSet}[4][1=, 2=, 3=, 4=]{\mthset[#4]{l#3}[#1][#2]}
\newcommandx{\mSet}[4][1=, 2=, 3=, 4=]{\mthset[#4]{m#3}[#1][#2]}
\newcommandx{\nSet}[4][1=, 2=, 3=, 4=]{\mthset[#4]{n#3}[#1][#2]}
\newcommandx{\oSet}[4][1=, 2=, 3=, 4=]{\mthset[#4]{o#3}[#1][#2]}
\newcommandx{\pSet}[4][1=, 2=, 3=, 4=]{\mthset[#4]{p#3}[#1][#2]}
\newcommandx{\qSet}[4][1=, 2=, 3=, 4=]{\mthset[#4]{q#3}[#1][#2]}
\newcommandx{\rSet}[4][1=, 2=, 3=, 4=]{\mthset[#4]{r#3}[#1][#2]}
\newcommandx{\sSet}[4][1=, 2=, 3=, 4=]{\mthset[#4]{s#3}[#1][#2]}
\newcommandx{\tSet}[4][1=, 2=, 3=, 4=]{\mthset[#4]{t#3}[#1][#2]}
\newcommandx{\uSet}[4][1=, 2=, 3=, 4=]{\mthset[#4]{u#3}[#1][#2]}
\newcommandx{\vSet}[4][1=, 2=, 3=, 4=]{\mthset[#4]{v#3}[#1][#2]}
\newcommandx{\wSet}[4][1=, 2=, 3=, 4=]{\mthset[#4]{w#3}[#1][#2]}
\newcommandx{\xSet}[4][1=, 2=, 3=, 4=]{\mthset[#4]{x#3}[#1][#2]}
\newcommandx{\ySet}[4][1=, 2=, 3=, 4=]{\mthset[#4]{y#3}[#1][#2]}
\newcommandx{\zSet}[4][1=, 2=, 3=, 4=]{\mthset[#4]{z#3}[#1][#2]}

\newcommandx{\AFun}[4][1=, 2=, 3=, 4=]{\mthfun[#4]{A#3}[#1][#2]}
\newcommandx{\BFun}[4][1=, 2=, 3=, 4=]{\mthfun[#4]{B#3}[#1][#2]}
\newcommandx{\CFun}[4][1=, 2=, 3=, 4=]{\mthfun[#4]{C#3}[#1][#2]}
\newcommandx{\DFun}[4][1=, 2=, 3=, 4=]{\mthfun[#4]{D#3}[#1][#2]}
\newcommandx{\EFun}[4][1=, 2=, 3=, 4=]{\mthfun[#4]{E#3}[#1][#2]}
\newcommandx{\FFun}[4][1=, 2=, 3=, 4=]{\mthfun[#4]{F#3}[#1][#2]}
\newcommandx{\GFun}[4][1=, 2=, 3=, 4=]{\mthfun[#4]{G#3}[#1][#2]}
\newcommandx{\HFun}[4][1=, 2=, 3=, 4=]{\mthfun[#4]{H#3}[#1][#2]}
\newcommandx{\IFun}[4][1=, 2=, 3=, 4=]{\mthfun[#4]{I#3}[#1][#2]}
\newcommandx{\JFun}[4][1=, 2=, 3=, 4=]{\mthfun[#4]{J#3}[#1][#2]}
\newcommandx{\KFun}[4][1=, 2=, 3=, 4=]{\mthfun[#4]{K#3}[#1][#2]}
\newcommandx{\LFun}[4][1=, 2=, 3=, 4=]{\mthfun[#4]{L#3}[#1][#2]}
\newcommandx{\MFun}[4][1=, 2=, 3=, 4=]{\mthfun[#4]{M#3}[#1][#2]}
\newcommandx{\NFun}[4][1=, 2=, 3=, 4=]{\mthfun[#4]{N#3}[#1][#2]}
\newcommandx{\OFun}[4][1=, 2=, 3=, 4=]{\mthfun[#4]{O#3}[#1][#2]}
\newcommandx{\PFun}[4][1=, 2=, 3=, 4=]{\mthfun[#4]{P#3}[#1][#2]}
\newcommandx{\QFun}[4][1=, 2=, 3=, 4=]{\mthfun[#4]{Q#3}[#1][#2]}
\newcommandx{\RFun}[4][1=, 2=, 3=, 4=]{\mthfun[#4]{R#3}[#1][#2]}
\newcommandx{\SFun}[4][1=, 2=, 3=, 4=]{\mthfun[#4]{S#3}[#1][#2]}
\newcommandx{\TFun}[4][1=, 2=, 3=, 4=]{\mthfun[#4]{T#3}[#1][#2]}
\newcommandx{\UFun}[4][1=, 2=, 3=, 4=]{\mthfun[#4]{U#3}[#1][#2]}
\newcommandx{\VFun}[4][1=, 2=, 3=, 4=]{\mthfun[#4]{V#3}[#1][#2]}
\newcommandx{\WFun}[4][1=, 2=, 3=, 4=]{\mthfun[#4]{W#3}[#1][#2]}
\newcommandx{\XFun}[4][1=, 2=, 3=, 4=]{\mthfun[#4]{X#3}[#1][#2]}
\newcommandx{\YFun}[4][1=, 2=, 3=, 4=]{\mthfun[#4]{Y#3}[#1][#2]}
\newcommandx{\ZFun}[4][1=, 2=, 3=, 4=]{\mthfun[#4]{Z#3}[#1][#2]}

\newcommandx{\aFun}[4][1=, 2=, 3=, 4=]{\mthfun[#4]{a#3}[#1][#2]}
\newcommandx{\bFun}[4][1=, 2=, 3=, 4=]{\mthfun[#4]{b#3}[#1][#2]}
\newcommandx{\cFun}[4][1=, 2=, 3=, 4=]{\mthfun[#4]{c#3}[#1][#2]}
\newcommandx{\dFun}[4][1=, 2=, 3=, 4=]{\mthfun[#4]{d#3}[#1][#2]}
\newcommandx{\eFun}[4][1=, 2=, 3=, 4=]{\mthfun[#4]{e#3}[#1][#2]}
\newcommandx{\fFun}[4][1=, 2=, 3=, 4=]{\mthfun[#4]{f#3}[#1][#2]}
\newcommandx{\gFun}[4][1=, 2=, 3=, 4=]{\mthfun[#4]{g#3}[#1][#2]}
\newcommandx{\hFun}[4][1=, 2=, 3=, 4=]{\mthfun[#4]{h#3}[#1][#2]}
\newcommandx{\iFun}[4][1=, 2=, 3=, 4=]{\mthfun[#4]{i#3}[#1][#2]}
\newcommandx{\jFun}[4][1=, 2=, 3=, 4=]{\mthfun[#4]{j#3}[#1][#2]}
\newcommandx{\kFun}[4][1=, 2=, 3=, 4=]{\mthfun[#4]{k#3}[#1][#2]}
\newcommandx{\lFun}[4][1=, 2=, 3=, 4=]{\mthfun[#4]{l#3}[#1][#2]}
\newcommandx{\mFun}[4][1=, 2=, 3=, 4=]{\mthfun[#4]{m#3}[#1][#2]}
\newcommandx{\nFun}[4][1=, 2=, 3=, 4=]{\mthfun[#4]{n#3}[#1][#2]}
\newcommandx{\oFun}[4][1=, 2=, 3=, 4=]{\mthfun[#4]{o#3}[#1][#2]}
\newcommandx{\pFun}[4][1=, 2=, 3=, 4=]{\mthfun[#4]{p#3}[#1][#2]}
\newcommandx{\qFun}[4][1=, 2=, 3=, 4=]{\mthfun[#4]{q#3}[#1][#2]}
\newcommandx{\rFun}[4][1=, 2=, 3=, 4=]{\mthfun[#4]{r#3}[#1][#2]}
\newcommandx{\sFun}[4][1=, 2=, 3=, 4=]{\mthfun[#4]{s#3}[#1][#2]}
\newcommandx{\tFun}[4][1=, 2=, 3=, 4=]{\mthfun[#4]{t#3}[#1][#2]}
\newcommandx{\uFun}[4][1=, 2=, 3=, 4=]{\mthfun[#4]{u#3}[#1][#2]}
\newcommandx{\vFun}[4][1=, 2=, 3=, 4=]{\mthfun[#4]{v#3}[#1][#2]}
\newcommandx{\wFun}[4][1=, 2=, 3=, 4=]{\mthfun[#4]{w#3}[#1][#2]}
\newcommandx{\xFun}[4][1=, 2=, 3=, 4=]{\mthfun[#4]{x#3}[#1][#2]}
\newcommandx{\yFun}[4][1=, 2=, 3=, 4=]{\mthfun[#4]{y#3}[#1][#2]}
\newcommandx{\zFun}[4][1=, 2=, 3=, 4=]{\mthfun[#4]{z#3}[#1][#2]}

\newcommandx{\ARel}[4][1=, 2=, 3=, 4=]{\mthrel[#4]{A#3}[#1][#2]}
\newcommandx{\BRel}[4][1=, 2=, 3=, 4=]{\mthrel[#4]{B#3}[#1][#2]}
\newcommandx{\CRel}[4][1=, 2=, 3=, 4=]{\mthrel[#4]{C#3}[#1][#2]}
\newcommandx{\DRel}[4][1=, 2=, 3=, 4=]{\mthrel[#4]{D#3}[#1][#2]}
\newcommandx{\ERel}[4][1=, 2=, 3=, 4=]{\mthrel[#4]{E#3}[#1][#2]}
\newcommandx{\FRel}[4][1=, 2=, 3=, 4=]{\mthrel[#4]{F#3}[#1][#2]}
\newcommandx{\GRel}[4][1=, 2=, 3=, 4=]{\mthrel[#4]{G#3}[#1][#2]}
\newcommandx{\HRel}[4][1=, 2=, 3=, 4=]{\mthrel[#4]{H#3}[#1][#2]}
\newcommandx{\IRel}[4][1=, 2=, 3=, 4=]{\mthrel[#4]{I#3}[#1][#2]}
\newcommandx{\JRel}[4][1=, 2=, 3=, 4=]{\mthrel[#4]{J#3}[#1][#2]}
\newcommandx{\KRel}[4][1=, 2=, 3=, 4=]{\mthrel[#4]{K#3}[#1][#2]}
\newcommandx{\LRel}[4][1=, 2=, 3=, 4=]{\mthrel[#4]{L#3}[#1][#2]}
\newcommandx{\MRel}[4][1=, 2=, 3=, 4=]{\mthrel[#4]{M#3}[#1][#2]}
\newcommandx{\NRel}[4][1=, 2=, 3=, 4=]{\mthrel[#4]{N#3}[#1][#2]}
\newcommandx{\ORel}[4][1=, 2=, 3=, 4=]{\mthrel[#4]{O#3}[#1][#2]}
\newcommandx{\PRel}[4][1=, 2=, 3=, 4=]{\mthrel[#4]{P#3}[#1][#2]}
\newcommandx{\QRel}[4][1=, 2=, 3=, 4=]{\mthrel[#4]{Q#3}[#1][#2]}
\newcommandx{\RRel}[4][1=, 2=, 3=, 4=]{\mthrel[#4]{R#3}[#1][#2]}
\newcommandx{\SRel}[4][1=, 2=, 3=, 4=]{\mthrel[#4]{S#3}[#1][#2]}
\newcommandx{\TRel}[4][1=, 2=, 3=, 4=]{\mthrel[#4]{T#3}[#1][#2]}
\newcommandx{\URel}[4][1=, 2=, 3=, 4=]{\mthrel[#4]{U#3}[#1][#2]}
\newcommandx{\VRel}[4][1=, 2=, 3=, 4=]{\mthrel[#4]{V#3}[#1][#2]}
\newcommandx{\WRel}[4][1=, 2=, 3=, 4=]{\mthrel[#4]{W#3}[#1][#2]}
\newcommandx{\XRel}[4][1=, 2=, 3=, 4=]{\mthrel[#4]{X#3}[#1][#2]}
\newcommandx{\YRel}[4][1=, 2=, 3=, 4=]{\mthrel[#4]{Y#3}[#1][#2]}
\newcommandx{\ZRel}[4][1=, 2=, 3=, 4=]{\mthrel[#4]{Z#3}[#1][#2]}

\newcommandx{\aRel}[4][1=, 2=, 3=, 4=]{\mthrel[#4]{a#3}[#1][#2]}
\newcommandx{\bRel}[4][1=, 2=, 3=, 4=]{\mthrel[#4]{b#3}[#1][#2]}
\newcommandx{\cRel}[4][1=, 2=, 3=, 4=]{\mthrel[#4]{c#3}[#1][#2]}
\newcommandx{\dRel}[4][1=, 2=, 3=, 4=]{\mthrel[#4]{d#3}[#1][#2]}
\newcommandx{\eRel}[4][1=, 2=, 3=, 4=]{\mthrel[#4]{e#3}[#1][#2]}
\newcommandx{\fRel}[4][1=, 2=, 3=, 4=]{\mthrel[#4]{f#3}[#1][#2]}
\newcommandx{\gRel}[4][1=, 2=, 3=, 4=]{\mthrel[#4]{g#3}[#1][#2]}
\newcommandx{\hRel}[4][1=, 2=, 3=, 4=]{\mthrel[#4]{h#3}[#1][#2]}
\newcommandx{\iRel}[4][1=, 2=, 3=, 4=]{\mthrel[#4]{i#3}[#1][#2]}
\newcommandx{\jRel}[4][1=, 2=, 3=, 4=]{\mthrel[#4]{j#3}[#1][#2]}
\newcommandx{\kRel}[4][1=, 2=, 3=, 4=]{\mthrel[#4]{k#3}[#1][#2]}
\newcommandx{\lRel}[4][1=, 2=, 3=, 4=]{\mthrel[#4]{l#3}[#1][#2]}
\newcommandx{\mRel}[4][1=, 2=, 3=, 4=]{\mthrel[#4]{m#3}[#1][#2]}
\newcommandx{\nRel}[4][1=, 2=, 3=, 4=]{\mthrel[#4]{n#3}[#1][#2]}
\newcommandx{\oRel}[4][1=, 2=, 3=, 4=]{\mthrel[#4]{o#3}[#1][#2]}
\newcommandx{\pRel}[4][1=, 2=, 3=, 4=]{\mthrel[#4]{p#3}[#1][#2]}
\newcommandx{\qRel}[4][1=, 2=, 3=, 4=]{\mthrel[#4]{q#3}[#1][#2]}
\newcommandx{\rRel}[4][1=, 2=, 3=, 4=]{\mthrel[#4]{r#3}[#1][#2]}
\newcommandx{\sRel}[4][1=, 2=, 3=, 4=]{\mthrel[#4]{s#3}[#1][#2]}
\newcommandx{\tRel}[4][1=, 2=, 3=, 4=]{\mthrel[#4]{t#3}[#1][#2]}
\newcommandx{\uRel}[4][1=, 2=, 3=, 4=]{\mthrel[#4]{u#3}[#1][#2]}
\newcommandx{\vRel}[4][1=, 2=, 3=, 4=]{\mthrel[#4]{v#3}[#1][#2]}
\newcommandx{\wRel}[4][1=, 2=, 3=, 4=]{\mthrel[#4]{w#3}[#1][#2]}
\newcommandx{\xRel}[4][1=, 2=, 3=, 4=]{\mthrel[#4]{x#3}[#1][#2]}
\newcommandx{\yRel}[4][1=, 2=, 3=, 4=]{\mthrel[#4]{y#3}[#1][#2]}
\newcommandx{\zRel}[4][1=, 2=, 3=, 4=]{\mthrel[#4]{z#3}[#1][#2]}

\newcommandx{\ASym}[4][1=, 2=, 3=, 4=]{\mthsym[#4]{A#3}[#1][#2]}
\newcommandx{\BSym}[4][1=, 2=, 3=, 4=]{\mthsym[#4]{B#3}[#1][#2]}
\newcommandx{\CSym}[4][1=, 2=, 3=, 4=]{\mthsym[#4]{C#3}[#1][#2]}
\newcommandx{\DSym}[4][1=, 2=, 3=, 4=]{\mthsym[#4]{D#3}[#1][#2]}
\newcommandx{\ESym}[4][1=, 2=, 3=, 4=]{\mthsym[#4]{E#3}[#1][#2]}
\newcommandx{\FSym}[4][1=, 2=, 3=, 4=]{\mthsym[#4]{F#3}[#1][#2]}
\newcommandx{\GSym}[4][1=, 2=, 3=, 4=]{\mthsym[#4]{G#3}[#1][#2]}
\newcommandx{\HSym}[4][1=, 2=, 3=, 4=]{\mthsym[#4]{H#3}[#1][#2]}
\newcommandx{\ISym}[4][1=, 2=, 3=, 4=]{\mthsym[#4]{I#3}[#1][#2]}
\newcommandx{\JSym}[4][1=, 2=, 3=, 4=]{\mthsym[#4]{J#3}[#1][#2]}
\newcommandx{\KSym}[4][1=, 2=, 3=, 4=]{\mthsym[#4]{K#3}[#1][#2]}
\newcommandx{\LSym}[4][1=, 2=, 3=, 4=]{\mthsym[#4]{L#3}[#1][#2]}
\newcommandx{\MSym}[4][1=, 2=, 3=, 4=]{\mthsym[#4]{M#3}[#1][#2]}
\newcommandx{\NSym}[4][1=, 2=, 3=, 4=]{\mthsym[#4]{N#3}[#1][#2]}
\newcommandx{\OSym}[4][1=, 2=, 3=, 4=]{\mthsym[#4]{O#3}[#1][#2]}
\newcommandx{\PSym}[4][1=, 2=, 3=, 4=]{\mthsym[#4]{P#3}[#1][#2]}
\newcommandx{\QSym}[4][1=, 2=, 3=, 4=]{\mthsym[#4]{Q#3}[#1][#2]}
\newcommandx{\RSym}[4][1=, 2=, 3=, 4=]{\mthsym[#4]{R#3}[#1][#2]}
\newcommandx{\SSym}[4][1=, 2=, 3=, 4=]{\mthsym[#4]{S#3}[#1][#2]}
\newcommandx{\TSym}[4][1=, 2=, 3=, 4=]{\mthsym[#4]{T#3}[#1][#2]}
\newcommandx{\USym}[4][1=, 2=, 3=, 4=]{\mthsym[#4]{U#3}[#1][#2]}
\newcommandx{\VSym}[4][1=, 2=, 3=, 4=]{\mthsym[#4]{V#3}[#1][#2]}
\newcommandx{\WSym}[4][1=, 2=, 3=, 4=]{\mthsym[#4]{W#3}[#1][#2]}
\newcommandx{\XSym}[4][1=, 2=, 3=, 4=]{\mthsym[#4]{X#3}[#1][#2]}
\newcommandx{\YSym}[4][1=, 2=, 3=, 4=]{\mthsym[#4]{Y#3}[#1][#2]}
\newcommandx{\ZSym}[4][1=, 2=, 3=, 4=]{\mthsym[#4]{Z#3}[#1][#2]}

\newcommandx{\aSym}[4][1=, 2=, 3=, 4=]{\mthsym[#4]{a#3}[#1][#2]}
\newcommandx{\bSym}[4][1=, 2=, 3=, 4=]{\mthsym[#4]{b#3}[#1][#2]}
\newcommandx{\cSym}[4][1=, 2=, 3=, 4=]{\mthsym[#4]{c#3}[#1][#2]}
\newcommandx{\dSym}[4][1=, 2=, 3=, 4=]{\mthsym[#4]{d#3}[#1][#2]}
\newcommandx{\eSym}[4][1=, 2=, 3=, 4=]{\mthsym[#4]{e#3}[#1][#2]}
\newcommandx{\fSym}[4][1=, 2=, 3=, 4=]{\mthsym[#4]{f#3}[#1][#2]}
\newcommandx{\gSym}[4][1=, 2=, 3=, 4=]{\mthsym[#4]{g#3}[#1][#2]}
\newcommandx{\hSym}[4][1=, 2=, 3=, 4=]{\mthsym[#4]{h#3}[#1][#2]}
\newcommandx{\iSym}[4][1=, 2=, 3=, 4=]{\mthsym[#4]{i#3}[#1][#2]}
\newcommandx{\jSym}[4][1=, 2=, 3=, 4=]{\mthsym[#4]{j#3}[#1][#2]}
\newcommandx{\kSym}[4][1=, 2=, 3=, 4=]{\mthsym[#4]{k#3}[#1][#2]}
\newcommandx{\lSym}[4][1=, 2=, 3=, 4=]{\mthsym[#4]{l#3}[#1][#2]}
\newcommandx{\mSym}[4][1=, 2=, 3=, 4=]{\mthsym[#4]{m#3}[#1][#2]}
\newcommandx{\nSym}[4][1=, 2=, 3=, 4=]{\mthsym[#4]{n#3}[#1][#2]}
\newcommandx{\oSym}[4][1=, 2=, 3=, 4=]{\mthsym[#4]{o#3}[#1][#2]}
\newcommandx{\pSym}[4][1=, 2=, 3=, 4=]{\mthsym[#4]{p#3}[#1][#2]}
\newcommandx{\qSym}[4][1=, 2=, 3=, 4=]{\mthsym[#4]{q#3}[#1][#2]}
\newcommandx{\rSym}[4][1=, 2=, 3=, 4=]{\mthsym[#4]{r#3}[#1][#2]}
\newcommandx{\sSym}[4][1=, 2=, 3=, 4=]{\mthsym[#4]{s#3}[#1][#2]}
\newcommandx{\tSym}[4][1=, 2=, 3=, 4=]{\mthsym[#4]{t#3}[#1][#2]}
\newcommandx{\uSym}[4][1=, 2=, 3=, 4=]{\mthsym[#4]{u#3}[#1][#2]}
\newcommandx{\vSym}[4][1=, 2=, 3=, 4=]{\mthsym[#4]{v#3}[#1][#2]}
\newcommandx{\wSym}[4][1=, 2=, 3=, 4=]{\mthsym[#4]{w#3}[#1][#2]}
\newcommandx{\xSym}[4][1=, 2=, 3=, 4=]{\mthsym[#4]{x#3}[#1][#2]}
\newcommandx{\ySym}[4][1=, 2=, 3=, 4=]{\mthsym[#4]{y#3}[#1][#2]}
\newcommandx{\zSym}[4][1=, 2=, 3=, 4=]{\mthsym[#4]{z#3}[#1][#2]}

\newcommandx{\AElm}[4][1=, 2=, 3=, 4=]{\mthelm[#4]{A#3}[#1][#2]}
\newcommandx{\BElm}[4][1=, 2=, 3=, 4=]{\mthelm[#4]{B#3}[#1][#2]}
\newcommandx{\CElm}[4][1=, 2=, 3=, 4=]{\mthelm[#4]{C#3}[#1][#2]}
\newcommandx{\DElm}[4][1=, 2=, 3=, 4=]{\mthelm[#4]{D#3}[#1][#2]}
\newcommandx{\EElm}[4][1=, 2=, 3=, 4=]{\mthelm[#4]{E#3}[#1][#2]}
\newcommandx{\FElm}[4][1=, 2=, 3=, 4=]{\mthelm[#4]{F#3}[#1][#2]}
\newcommandx{\GElm}[4][1=, 2=, 3=, 4=]{\mthelm[#4]{G#3}[#1][#2]}
\newcommandx{\HElm}[4][1=, 2=, 3=, 4=]{\mthelm[#4]{H#3}[#1][#2]}
\newcommandx{\IElm}[4][1=, 2=, 3=, 4=]{\mthelm[#4]{I#3}[#1][#2]}
\newcommandx{\JElm}[4][1=, 2=, 3=, 4=]{\mthelm[#4]{J#3}[#1][#2]}
\newcommandx{\KElm}[4][1=, 2=, 3=, 4=]{\mthelm[#4]{K#3}[#1][#2]}
\newcommandx{\LElm}[4][1=, 2=, 3=, 4=]{\mthelm[#4]{L#3}[#1][#2]}
\newcommandx{\MElm}[4][1=, 2=, 3=, 4=]{\mthelm[#4]{M#3}[#1][#2]}
\newcommandx{\NElm}[4][1=, 2=, 3=, 4=]{\mthelm[#4]{N#3}[#1][#2]}
\newcommandx{\OElm}[4][1=, 2=, 3=, 4=]{\mthelm[#4]{O#3}[#1][#2]}
\newcommandx{\PElm}[4][1=, 2=, 3=, 4=]{\mthelm[#4]{P#3}[#1][#2]}
\newcommandx{\QElm}[4][1=, 2=, 3=, 4=]{\mthelm[#4]{Q#3}[#1][#2]}
\newcommandx{\RElm}[4][1=, 2=, 3=, 4=]{\mthelm[#4]{R#3}[#1][#2]}
\newcommandx{\SElm}[4][1=, 2=, 3=, 4=]{\mthelm[#4]{S#3}[#1][#2]}
\newcommandx{\TElm}[4][1=, 2=, 3=, 4=]{\mthelm[#4]{T#3}[#1][#2]}
\newcommandx{\UElm}[4][1=, 2=, 3=, 4=]{\mthelm[#4]{U#3}[#1][#2]}
\newcommandx{\VElm}[4][1=, 2=, 3=, 4=]{\mthelm[#4]{V#3}[#1][#2]}
\newcommandx{\WElm}[4][1=, 2=, 3=, 4=]{\mthelm[#4]{W#3}[#1][#2]}
\newcommandx{\XElm}[4][1=, 2=, 3=, 4=]{\mthelm[#4]{X#3}[#1][#2]}
\newcommandx{\YElm}[4][1=, 2=, 3=, 4=]{\mthelm[#4]{Y#3}[#1][#2]}
\newcommandx{\ZElm}[4][1=, 2=, 3=, 4=]{\mthelm[#4]{Z#3}[#1][#2]}

\newcommandx{\aElm}[4][1=, 2=, 3=, 4=]{\mthelm[#4]{a#3}[#1][#2]}
\newcommandx{\bElm}[4][1=, 2=, 3=, 4=]{\mthelm[#4]{b#3}[#1][#2]}
\newcommandx{\cElm}[4][1=, 2=, 3=, 4=]{\mthelm[#4]{c#3}[#1][#2]}
\newcommandx{\dElm}[4][1=, 2=, 3=, 4=]{\mthelm[#4]{d#3}[#1][#2]}
\newcommandx{\eElm}[4][1=, 2=, 3=, 4=]{\mthelm[#4]{e#3}[#1][#2]}
\newcommandx{\fElm}[4][1=, 2=, 3=, 4=]{\mthelm[#4]{f#3}[#1][#2]}
\newcommandx{\gElm}[4][1=, 2=, 3=, 4=]{\mthelm[#4]{g#3}[#1][#2]}
\newcommandx{\hElm}[4][1=, 2=, 3=, 4=]{\mthelm[#4]{h#3}[#1][#2]}
\newcommandx{\iElm}[4][1=, 2=, 3=, 4=]{\mthelm[#4]{i#3}[#1][#2]}
\newcommandx{\jElm}[4][1=, 2=, 3=, 4=]{\mthelm[#4]{j#3}[#1][#2]}
\newcommandx{\kElm}[4][1=, 2=, 3=, 4=]{\mthelm[#4]{k#3}[#1][#2]}
\newcommandx{\lElm}[4][1=, 2=, 3=, 4=]{\mthelm[#4]{l#3}[#1][#2]}
\newcommandx{\mElm}[4][1=, 2=, 3=, 4=]{\mthelm[#4]{m#3}[#1][#2]}
\newcommandx{\nElm}[4][1=, 2=, 3=, 4=]{\mthelm[#4]{n#3}[#1][#2]}
\newcommandx{\oElm}[4][1=, 2=, 3=, 4=]{\mthelm[#4]{o#3}[#1][#2]}
\newcommandx{\pElm}[4][1=, 2=, 3=, 4=]{\mthelm[#4]{p#3}[#1][#2]}
\newcommandx{\qElm}[4][1=, 2=, 3=, 4=]{\mthelm[#4]{q#3}[#1][#2]}
\newcommandx{\rElm}[4][1=, 2=, 3=, 4=]{\mthelm[#4]{r#3}[#1][#2]}
\newcommandx{\sElm}[4][1=, 2=, 3=, 4=]{\mthelm[#4]{s#3}[#1][#2]}
\newcommandx{\tElm}[4][1=, 2=, 3=, 4=]{\mthelm[#4]{t#3}[#1][#2]}
\newcommandx{\uElm}[4][1=, 2=, 3=, 4=]{\mthelm[#4]{u#3}[#1][#2]}
\newcommandx{\vElm}[4][1=, 2=, 3=, 4=]{\mthelm[#4]{v#3}[#1][#2]}
\newcommandx{\wElm}[4][1=, 2=, 3=, 4=]{\mthelm[#4]{w#3}[#1][#2]}
\newcommandx{\xElm}[4][1=, 2=, 3=, 4=]{\mthelm[#4]{x#3}[#1][#2]}
\newcommandx{\yElm}[4][1=, 2=, 3=, 4=]{\mthelm[#4]{y#3}[#1][#2]}
\newcommandx{\zElm}[4][1=, 2=, 3=, 4=]{\mthelm[#4]{z#3}[#1][#2]}

\newcommand{\apriori}
	{\txtabr{a priori}\xspace}

\newcommand{\eg}
	{\txtabr{e.g.}\xspace}

\newcommand{\ie}
	{\txtabr{i.e.}\xspace}

\newcommand{\vs}
	{\txtabr{vs.}\xspace}

\newcommand{\wrt}
	{\txtabr{w.r.t.}\xspace}

\newcommand{\Wlogx}
	{\txtabr{W.l.o.g.}\xspace}

\newcommandx{\defeq}
	{\ensuremath{\triangleq}}
\newcommandx{\seteq}
	{\ensuremath{:=}}

\newcommand{\cmodels}
	{\:\mthempty{\models}}

\renewcommand{\implies}
	{\ensuremath{\Rightarrow}}

\newcommand{\der}[1]
	{\mthempty{\widehat{#1}}}

\newcommand{\tuple}[1]
	{\ensuremath{\!\argint{\langle}{#1}{\rangle}}}

\newcommand{\tupleb}[2]
	{\tuple{\argb{#1}{#2}}}
\newcommand{\tuplec}[3]
	{\tuple{\argc{#1}{#2}{#3}}}
\newcommand{\tupled}[4]
	{\tuple{\argd{#1}{#2}{#3}{#4}}}
\newcommand{\tuplee}[5]
	{\tuple{\arge{#1}{#2}{#3}{#4}{#5}}}
\newcommand{\tuplef}[6]
	{\tuple{\argf{#1}{#2}{#3}{#4}{#5}{#6}}}
\newcommand{\tupleg}[7]
	{\tuple{\argg{#1}{#2}{#3}{#4}{#5}{#6}{#7}}}
\newcommand{\tupleh}[8]
	{\tuple{\argh{#1}{#2}{#3}{#4}{#5}{#6}{#7}{#8}}}
\newcommand{\tuplei}[9]
	{\tuple{\argi{#1}{#2}{#3}{#4}{#5}{#6}{#7}{#8}{#9}}}

\newcommand{\tuplebx}[2]
	{%
	\def\defarga{#1}%
	\def\defargb{#2}%
	\argsubsup{\tupleauxbx}%
	}
\newcommand{\tuplecx}[3]
	{%
	\def\defarga{#1}%
	\def\defargb{#2}%
	\def\defargc{#3}%
	\argsubsup{\tupleauxcx}%
	}
\newcommand{\tupledx}[4]
	{%
	\def\defarga{#1}%
	\def\defargb{#2}%
	\def\defargc{#3}%
	\def\defargd{#4}%
	\argsubsup{\tupleauxdx}%
	}
\newcommand{\tupleex}[5]
	{%
	\def\defarga{#1}%
	\def\defargb{#2}%
	\def\defargc{#3}%
	\def\defargd{#4}%
	\def\defarge{#5}%
	\argsubsup{\tupleauxex}%
	}
\newcommand{\tuplefx}[6]
	{%
	\def\defarga{#1}%
	\def\defargb{#2}%
	\def\defargc{#3}%
	\def\defargd{#4}%
	\def\defarge{#5}%
	\def\defargf{#6}%
	\argsubsup{\tupleauxfx}%
	}
\newcommand{\tuplegx}[7]
	{%
	\def\defarga{#1}%
	\def\defargb{#2}%
	\def\defargc{#3}%
	\def\defargd{#4}%
	\def\defarge{#5}%
	\def\defargf{#6}%
	\def\defargg{#7}%
	\argsubsup{\tupleauxgx}%
	}

\newcommandx{\tupleauxbx}[2][1=, 2=]
	{%
	\tupleb
		{\argdef{#1}{\defarga[\argsubscript][\argsuperscript]}}
		{\argdef{#2}{\defargb[\argsubscript][\argsuperscript]}}%
	}
\newcommandx{\tupleauxcx}[3][1=, 2=, 3=]
	{%
	\tuplec
		{\argdef{#1}{\defarga[\argsubscript][\argsuperscript]}}
		{\argdef{#2}{\defargb[\argsubscript][\argsuperscript]}}
		{\argdef{#3}{\defargc[\argsubscript][\argsuperscript]}}%
	}
\newcommandx{\tupleauxdx}[4][1=, 2=, 3=, 4=]
	{%
	\tupled
		{\argdef{#1}{\defarga[\argsubscript][\argsuperscript]}}
		{\argdef{#2}{\defargb[\argsubscript][\argsuperscript]}}
		{\argdef{#3}{\defargc[\argsubscript][\argsuperscript]}}
		{\argdef{#4}{\defargd[\argsubscript][\argsuperscript]}}%
	}
\newcommandx{\tupleauxex}[5][1=, 2=, 3=, 4=, 5=]
	{%
	\tuplee
		{\argdef{#1}{\defarga[\argsubscript][\argsuperscript]}}
		{\argdef{#2}{\defargb[\argsubscript][\argsuperscript]}}
		{\argdef{#3}{\defargc[\argsubscript][\argsuperscript]}}
		{\argdef{#4}{\defargd[\argsubscript][\argsuperscript]}}
		{\argdef{#5}{\defarge[\argsubscript][\argsuperscript]}}%
	}
\newcommandx{\tupleauxfx}[6][1=, 2=, 3=, 4=, 5=, 6=]
	{%
	\tuplef
		{\argdef{#1}{\defarga[\argsubscript][\argsuperscript]}}
		{\argdef{#2}{\defargb[\argsubscript][\argsuperscript]}}
		{\argdef{#3}{\defargc[\argsubscript][\argsuperscript]}}
		{\argdef{#4}{\defargd[\argsubscript][\argsuperscript]}}
		{\argdef{#5}{\defarge[\argsubscript][\argsuperscript]}}
		{\argdef{#6}{\defargf[\argsubscript][\argsuperscript]}}%
	}
\newcommandx{\tupleauxgx}[7][1=, 2=, 3=, 4=, 5=, 6=, 7=]
	{%
	\tupleg
		{\argdef{#1}{\defarga[\argsubscript][\argsuperscript]}}
		{\argdef{#2}{\defargb[\argsubscript][\argsuperscript]}}
		{\argdef{#3}{\defargc[\argsubscript][\argsuperscript]}}
		{\argdef{#4}{\defargd[\argsubscript][\argsuperscript]}}
		{\argdef{#5}{\defarge[\argsubscript][\argsuperscript]}}
		{\argdef{#6}{\defargf[\argsubscript][\argsuperscript]}}
		{\argdef{#7}{\defargg[\argsubscript][\argsuperscript]}}%
	}
\newcommandx{\tupleauxhx}[8][1=, 2=, 3=, 4=, 5=, 6=, 7=, 8=]
	{%
	\tupleh
		{\argdef{#1}{\defarga[\argsubscript][\argsuperscript]}}
		{\argdef{#2}{\defargb[\argsubscript][\argsuperscript]}}
		{\argdef{#3}{\defargc[\argsubscript][\argsuperscript]}}
		{\argdef{#4}{\defargd[\argsubscript][\argsuperscript]}}
		{\argdef{#5}{\defarge[\argsubscript][\argsuperscript]}}
		{\argdef{#6}{\defargf[\argsubscript][\argsuperscript]}}
		{\argdef{#7}{\defargg[\argsubscript][\argsuperscript]}}
		{\argdef{#8}{\defargh[\argsubscript][\argsuperscript]}}%
	}
\newcommandx{\tupleauxix}[9][1=, 2=, 3=, 4=, 5=, 6=, 7=, 8=, 9=]
	{%
	\tuplei
		{\argdef{#1}{\defarga[\argsubscript][\argsuperscript]}}
		{\argdef{#2}{\defargb[\argsubscript][\argsuperscript]}}
		{\argdef{#3}{\defargc[\argsubscript][\argsuperscript]}}
		{\argdef{#4}{\defargd[\argsubscript][\argsuperscript]}}
		{\argdef{#5}{\defarge[\argsubscript][\argsuperscript]}}
		{\argdef{#6}{\defargf[\argsubscript][\argsuperscript]}}
		{\argdef{#7}{\defargg[\argsubscript][\argsuperscript]}}
		{\argdef{#8}{\defargh[\argsubscript][\argsuperscript]}}
		{\argdef{#9}{\defargi[\argsubscript][\argsuperscript]}}%
	}

\newcommand{\set}[2]
	{\ensuremath{\argint{\{}{\argext{#1}{\allowbreak:\allowbreak}{#2}}{\}}}}

\newcommand{\pow}[1]
	{\ensuremath{2^{#1}}}

\newcommand{\card}[1]
	{\mthempty{\argint{\vert}{#1}{\vert}}}

\newcommand{\dom}
	{\mthargfun{dom}}

\newcommand{\rng}
	{\mthargfun{rng}}

\newcommand{\cmp}
	{\ensuremath{\circ}}

\newcommand{\rst}
	{\mthempty{\upharpoonright}}

\newcommandx{\pto}[2][1=, 2=]
	{\ensuremath{\rightharpoonup}}

\newcommandx{\cto}[2][1=, 2=]
	{\:\mthempty{\to}[#1][#2]\:}
\newcommandx{\cpto}[2][1=, 2=]
	{\:\mthempty{\pto}[#1][#2]\:}

\newcommand{\emptyfun}
	{\mthempty{\varnothing}}

\newcommand{\AOmicron}
	{\mthargset{O}}

\newcommand{\numco}[2]
	{\mthempty{[\argb{#1}{#2}[\:\!}}

\newcommand{\argset}{Ar}
\newcommandx{\ArgSet}[3][1=, 2=, 3=]
	{\mthset{\argset#3}[#1][#2]}

\newcommand{\argsym}{a}
\newcommandx{\argSym}[3][1=, 2=, 3=]
	{\mthsym{\argsym#3}[#1][#2]}

\newcommand{\argelm}{a}
\newcommandx{\argElm}[3][1=, 2=, 3=]
	{\mthelm{\argelm#3}[#1][#2]}

\newcommand{\relset}{Rl}
\newcommandx{\RelSet}[3][1=, 2=, 3=]
	{\mthset{\relset#3}[#1][#2]}

\newcommand{\relsym}{r}
\newcommandx{\relSym}[3][1=, 2=, 3=]
	{\mthsym{\relsym#3}[#1][#2]}

\newcommand{\relelm}{r}
\newcommandx{\relElm}[3][1=, 2=, 3=]
	{\mthelm{\relelm#3}[#1][#2]}

\newcommand{\argfun}{ar}
\newcommandx{\argFun}[4][1=, 2=, 3=, 4=]
	{\mthargfun{\argfun#4}[#1][#2]{#3}}

\newcommand{\lansig}{LS}
\newcommandx{\LanSig}[5][1=, 2=, 3=, 4=, 5=]
	{\txtargname{\lansig#5{\small\argint{$[$}{#1}{$]$}}}[#2][#3]{#4}\xspace}

\newcommand{\lansigname}{L}
\newcommand{\LanSigName}
	{\mthname{\lansigname}}

\newcommand{\lansigcls}{LS}
\newcommandx{\LanSigCls}[5][1=, 2=, 3=, 4=, 5=]
	{\mthset[#5]{\lansigcls#4\text{\txtname{\small\argint{$[$}{#1}{$]$}}}}[#2]%
	[#3]}

\newcommand{\LanSigStr}
	{\tuplecx{\ArgSet}{\RelSet}{\argFun}}

\newcommand{\domset}{Dm}
\newcommandx{\DomSet}[3][1=, 2=, 3=]
	{\mthset{\domset#3}[#1][#2]}

\newcommand{\domsym}{d}
\newcommandx{\domSym}[3][1=, 2=, 3=]
	{\mthsym{\domsym#3}[#1][#2]}

\newcommand{\domelm}{d}
\newcommandx{\domElm}[3][1=, 2=, 3=]
	{\mthelm{\domelm#3}[#1][#2]}

\newcommand{\relfun}{rl}
\newcommandx{\relFun}[4][1=, 2=, 3=, 4=]
	{\mthargfun{\relfun#4}[#1][#2]{#3}}

\newcommand{\relstr}{RS}
\newcommandx{\RelStr}[5][1=, 2=, 3=, 4=, 5=]
	{\txtargname{\relstr#5{\small\argint{$[$}{#1}{$]$}}}[#2][#3]{#4}\xspace}

\newcommand{\relstrname}{R}
\newcommand{\RelStrName}
	{\mthname{\relstrname}}

\newcommand{\relstrcls}{RS}
\newcommandx{\RelStrCls}[5][1=, 2=, 3=, 4=, 5=]
	{\mthset[#5]{\relstrcls#4\text{\txtname{\small\argint{$[$}{#1}{$]$}}}}[#2]%
	[#3]}

\newcommand{\RelStrStr}
	{\tuplebx{\DomSet}{\relFun}}

\newcommandx{\ordFun}[3][1=, 2=, 3=]
	{\mthempty{\argint{\left\vert}{#3}{\right\vert}}[#1][#2]}

\newcommandx{\sizFun}[3][1=, 2=, 3=]
	{\mthempty{\argint{\left\Vert}{#3}{\right\Vert}}[#1][#2]}

\newcommand{\verset}{Vr}
\newcommandx{\VerSet}[3][1=, 2=, 3=]
	{\mthset{\verset#3}[#1][#2]}

\newcommand{\versym}{v}
\newcommandx{\verSym}[3][1=, 2=, 3=]
	{\mthsym{\versym#3}[#1][#2]}

\newcommand{\verelm}{v}
\newcommandx{\verElm}[3][1=, 2=, 3=]
	{\mthelm{\verelm#3}[#1][#2]}

\newcommand{\edgrel}{Ed}
\newcommandx{\EdgRel}[3][1=, 2=, 3=]
	{\mthrel{\edgrel#3}[#1][#2]}

\newcommand{\edgsym}{e}
\newcommandx{\edgSym}[3][1=, 2=, 3=]
	{\mthsym{\edgsym#3}[#1][#2]}

\newcommand{\edgelm}{e}
\newcommandx{\edgElm}[3][1=, 2=, 3=]
	{\mthelm{\edgelm#3}[#1][#2]}

\newcommand{\orgfun}{or}
\newcommandx{\orgFun}[4][1=, 2=, 3=, 4=]
	{\mthargfun{\orgfun#4}[#1][#2]{#3}}

\newcommand{\desfun}{ds}
\newcommandx{\desFun}[4][1=, 2=, 3=, 4=]
	{\mthargfun{\desfun#4}[#1][#2]{#3}}

\newcommand{\grp}{Gr}
\newcommandx{\Grp}[5][1=, 2=, 3=, 4=, 5=]
	{\txtargname{\grp#5{\small\argint{$[$}{#1}{$]$}}}[#2][#3]{#4}\xspace}

\newcommand{\grpcls}{Gr}
\newcommandx{\GrpCls}[5][1=, 2=, 3=, 4=, 5=]
	{\mthset[#5]{\grpcls#4\text{\small\txtname{\argint{$[$}{#1}{$]$}}}}[#2][#3]}

\newcommand{\GrpStr}
	{\tuplebx{\VerSet}{\EdgRel}}

\newcommand{\pthset}{Pth}
\newcommandx{\PthSet}[3][1=, 2=, 3=]
	{\mthset{\pthset#3}[#1][#2]}

\newcommand{\pthsym}{\pi}
\newcommandx{\pthSym}[3][1=, 2=, 3=]
	{\mthsym{\pthsym#3}[#1][#2]}

\newcommand{\pthelm}{\pi}
\newcommandx{\pthElm}[3][1=, 2=, 3=]
	{\mthelm{\pthelm#3}[#1][#2]}

\newcommand{\colset}{Cl}
\newcommandx{\ColSet}[3][1=, 2=, 3=]
	{\mthset{\colset#3}[#1][#2]}

\newcommand{\colsym}{c}
\newcommandx{\colSym}[3][1=, 2=, 3=]
	{\mthsym{\colsym#3}[#1][#2]}

\newcommand{\colelm}{c}
\newcommandx{\colElm}[3][1=, 2=, 3=]
	{\mthelm{\colelm#3}[#1][#2]}

\newcommand{\colfun}{cl}
\newcommandx{\colFun}[4][1=, 2=, 3=, 4=]
	{\mthargfun{\colfun#4}[#1][#2]{#3}}

\newcommandx{\ColGrpCls}[5][1=, 2=, 3=, 4=, 5=]
	{\mthset[#5]{C}\!\GrpCls[#1][#2][#3][#4][#5]}

\newcommand{\wghset}{Wg}
\newcommandx{\WghSet}[3][1=, 2=, 3=]
	{\mthset{\wghset#3}[#1][#2]}

\newcommand{\wghsym}{w}
\newcommandx{\wghSym}[3][1=, 2=, 3=]
	{\mthsym{\wghsym#3}[#1][#2]}

\newcommand{\wghelm}{w}
\newcommandx{\wghElm}[3][1=, 2=, 3=]
	{\mthelm{\wghelm#3}[#1][#2]}

\newcommand{\wghfun}{wg}
\newcommandx{\wghFun}[4][1=, 2=, 3=, 4=]
	{\mthargfun{\wghfun#4}[#1][#2]{#3}}

\newcommandx{\WghGrpCls}[5][1=, 2=, 3=, 4=, 5=]
	{\mthset[#5]{W}\!\GrpCls[#1][#2][#3][#4][#5]}

\newcommand{\apset}{AP}
\newcommandx{\APSet}[3][1=, 2=, 3=]
	{\mthset{\apset#3}[#1][#2]}

\newcommand{\apsym}{p}
\newcommandx{\apSym}[3][1=, 2=, 3=]
	{\mthsym{\apsym#3}[#1][#2]}

\newcommand{\apelm}{p}
\newcommandx{\apElm}[3][1=, 2=, 3=]
	{\mthelm{\apelm#3}[#1][#2]}

\newcommand{\worset}{W}
\newcommandx{\WorSet}[3][1=, 2=, 3=]
	{\mthset{\worset#3}[#1][#2]}

\newcommand{\worsym}{w}
\newcommandx{\worSym}[3][1=, 2=, 3=]
	{\mthsym{\worsym#3}[#1][#2]}

\newcommand{\worelm}{w}
\newcommandx{\worElm}[3][1=, 2=, 3=]
	{\mthelm{\worelm#3}[#1][#2]}

\newcommand{\trnrel}{R}
\newcommandx{\TrnRel}[3][1=, 2=, 3=]
	{\mthrel{\trnrel#3}[#1][#2]}

\newcommand{\trnsym}{r}
\newcommandx{\trnSym}[3][1=, 2=, 3=]
	{\mthsym{\trnsym#3}[#1][#2]}

\newcommand{\trnelm}{r}
\newcommandx{\trnElm}[3][1=, 2=, 3=]
	{\mthelm{\trnelm#3}[#1][#2]}

\newcommand{\labfun}{L}
\newcommandx{\labFun}[4][1=, 2=, 3=, 4=]
	{\mthargfun{\labfun#4}[#1][#2]{#3}}

\newcommand{\krpstr}{KS}
\newcommandx{\KrpStr}[5][1=, 2=, 3=, 4=, 5=]
	{\txtargname{\krpstr#5{\small\argint{$[$}{#1}{$]$}}}[#2][#3]{#4}\xspace}

\newcommand{\krpstrcls}{KS}
\newcommandx{\KrpStrCls}[5][1=, 2=, 3=, 4=, 5=]
	{\mthset[#5]{\krpstrcls#4\text{\small\txtname{\argint{$[$}{#1}{$]$}}}}[#2]%
	[#3]}

\newcommand{\trkset}{Trk}
\newcommandx{\TrkSet}[3][1=, 2=, 3=]
	{\mthset{\trkset#3}[#1][#2]}

\newcommand{\trksym}{\rho}
\newcommandx{\trkSym}[3][1=, 2=, 3=]
	{\mthsym{\trksym#3}[#1][#2]}

\newcommand{\trkelm}{\rho}
\newcommandx{\trkElm}[3][1=, 2=, 3=]
	{\mthelm{\trkelm#3}[#1][#2]}

\newcommand{\krptree}{KT}
\newcommandx{\KrpTree}[5][1=, 2=, 3=, 4=, 5=]
	{\txtargname{\krptree#5{\small\argint{$[$}{#1}{$]$}}}[#2][#3]{#4}\xspace}

\newcommand{\krptreecls}{KT}
\newcommandx{\KrpTreeCls}[5][1=, 2=, 3=, 4=, 5=]
	{\mthset[#5]{\krptreecls#4\text{\small\txtname{\argint{$[$}{#1}{$]$}}}}[#2]%
	[#3]}

\newcommand{\dirset}{Dir}
\newcommandx{\DirSet}[3][1=, 2=, 3=]
	{\mthset{\dirset#3}[#1][#2]}

\newcommand{\dirsym}{d}
\newcommandx{\dirSym}[3][1=, 2=, 3=]
	{\mthsym{\dirsym#3}[#1][#2]}

\newcommand{\direlm}{d}
\newcommandx{\dirElm}[3][1=, 2=, 3=]
	{\mthelm{\direlm#3}[#1][#2]}

\newcommand{\unwfun}{unw}
\newcommandx{\unwFun}[4][1=, 2=, 3=, 4=]
	{\mthargfun{\unwfun#4}[#1][#2]{#3}}

\newcommand{\congamstr}{CGS}
\newcommandx{\ConGamStr}[5][1=, 2=, 3=, 4=, 5=]
	{\txtargname{\congamstr#5{\small\argint{$[$}{#1}{$]$}}}[#2][#3]{#4}\xspace}

\newcommand{\gamkin}{2PT}

\newcommand{\plrset}{Pl}
\newcommandx{\PlrSet}[3][1=, 2=, 3=]
	{\mthset{\plrset#3}[#1][#2]}

\newcommand{\plrsym}{p}
\newcommandx{\plrSym}[3][1=, 2=, 3=]
	{\mthsym{\plrsym#3}[#1][#2]}

\newcommand{\plrelm}{p}
\newcommandx{\plrElm}[3][1=, 2=, 3=]
	{\mthelm{\plrelm#3}[#1][#2]}

\newcommand{\agnset}{Ag}
\newcommandx{\AgnSet}[3][1=, 2=, 3=]
	{\mthset{\agnset#3}[#1][#2]}

\newcommand{\agnsym}{a}
\newcommandx{\agnSym}[3][1=, 2=, 3=]
	{\mthsym{\agnsym#3}[#1][#2]}

\newcommand{\agnelm}{a}
\newcommandx{\agnElm}[3][1=, 2=, 3=]
	{\mthelm{\agnelm#3}[#1][#2]}

\newcommand{\movset}{Mv}
\newcommandx{\MovSet}[3][1=, 2=, 3=]
	{\mthset{\movset#3}[#1][#2]}

\newcommand{\movrel}{Mv}
\newcommandx{\MovRel}[3][1=, 2=, 3=]
	{\mthrel{\movrel#3}[#1][#2]}

\newcommand{\movsym}{m}
\newcommandx{\movSym}[3][1=, 2=, 3=]
	{\mthsym{\movsym#3}[#1][#2]}

\newcommand{\movelm}{m}
\newcommandx{\movElm}[3][1=, 2=, 3=]
	{\mthelm{\movelm#3}[#1][#2]}

\newcommand{\actset}{Ac}
\newcommandx{\ActSet}[3][1=, 2=, 3=]
	{\mthset{\actset#3}[#1][#2]}

\newcommand{\actrel}{Ac}
\newcommandx{\ActRel}[3][1=, 2=, 3=]
	{\mthrel{\actrel#3}[#1][#2]}

\newcommand{\actsym}{c}
\newcommandx{\actSym}[3][1=, 2=, 3=]
	{\mthsym{\actsym#3}[#1][#2]}

\newcommand{\actelm}{c}
\newcommandx{\actElm}[3][1=, 2=, 3=]
	{\mthelm{\actelm#3}[#1][#2]}

\newcommand{\decset}{Dc}
\newcommandx{\DecSet}[3][1=, 2=, 3=]
	{\mthset{\decset#3}[#1][#2]}

\newcommand{\decsym}{\delta}
\newcommandx{\decSym}[4][1=, 2=, 3=, 4=]
	{\mthargfun{\decsym#4}[#1][#2]{#3}}

\newcommand{\decelm}{\delta}
\newcommandx{\decElm}[4][1=, 2=, 3=, 4=]
	{\mthargfun{\decelm#4}[#1][#2]{#3}}

\newcommand{\posset}{Ps}
\newcommandx{\PosSet}[3][1=, 2=, 3=]
	{\mthset{\posset#3}[#1][#2]}

\newcommand{\possym}{v}
\newcommandx{\posSym}[3][1=, 2=, 3=]
	{\mthsym{\possym#3}[#1][#2]}

\newcommand{\poselm}{v}
\newcommandx{\posElm}[3][1=, 2=, 3=]
	{\mthelm{\poselm#3}[#1][#2]}

\newcommand{\sttset}{St}
\newcommandx{\SttSet}[3][1=, 2=, 3=]
	{\mthset{\sttset#3}[#1][#2]}

\newcommand{\sttsym}{s}
\newcommandx{\sttSym}[3][1=, 2=, 3=]
	{\mthsym{\sttsym#3}[#1][#2]}

\newcommand{\sttelm}{s}
\newcommandx{\sttElm}[3][1=, 2=, 3=]
	{\mthelm{\sttelm#3}[#1][#2]}

\newcommand{\plrfun}{pl}
\newcommandx{\plrFun}[4][1=, 2=, 3=, 4=]
	{\mthargfun{\plrfun#4}[#1][#2]{#3}}

\newcommand{\agnfun}{ag}
\newcommandx{\agnFun}[4][1=, 2=, 3=, 4=]
	{\mthargfun{\agnfun#4}[#1][#2]{#3}}

\newcommand{\movfun}{mv}
\newcommandx{\movFun}[4][1=, 2=, 3=, 4=]
	{\mthargfun{\movfun#4}[#1][#2]{#3}}

\newcommand{\actfun}{ac}
\newcommandx{\actFun}[4][1=, 2=, 3=, 4=]
	{\mthargfun{\actfun#4}[#1][#2]{#3}}

\newcommand{\decfun}{dc}
\newcommandx{\decFun}[4][1=, 2=, 3=, 4=]
	{\mthargfun{\decfun#4}[#1][#2]{#3}}

\newcommand{\trnfun}{tr}
\newcommandx{\trnFun}[4][1=, 2=, 3=, 4=]
	{\mthargfun{\trnfun#4}[#1][#2]{#3}}

\newcommand{\arn}{Ar}
\newcommandx{\Arn}[5][1=, 2=, 3=, 4=, 5=]
	{\txtargname{\arn#5{\small\argint{$[$}{#1}{$]$}}}[#2][#3]{#4}\xspace}

\newcommand{\arnname}{A}
\newcommand{\ArnName}
	{\mthname{\arnname}}

\newcommand{\arncls}{Ar}
\newcommandx{\ArnCls}[5][1=, 2=, 3=, 4=, 5=]
	{\mthset[#5]{\arncls#4\text{\small\txtname{\argint{$[$}{#1}{$]$}}}}[ #2][ #3]}

\newcommand{\ArnStr}[1][]
	{%
	\IfStrEqCase{\argdef{#1}{\gamkin}}
		{%
		{2PT}
			{\tuplecx{\PosSet[0]}{\PosSet[1]}{\MovRel}}%
		{MPC0}
			{\tupledx{\PlrSet}{\MovSet}{\PosSet}{\trnFun}}%
		{MPC1}
			{\tupleex{\PlrSet}{\MovSet}{\PosSet}{\decFun}{\trnFun}}%
		{MPC2}
			{\tuplefx{\PlrSet}{\MovSet}{\PosSet}{\plrFun}{\movFun}{\trnFun}}%
		{MPC3}
			{\tuplegx{\PlrSet}{\MovSet}{\PosSet}{\plrFun}{\movFun}{\decFun}{\trnFun}}%
		{2AT}
			{\tuplecx{\SttSet[0]}{\SttSet[1]}{\ActRel}}%
		{MAC0}
			{\tupledx{\AgnSet}{\ActSet}{\SttSet}{\trnFun}}%
		{MAC1}
			{\tupleex{\AgnSet}{\ActSet}{\SttSet}{\decFun}{\trnFun}}%
		{MAC2}
			{\tuplefx{\AgnSet}{\ActSet}{\SttSet}{\agnFun}{\actFun}{\trnFun}}%
		{MAC3}
			{\tuplegx{\AgnSet}{\ActSet}{\SttSet}{\agnFun}{\actFun}{\decFun}{\trnFun}}%
		}
		[\ensuremath{\clubsuit}]%
	}

\newcommand{\hstset}{Hst}
\newcommandx{\HstSet}[3][1=, 2=, 3=]
	{\mthset{\hstset#3}[#1][#2]}

\newcommand{\hstsym}{\rho}
\newcommandx{\hstSym}[3][1=, 2=, 3=]
	{\mthsym{\hstsym#3}[#1][#2]}

\newcommand{\hstelm}{\rho}
\newcommandx{\hstElm}[3][1=, 2=, 3=]
	{\mthelm{\hstelm#3}[#1][#2]}

\newcommand{\strset}{Str}
\newcommandx{\StrSet}[3][1=, 2=, 3=]
	{\mthset{\strset#3}[#1][#2]}

\newcommand{\strsym}{\sigma}
\newcommandx{\strSym}[4][1=, 2=, 3=, 4=]
	{\mthargfun{\strsym#4}[#1][#2]{#3}}

\newcommand{\strelm}{\sigma}
\newcommandx{\strElm}[4][1=, 2=, 3=, 4=]
	{\mthargfun{\strelm#4}[#1][#2]{#3}}

\newcommand{\prfset}{Prf}
\newcommandx{\PrfSet}[3][1=, 2=, 3=]
	{\mthset{\prfset#3}[#1][#2]}

\newcommand{\prfsym}{\xi}
\newcommandx{\prfSym}[4][1=, 2=, 3=, 4=]
	{\mthargfun{\prfsym#4}[#1][#2]{#3}}

\newcommandx{\prfElm}[4][1=, 2=, 3=, 4=]
	{\mthargfun{\prfsym#4}[#1][#2]{#3}}

\newcommand{\playfun}{play}
\newcommandx{\playFun}[4][1=, 2=, 3=, 4=]
	{\mthargfun{\playfun#4}[#1][#2]{#3}}

\newcommandx{\ColArnCls}[5][1=, 2=, 3=, 4=, 5=]
	{\mthset[#5]{C}\!\ArnCls[#1][#2][#3][#4][#5]}

\newcommandx{\WghArnCls}[5][1=, 2=, 3=, 4=, 5=]
	{\mthset[#5]{W}\!\ArnCls[#1][#2][#3][#4][#5]}

\newcommand{\prdset}{Pr}
\newcommandx{\PrdSet}[3][1=, 2=, 3=]
	{\mthset{\prdset#3}[#1][#2]}

\newcommand{\prdsym}{p}
\newcommandx{\prdSym}[3][1=, 2=, 3=]
	{\mthsym{\prdsym#3}[#1][#2]}

\newcommand{\prdelm}{p}
\newcommandx{\prdElm}[3][1=, 2=, 3=]
	{\mthelm{\prdelm#3}[#1][#2]}

\newcommand{\prdfun}{pr}
\newcommandx{\prdFun}[4][1=, 2=, 3=, 4=]
	{\mthargfun{\prdfun#4}[#1][#2]{#3}}

\newcommand{\extname}{E}
\newcommand{\ExtName}
	{\mthname{\extname}}

\newcommand{\extcls}{Ex}
\newcommandx{\ExtCls}[5][1=, 2=, 3=, 4=, 5=]
	{\mthset[#5]{\extcls#4\text{\small\txtname{\argint{$[$}{#1}{$]$}}}}[#2][#3]}

\newcommand{\tarset}{Tr}
\newcommandx{\TarSet}[3][1=, 2=, 3=]
	{\mthset{\tarset#3}[#1][#2]}

\newcommand{\tarsym}{t}
\newcommandx{\tarSym}[3][1=, 2=, 3=]
	{\mthsym{\tarsym#3}[#1][#2]}

\newcommand{\tarelm}{t}
\newcommandx{\tarElm}[3][1=, 2=, 3=]
	{\mthelm{\tarelm#3}[#1][#2]}

\newcommand{\schrel}{\models}
\newcommandx{\schRel}[4][1=, 2=, 3=, 4=]
	{\mthrel{\schrel#3}[#1][#2]}

\newcommand{\schcls}{Sc}
\newcommandx{\SchCls}[5][1=, 2=, 3=, 4=, 5=]
	{\mthset[#5]{\schcls#4\text{\small\txtname{\argint{$[$}{#1}{$]$}}}}[#2][#3]}

\newcommand{\winset}{Wn}
\newcommandx{\WinSet}[3][1=, 2=, 3=]
	{\mthset{\winset#3}[#1][#2]}

\newcommand{\gamcls}{Gm}
\newcommandx{\GamCls}[5][1=, 2=, 3=, 4=, 5=]
	{\mthset[#5]{\gamcls#4\text{\small\txtname{\argint{$[$}{#1}{$]$}}}}[#2][#3]}

\newcommandx{\GamStr}[3][1=, 2=, 3=]
	{%
	\StrLeft{\argdef{#1}{\gamkin}}{2}[\optgamkin]%
	\def\defposelm{#2}%
	\def\deftarelm{#3}%
	\argsubsup{\GamStrAux}%
	}
\newcommandx{\GamStrAux}[2][1=, 2=]
	{%
	\IfStrEqCase{\optgamkin}
		{%
		{2P}
			{%
			\tuplec
				{\argdef{#1}{\ArnName[\argsubscript][\argsuperscript]}}
				{\argdef{\defposelm}{\posElm[\argsubscript][\argsuperscript]}}
				{\argdef{#2}{\WinSet[\argsubscript][\argsuperscript]}}%
			}%
		{MP}
			{%
			\tuplec
				{\argdef{#1}{\ExtName[\argsubscript][\argsuperscript]}}
				{\argdef{\defposelm}{\posElm[\argsubscript][\argsuperscript]}}
				{\argdef{\deftarelm}{\tarElm[\argsubscript][\argsuperscript]}}%
			}%
		{2A}
			{%
			\tuplec
				{\argdef{#1}{\ArnName[\argsubscript][\argsuperscript]}}
				{\argdef{\defposelm}{\sttElm[\argsubscript][\argsuperscript]}}
				{\argdef{#2}{\WinSet[\argsubscript][\argsuperscript]}}%
			}%
		{MA}
			{%
			\tuplec
				{\argdef{#1}{\ExtName[\argsubscript][\argsuperscript]}}
				{\argdef{\defposelm}{\sttElm[\argsubscript][\argsuperscript]}}
				{\argdef{\deftarelm}{\tarElm[\argsubscript][\argsuperscript]}}%
			}%
		}
		[\ensuremath{\clubsuit}]%
	}

\newcommand{\trntabkin}{D}

\newcommand{\symset}{Sm}
\newcommandx{\SymSet}[3][1=, 2=, 3=]
	{\mthset{\symset#3}[#1][#2]}

\newcommand{\symsym}{\ell}
\newcommandx{\symSym}[3][1=, 2=, 3=]
	{\mthsym{\symsym#3}[#1][#2]}

\newcommand{\symelm}{\ell}
\newcommandx{\symElm}[3][1=, 2=, 3=]
	{\mthelm{\symelm#3}[#1][#2]}

\newcommand{\DSttSet}[1][]
	{\SttSet[\Delta#1]}

\newcommand{\ESttSet}[1][]
	{\SttSet[\exists#1]}

\newcommand{\ASttSet}[1][]
	{\SttSet[\forall#1]}

\newcommand{\trntab}{tt}
\newcommandx{\TrnTab}[5][1=, 2=, 3=, 4=, 5=]
	{\txtargname{\trntab#5{\small\argint{$[$}{#1}{$]$}}}[#2][#3]{#4}\xspace}

\newcommand{\trntabcls}{TT}
\newcommandx{\TrnTabCls}[5][1=, 2=, 3=, 4=, 5=]
	{\mthset[#5]{\trntabcls#4\text{\txtname{\small\argint{$[$}{#1}{$]$}}}}[#2]%
	[#3]}

\newcommand{\TrnTabStr}[1][]
	{%
	\IfStrEqCase{\argdef{#1}{\trntabkin}}
		{%
		{D}{\tuplecx{\SymSet}{\SttSet}{\trnFun}}%
		{N}{\tupledx{\SymSet}{\DSttSet}{\ESttSet}{\trnFun}}%
		{U}{\tupledx{\SymSet}{\DSttSet}{\ASttSet}{\trnFun}}%
		{A}{\tupleex{\SymSet}{\DSttSet}{\ESttSet}{\ASttSet}{\trnFun}}%
		}
		[\ensuremath{\clubsuit}]%
	}

\newcommandx{\FOL}[5][1=, 2=, 3=, 4=, 5=]
	{\txtargname{FOL#5{\small\argint{$[$}{#1}{$]$}}}[#2][#3]{#4}\xspace}

\newcommand{\OBFOL}[1][]
	{\FOL[\argb{1b}{#1}]}

\newcommand{\CBFOL}[1][]
	{\FOL[\argb{cb}{#1}]}

\newcommand{\DBFOL}[1][]
	{\FOL[\argb{db}{#1}]}

\newcommand{\BBFOL}[1][]
	{\FOL[\argb{bb}{#1}]}

\newcommand{\CGFOL}[1][]
	{\FOL[\argb{cg}{#1}]}

\newcommand{\GNFOL}[1][]
	{\FOL[\argb{gn}{#1}]}

\newcommand{\Qnt}
	{\mthsym{Qn\:}}

\newcommand{\Opr}
	{\mthsym{Op\:}}

\newcommand{\Tt}
	{\mthsym{t}}

\newcommand{\Ff}
	{\mthsym{f}}

\newcommand{\varset}{Vr}
\newcommandx{\VarSet}[3][1=, 2=, 3=]
	{\mthset{\varset#3}[#1][#2]}

\newcommand{\varsym}{x}
\newcommandx{\varSym}[3][1=, 2=, 3=]
	{\mthsym{\varsym#3}[#1][#2]}

\newcommand{\varelm}{x}
\newcommandx{\varElm}[3][1=, 2=, 3=]
	{\mthelm{\varelm#3}[#1][#2]}

\newcommand{\varfun}{vr}
\newcommandx{\varFun}[4][1=, 2=, 3=, 4=]
	{\mthargfun{\varfun#4}[#1][#2]{#3}}

\newcommand{\freeFun}
	{\mthargfun{free}}

\newcommand{\qntset}{Qn}
\newcommandx{\QntSet}[3][1=, 2=, 3=]
	{\mthset{\qntset#3}[#1][#2]}

\newcommand{\qntsym}{\wp}
\newcommandx{\qntSym}[3][1=, 2=, 3=]
	{\mthsym{\qntsym#3}[#1][#2]}

\newcommand{\qntelm}{\wp}
\newcommandx{\qntElm}[3][1=, 2=, 3=]
	{\mthelm{\qntelm#3}[#1][#2]}

\newcommand{\bndset}{Bn}
\newcommandx{\BndSet}[3][1=, 2=, 3=]
	{\mthset{\bndset#3}[#1][#2]}

\newcommand{\bndsym}{\flat}
\newcommandx{\bndSym}[3][1=, 2=, 3=]
	{\mthsym{\bndsym#3}[#1][#2]}

\newcommand{\bndelm}{\flat}
\newcommandx{\bndElm}[3][1=, 2=, 3=]
	{\mthelm{\bndelm#3}[#1][#2]}

\newcommand{\depset}{\Delta}
\newcommandx{\DepSet}[3][1=, 2=, 3=]
	{\mthset{\depset#3}[#1][#2]}

\newcommand{\asgset}{Asg}
\newcommandx{\AsgSet}[3][1=, 2=, 3=]
	{\mthset{\asgset#3}[#1][#2]}

\newcommand{\asgfun}{\chi}
\newcommandx{\asgFun}[4][1=, 2=, 3=, 4=]
	{\mthargfun{\asgfun#4}[#1][#2]{#3}}

\newcommand{\smset}{SM}
\newcommandx{\SMSet}[3][1=, 2=, 3=]
	{\mthset{\smset#3}[#1][#2]}

\newcommand{\smfun}{\delta}
\newcommandx{\smFun}[4][1=, 2=, 3=, 4=]
	{\mthargfun{\smfun#4}[#1][#2]{#3}}

\newcommand{\cmset}{CM}
\newcommandx{\CMSet}[3][1=, 2=, 3=]
	{\mthset{\cmset#3}[#1][#2]}

\newcommand{\cmfun}{\gamma}
\newcommandx{\cmFun}[4][1=, 2=, 3=, 4=]
	{\mthargfun{\cmfun#4}[#1][#2]{#3}}

\newcommand{\schset}{Sch}
\newcommandx{\SchSet}[3][1=, 2=, 3=]
	{\mthset{\schset#3}[#1][#2]}

\newcommand{\schsym}{\sigma}
\newcommandx{\schSym}[3][1=, 2=, 3=]
	{\mthsym{\schsym#3}[#1][#2]}

\newcommand{\schelm}{\sigma}
\newcommandx{\schElm}[3][1=, 2=, 3=]
	{\mthelm{\schelm#3}[#1][#2]}

\newcommand{\entset}{Ent}
\newcommandx{\EntSet}[4][1=, 2=, 3=, 4=]
	{\mthset{\entset#4}[#1][#2]{#3}}

\newcommand{\entfun}{ent}
\newcommandx{\entFun}[4][1=, 2=, 3=, 4=]
	{\mthargfun{\entfun#4}[#1][#2]{#3}}

\newcommandx{\SOL}[5][1=, 2=, 3=, 4=, 5=]
	{\txtargname{SOL#5{\small\argint{$[$}{#1}{$]$}}}[#2][#3]{#4}\xspace}

\newcommandx{\TL}[5][1=, 2=, 3=, 4=, 5=]
	{\txtargname{TL#5{\small\argint{$[$}{#1}{$]$}}}[#2][#3]{#4}\xspace}

\newcommandx{\PL}[5][1=, 2=, 3=, 4=, 5=]
	{\txtargname{PL#5{\small\argint{$[$}{#1}{$]$}}}[#2][#3]{#4}\xspace}

\newcommand{\fvarset}{FVr}
\newcommandx{\FVarSet}[3][1=, 2=, 3=]
	{\mthset{\fvarset#3}[#1][#2]}

\newcommand{\fvarsym}{x}
\newcommandx{\fvarSym}[3][1=, 2=, 3=]
	{\mthsym{\fvarsym#3}[#1][#2]}

\newcommand{\fvarelm}{x}
\newcommandx{\fvarElm}[3][1=, 2=, 3=]
	{\mthelm{\fvarelm#3}[#1][#2]}

\newcommand{\fvarfun}{fvr}
\newcommandx{\fvarFun}[4][1=, 2=, 3=, 4=]
	{\mthargfun{\fvarfun#4}[#1][#2]{#3}}

\newcommand{\svarset}{SVr}
\newcommandx{\SVarSet}[3][1=, 2=, 3=]
	{\mthset{\svarset#3}[#1][#2]}

\newcommand{\svarsym}{X}
\newcommandx{\svarSym}[3][1=, 2=, 3=]
	{\mthsym{\svarsym#3}[#1][#2]}

\newcommand{\svarelm}{X}
\newcommandx{\svarElm}[3][1=, 2=, 3=]
	{\mthelm{\svarelm#3}[#1][#2]}

\newcommand{\svarfun}{svr}
\newcommandx{\svarFun}[4][1=, 2=, 3=, 4=]
	{\mthargfun{\svarfun#4}[#1][#2]{#3}}

\newcommandx{\MuCalculus}[5][1=, 2=, 3=, 4=, 5=]
	{\txtargname{$\mu$Calculus#5{\small\argint{$[$}{#1}{$]$}}}[#2][#3]{#4}\xspace}

\newcommandx{\LTL}[5][1=, 2=, 3=, 4=, 5=]
	{\txtargname{LTL#5{\small\argint{$[$}{#1}{$]$}}}[#2][#3]{#4}\xspace}

\newcommandx{\PTL}[5][1=, 2=, 3=, 4=, 5=]
	{\txtargname{PTL#5{\small\argint{$[$}{#1}{$]$}}}[#2][#3]{#4}\xspace}

\newcommandx{\CTL}[5][1=, 2=, 3=, 4=, 5=]
	{\txtargname{CTL#5{\small\argint{$[$}{#1}{$]$}}}[#2][#3]{#4}\xspace}

\newcommandx{\CTLP}[5][1=, 2=, 3=, 4=, 5=]
	{\txtargname{CTL$^{+}$#5{\small\argint{$[$}{#1}{$]$}}}[#2][#3]{#4}\xspace}

\newcommandx{\CTLS}[5][1=, 2=, 3=, 4=, 5=]
	{\txtargname{CTL$^{\star}$#5{\small\argint{$[$}{#1}{$]$}}}[#2][#3]{#4}\xspace}

\newcommandx{\STL}[5][1=, 2=, 3=, 4=, 5=]
	{\txtargname{STL#5{\small\argint{$[$}{#1}{$]$}}}[#2][#3]{#4}\xspace}

\newcommandx{\STLP}[5][1=, 2=, 3=, 4=, 5=]
	{\txtargname{STL$^{+}$#5{\small\argint{$[$}{#1}{$]$}}}[#2][#3]{#4}\xspace}

\newcommandx{\STLS}[5][1=, 2=, 3=, 4=, 5=]
	{\txtargname{STL$^{\star}$#5{\small\argint{$[$}{#1}{$]$}}}[#2][#3]{#4}\xspace}

\newcommandx{\ATL}[5][1=, 2=, 3=, 4=, 5=]
	{\txtargname{ATL#5{\small\argint{$[$}{#1}{$]$}}}[#2][#3]{#4}\xspace}

\newcommandx{\ATLP}[5][1=, 2=, 3=, 4=, 5=]
	{\txtargname{ATL$^{+}$#5{\small\argint{$[$}{#1}{$]$}}}[#2][#3]{#4}\xspace}

\newcommandx{\ATLS}[5][1=, 2=, 3=, 4=, 5=]
	{\txtargname{ATL$^{\star}$#5{\small\argint{$[$}{#1}{$]$}}}[#2][#3]{#4}\xspace}

\newcommandx{\SL}[5][1=, 2=, 3=, 4=, 5=]
	{\txtargname{SL#5{\small\argint{$[$}{#1}{$]$}}}[#2][#3]{#4}\xspace}

\newcommand{\OGSL}[1][]
	{\SL[\argb{1g}{#1}]}

\newcommand{\EExs}[1]
	{\ensuremath{%
	\argint{\mbox{$\langle\!\langle$}}{#1}{\mbox{$\rangle\!\rangle$}}%
	}}

\newcommand{\AAll}[1]
	{\ensuremath{\argint{\mbox{$[\:\!\![$}}{#1}{\mbox{$]\:\!\!]$}}}}

\providecommandx{\EF}[5][1=, 2=, 3=, 4=, 5=]
	{\txtargname{EF#5{\small\argint{$[$}{#1}{$]$}}}[#2][#3]{#4}\xspace}

\providecommandx{\SG}[5][1=, 2=, 3=, 4=, 5=]
	{\txtargname{SG#5{\small\argint{$[$}{#1}{$]$}}}[#2][#3]{#4}\xspace}

\newcommandx{\LogTime}[4][1=, 2=, 3=, 4=]
	{\txtargname{LogTime#4}[#2][#3]{#1}\xspace}
\newcommandx{\LogTimeH}[4][1=, 2=, 3=, 4=]
	{\LogTime[#1][#2][#3][#4]-\HComp}
\newcommandx{\LogTimeE}[4][1=, 2=, 3=, 4=]
	{\LogTime[#1][#2][#3][#4]-\EComp}
\newcommandx{\LogTimeC}[4][1=, 2=, 3=, 4=]
	{\LogTime[#1][#2][#3][#4]-\CComp}

\newcommand{\NLogTime}
	{\txtname{N}\LogTime}
\newcommandx{\NLogTimeH}[4][1=, 2=, 3=, 4=]
	{\NLogTime[#1][#2][#3][#4]-\HComp}
\newcommandx{\NLogTimeE}[4][1=, 2=, 3=, 4=]
	{\NLogTime[#1][#2][#3][#4]-\EComp}
\newcommandx{\NLogTimeC}[4][1=, 2=, 3=, 4=]
	{\NLogTime[#1][#2][#3][#4]-\CComp}

\newcommand{\CoNLogTime}
	{\txtname{Co}\NLogTime}
\newcommandx{\CoNLogTimeH}[4][1=, 2=, 3=, 4=]
	{\CoNLogTime[#1][#2][#3][#4]-\HComp}
\newcommandx{\CoNLogTimeE}[4][1=, 2=, 3=, 4=]
	{\CoNLogTime[#1][#2][#3][#4]-\EComp}
\newcommandx{\CoNLogTimeC}[4][1=, 2=, 3=, 4=]
	{\CoNLogTime[#1][#2][#3][#4]-\CComp}

\newcommandx{\ALogTime}
	{\txtname{A}\LogTime}
\newcommandx{\ALogTimeH}[4][1=, 2=, 3=, 4=]
	{\ALogTime[#1][#2][#3][#4]-\HComp}
\newcommandx{\ALogTimeE}[4][1=, 2=, 3=, 4=]
	{\ALogTime[#1][#2][#3][#4]-\EComp}
\newcommandx{\ALogTimeC}[4][1=, 2=, 3=, 4=]
	{\ALogTime[#1][#2][#3][#4]-\CComp}

\newcommandx{\LogSpace}[4][1=, 2=, 3=, 4=]
	{\txtargname{LogSpace#4}[#2][#3]{#1}\xspace}
\newcommandx{\LogSpaceH}[4][1=, 2=, 3=, 4=]
	{\LogSpace[#1][#2][#3][#4]-\HComp}
\newcommandx{\LogSpaceE}[4][1=, 2=, 3=, 4=]
	{\LogSpace[#1][#2][#3][#4]-\EComp}
\newcommandx{\LogSpaceC}[4][1=, 2=, 3=, 4=]
	{\LogSpace[#1][#2][#3][#4]-\CComp}

\newcommandx{\NLogSpace}
	{\txtname{N}\LogSpace}
\newcommandx{\NLogSpaceH}[4][1=, 2=, 3=, 4=]
	{\NLogSpace[#1][#2][#3][#4]-\HComp}
\newcommandx{\NLogSpaceE}[4][1=, 2=, 3=, 4=]
	{\NLogSpace[#1][#2][#3][#4]-\EComp}
\newcommandx{\NLogSpaceC}[4][1=, 2=, 3=, 4=]
	{\NLogSpace[#1][#2][#3][#4]-\CComp}

\newcommandx{\CoNLogSpace}
	{\txtname{Co}\NLogSpace}
\newcommandx{\CoNLogSpaceH}[4][1=, 2=, 3=, 4=]
	{\CoNLogSpace[#1][#2][#3][#4]-\HComp}
\newcommandx{\CoNLogSpaceE}[4][1=, 2=, 3=, 4=]
	{\CoNLogSpace[#1][#2][#3][#4]-\EComp}
\newcommandx{\CoNLogSpaceC}[4][1=, 2=, 3=, 4=]
	{\CoNLogSpace[#1][#2][#3][#4]-\CComp}

\newcommandx{\ALogSpace}
	{\txtname{A}\LogSpace}
\newcommandx{\ALogSpaceH}[4][1=, 2=, 3=, 4=]
	{\ALogSpace[#1][#2][#3][#4]-\HComp}
\newcommandx{\ALogSpaceE}[4][1=, 2=, 3=, 4=]
	{\ALogSpace[#1][#2][#3][#4]-\EComp}
\newcommandx{\ALogSpaceC}[4][1=, 2=, 3=, 4=]
	{\ALogSpace[#1][#2][#3][#4]-\CComp}

\newcommandx{\PTime}[4][1=, 2=, 3=, 4=]
	{\txtargname{PTime#4}[#2][#3]{#1}\xspace}
\newcommandx{\PTimeH}[4][1=, 2=, 3=, 4=]
	{\PTime[#1][#2][#3][#4]-\HComp}
\newcommandx{\PTimeE}[4][1=, 2=, 3=, 4=]
	{\PTime[#1][#2][#3][#4]-\EComp}
\newcommandx{\PTimeC}[4][1=, 2=, 3=, 4=]
	{\PTime[#1][#2][#3][#4]-\CComp}

\newcommandx{\UPTime}
	{\txtname{U}\PTime}
\newcommandx{\UPTimeH}[4][1=, 2=, 3=, 4=]
	{\UPTime[#1][#2][#3][#4]-\HComp}
\newcommandx{\UPTimeE}[4][1=, 2=, 3=, 4=]
	{\UPTime[#1][#2][#3][#4]-\EComp}
\newcommandx{\UPTimeC}[4][1=, 2=, 3=, 4=]
	{\UPTime[#1][#2][#3][#4]-\CComp}

\newcommandx{\CoUPTime}
	{\txtname{Co}\UPTime}
\newcommandx{\CoUPTimeH}[4][1=, 2=, 3=, 4=]
	{\CoUPTime[#1][#2][#3][#4]-\HComp}
\newcommandx{\CoUPTimeE}[4][1=, 2=, 3=, 4=]
	{\CoUPTime[#1][#2][#3][#4]-\EComp}
\newcommandx{\CoUPTimeC}[4][1=, 2=, 3=, 4=]
	{\CoUPTime[#1][#2][#3][#4]-\CComp}

\newcommandx{\NPTime}
	{\txtname{N}\PTime}
\newcommandx{\NPTimeH}[4][1=, 2=, 3=, 4=]
	{\NPTime[#1][#2][#3][#4]-\HComp}
\newcommandx{\NPTimeE}[4][1=, 2=, 3=, 4=]
	{\NPTime[#1][#2][#3][#4]-\EComp}
\newcommandx{\NPTimeC}[4][1=, 2=, 3=, 4=]
	{\NPTime[#1][#2][#3][#4]-\CComp}

\newcommandx{\CoNPTime}
	{\txtname{Co}\NPTime}
\newcommandx{\CoNPTimeH}[4][1=, 2=, 3=, 4=]
	{\CoNPTime[#1][#2][#3][#4]-\HComp}
\newcommandx{\CoNPTimeE}[4][1=, 2=, 3=, 4=]
	{\CoNPTime[#1][#2][#3][#4]-\EComp}
\newcommandx{\CoNPTimeC}[4][1=, 2=, 3=, 4=]
	{\CoNPTime[#1][#2][#3][#4]-\CComp}

\newcommandx{\APTime}
	{\txtname{A}\PTime}
\newcommandx{\APTimeH}[4][1=, 2=, 3=, 4=]
	{\APTime[#1][#2][#3][#4]-\HComp}
\newcommandx{\APTimeE}[4][1=, 2=, 3=, 4=]
	{\APTime[#1][#2][#3][#4]-\EComp}
\newcommandx{\APTimeC}[4][1=, 2=, 3=, 4=]
	{\APTime[#1][#2][#3][#4]-\CComp}

\newcommandx{\PSpace}[4][1=, 2=, 3=, 4=]
	{\txtargname{PSpace#4}[#2][#3]{#1}\xspace}
\newcommandx{\PSpaceH}[4][1=, 2=, 3=, 4=]
	{\PSpace[#1][#2][#3][#4]-\HComp}
\newcommandx{\PSpaceE}[4][1=, 2=, 3=, 4=]
	{\PSpace[#1][#2][#3][#4]-\EComp}
\newcommandx{\PSpaceC}[4][1=, 2=, 3=, 4=]
	{\PSpace[#1][#2][#3][#4]-\CComp}

\newcommandx{\NPSpace}
	{\txtname{N}\PSpace}
\newcommandx{\NPSpaceH}[4][1=, 2=, 3=, 4=]
	{\NPSpace[#1][#2][#3][#4]-\HComp}
\newcommandx{\NPSpaceE}[4][1=, 2=, 3=, 4=]
	{\NPSpace[#1][#2][#3][#4]-\EComp}
\newcommandx{\NPSpaceC}[4][1=, 2=, 3=, 4=]
	{\NPSpace[#1][#2][#3][#4]-\CComp}

\newcommandx{\CoNPSpace}
	{\txtname{Co}\NPSpace}
\newcommandx{\CoNPSpaceH}[4][1=, 2=, 3=, 4=]
	{\CoNPSpace[#1][#2][#3][#4]-\HComp}
\newcommandx{\CoNPSpaceE}[4][1=, 2=, 3=, 4=]
	{\CoNPSpace[#1][#2][#3][#4]-\EComp}
\newcommandx{\CoNPSpaceC}[4][1=, 2=, 3=, 4=]
	{\CoNPSpace[#1][#2][#3][#4]-\CComp}

\newcommandx{\APSpace}
	{\txtname{A}\PSpace}
\newcommandx{\APSpaceH}[4][1=, 2=, 3=, 4=]
	{\APSpace[#1][#2][#3][#4]-\HComp}
\newcommandx{\APSpaceE}[4][1=, 2=, 3=, 4=]
	{\APSpace[#1][#2][#3][#4]-\EComp}
\newcommandx{\APSpaceC}[4][1=, 2=, 3=, 4=]
	{\APSpace[#1][#2][#3][#4]-\CComp}

\newcommandx{\ExpTime}[4][1=, 2=, 3=, 4=]
	{\txtargname{ExpTime#4}[#2][#3]{#1}\xspace}
\newcommandx{\ExpTimeH}[4][1=, 2=, 3=, 4=]
	{\ExpTime[#1][#2][#3][#4]-\HComp}
\newcommandx{\ExpTimeE}[4][1=, 2=, 3=, 4=]
	{\ExpTime[#1][#2][#3][#4]-\EComp}
\newcommandx{\ExpTimeC}[4][1=, 2=, 3=, 4=]
	{\ExpTime[#1][#2][#3][#4]-\CComp}

\newcommandx{\NExpTime}
	{\txtname{N}\ExpTime}
\newcommandx{\NExpTimeH}[4][1=, 2=, 3=, 4=]
	{\NExpTime[#1][#2][#3][#4]-\HComp}
\newcommandx{\NExpTimeE}[4][1=, 2=, 3=, 4=]
	{\NExpTime[#1][#2][#3][#4]-\EComp}
\newcommandx{\NExpTimeC}[4][1=, 2=, 3=, 4=]
	{\NExpTime[#1][#2][#3][#4]-\CComp}

\newcommandx{\CoNExpTime}
	{\txtname{Co}\NExpTime}
\newcommandx{\CoNExpTimeH}[4][1=, 2=, 3=, 4=]
	{\CoNExpTime[#1][#2][#3][#4]-\HComp}
\newcommandx{\CoNExpTimeE}[4][1=, 2=, 3=, 4=]
	{\CoNExpTime[#1][#2][#3][#4]-\EComp}
\newcommandx{\CoNExpTimeC}[4][1=, 2=, 3=, 4=]
	{\CoNExpTime[#1][#2][#3][#4]-\CComp}

\newcommandx{\AExpTime}
	{\txtname{A}\ExpTime}
\newcommandx{\AExpTimeH}[4][1=, 2=, 3=, 4=]
	{\AExpTime[#1][#2][#3][#4]-\HComp}
\newcommandx{\AExpTimeE}[4][1=, 2=, 3=, 4=]
	{\AExpTime[#1][#2][#3][#4]-\EComp}
\newcommandx{\AExpTimeC}[4][1=, 2=, 3=, 4=]
	{\AExpTime[#1][#2][#3][#4]-\CComp}

\newcommandx{\ExpSpace}[4][1=, 2=, 3=, 4=]
	{\txtargname{ExpSpace#4}[#2][#3]{#1}\xspace}
\newcommandx{\ExpSpaceH}[4][1=, 2=, 3=, 4=]
	{\ExpSpace[#1][#2][#3][#4]-\HComp}
\newcommandx{\ExpSpaceE}[4][1=, 2=, 3=, 4=]
	{\ExpSpace[#1][#2][#3][#4]-\EComp}
\newcommandx{\ExpSpaceC}[4][1=, 2=, 3=, 4=]
	{\ExpSpace[#1][#2][#3][#4]-\CComp}

\newcommandx{\NExpSpace}
	{\txtname{N}\ExpSpace}
\newcommandx{\NExpSpaceH}[4][1=, 2=, 3=, 4=]
	{\NExpSpace[#1][#2][#3][#4]-\HComp}
\newcommandx{\NExpSpaceE}[4][1=, 2=, 3=, 4=]
	{\NExpSpace[#1][#2][#3][#4]-\EComp}
\newcommandx{\NExpSpaceC}[4][1=, 2=, 3=, 4=]
	{\NExpSpace[#1][#2][#3][#4]-\CComp}

\newcommandx{\CoNExpSpace}
	{\txtname{Co}\NExpSpace}
\newcommandx{\CoNExpSpaceH}[4][1=, 2=, 3=, 4=]
	{\CoNExpSpace[#1][#2][#3][#4]-\HComp}
\newcommandx{\CoNExpSpaceE}[4][1=, 2=, 3=, 4=]
	{\CoNExpSpace[#1][#2][#3][#4]-\EComp}
\newcommandx{\CoNExpSpaceC}[4][1=, 2=, 3=, 4=]
	{\CoNExpSpace[#1][#2][#3][#4]-\CComp}

\newcommandx{\AExpSpace}
	{\txtname{A}\ExpSpace}
\newcommandx{\AExpSpaceH}[4][1=, 2=, 3=, 4=]
	{\AExpSpace[#1][#2][#3][#4]-\HComp}
\newcommandx{\AExpSpaceE}[4][1=, 2=, 3=, 4=]
	{\AExpSpace[#1][#2][#3][#4]-\EComp}
\newcommandx{\AExpSpaceC}[4][1=, 2=, 3=, 4=]
	{\AExpSpace[#1][#2][#3][#4]-\CComp}

\newcommandx{\NonElmTime}[4][1=, 2=, 3=, 4=]
	{\txtargname{NonElementaryTime#4}[#2][#3]{#1}\xspace}
\newcommandx{\NonElmTimeH}[4][1=, 2=, 3=, 4=]
	{\NonElmTime[#1][#2][#3][#4]-\HComp}
\newcommandx{\NonElmTimeE}[4][1=, 2=, 3=, 4=]
	{\NonElmTime[#1][#2][#3][#4]-\EComp}
\newcommandx{\NonElmTimeC}[4][1=, 2=, 3=, 4=]
	{\NonElmTime[#1][#2][#3][#4]-\CComp}

\newcommandx{\NonElmSpace}[4][1=, 2=, 3=, 4=]
	{\txtargname{NonElementarySpace#4}[#2][#3]{#1}\xspace}
\newcommandx{\NonElmSpaceH}[4][1=, 2=, 3=, 4=]
	{\NonElmSpace[#1][#2][#3][#4]-\HComp}
\newcommandx{\NonElmSpaceE}[4][1=, 2=, 3=, 4=]
	{\NonElmSpace[#1][#2][#3][#4]-\EComp}
\newcommandx{\NonElmSpaceC}[4][1=, 2=, 3=, 4=]
	{\NonElmSpace[#1][#2][#3][#4]-\CComp}

\newcommandx{\DLHier}[4][2=, 3=, 4=]
	{\mthargset[0]{\Delta#4}[#1][#3]{#2}\xspace}
\newcommandx{\DLHierH}[4][2=, 3=, 4=]
	{\DLHier{#1}[#2][#3][#4]-\HComp}
\newcommandx{\DLHierE}[4][2=, 3=, 4=]
	{\DLHier{#1}[#2][#3][#4]-\EComp}
\newcommandx{\DLHierC}[4][2=, 3=, 4=]
	{\DLHier{#1}[#2][#3][#4]-\CComp}

\newcommandx{\ELHier}[4][2=, 3=, 4=]
	{\mthargset[0]{\Sigma#4}[#1][#3]{#2}\xspace}
\newcommandx{\ELHierH}[4][2=, 3=, 4=]
	{\ELHier{#1}[#2][#3][#4]-\HComp}
\newcommandx{\ELHierE}[4][2=, 3=, 4=]
	{\ELHier{#1}[#2][#3][#4]-\EComp}
\newcommandx{\ELHierC}[4][2=, 3=, 4=]
	{\ELHier{#1}[#2][#3][#4]-\CComp}

\newcommandx{\ULHier}[4][2=, 3=, 4=]
	{\mthargset[0]{\Pi#4}[#1][#3]{#2}\xspace}
\newcommandx{\ULHierH}[4][2=, 3=, 4=]
	{\ULHier{#1}[#2][#3][#4]-\HComp}
\newcommandx{\ULHierE}[4][2=, 3=, 4=]
	{\ULHier{#1}[#2][#3][#4]-\EComp}
\newcommandx{\ULHierC}[4][2=, 3=, 4=]
	{\ULHier{#1}[#2][#3][#4]-\CComp}

\newcommandx{\DBHier}[4][2=, 3=, 4=]
	{\mthargset[3]{\Delta#4}[#1][#3]{#2}\xspace}
\newcommandx{\DBHierH}[4][2=, 3=, 4=]
	{\DBHier{#1}[#2][#3][#4]-\HComp}
\newcommandx{\DBHierE}[4][2=, 3=, 4=]
	{\DBHier{#1}[#2][#3][#4]-\EComp}
\newcommandx{\DBHierC}[4][2=, 3=, 4=]
	{\DBHier{#1}[#2][#3][#4]-\CComp}

\newcommandx{\EBHier}[4][2=, 3=, 4=]
	{\mthargset[3]{\Sigma#4}[#1][#3]{#2}\xspace}
\newcommandx{\EBHierH}[4][2=, 3=, 4=]
	{\EBHier{#1}[#2][#3][#4]-\HComp}
\newcommandx{\EBHierE}[4][2=, 3=, 4=]
	{\EBHier{#1}[#2][#3][#4]-\EComp}
\newcommandx{\EBHierC}[4][2=, 3=, 4=]
	{\EBHier{#1}[#2][#3][#4]-\CComp}

\newcommandx{\UBHier}[4][2=, 3=, 4=]
	{\mthargset[3]{\Pi#4}[#1][#3]{#2}\xspace}
\newcommandx{\UBHierH}[4][2=, 3=, 4=]
	{\UBHier{#1}[#2][#3][#4]-\HComp}
\newcommandx{\UBHierE}[4][2=, 3=, 4=]
	{\UBHier{#1}[#2][#3][#4]-\EComp}
\newcommandx{\UBHierC}[4][2=, 3=, 4=]
	{\UBHier{#1}[#2][#3][#4]-\CComp}

\newcommandx{\DPolHier}[4][2=, 3=, 4=]
	{\DLHier{#1}[#2][\argb{\mathrm{P}}{#3}][#4]}
\newcommandx{\DPolHierH}[4][2=, 3=, 4=]
	{\DPolHier{#1}[#2][#3][#4]-\HComp}
\newcommandx{\DPolHierE}[4][2=, 3=, 4=]
	{\DPolHier{#1}[#2][#3][#4]-\EComp}
\newcommandx{\DPolHierC}[4][2=, 3=, 4=]
	{\DPolHier{#1}[#2][#3][#4]-\CComp}

\newcommandx{\EPolHier}[4][2=, 3=, 4=]
	{\ELHier{#1}[#2][\argb{\mathrm{P}}{#3}][#4]}
\newcommandx{\EPolHierH}[4][2=, 3=, 4=]
	{\EPolHier{#1}[#2][#3][#4]-\HComp}
\newcommandx{\EPolHierE}[4][2=, 3=, 4=]
	{\EPolHier{#1}[#2][#3][#4]-\EComp}
\newcommandx{\EPolHierC}[4][2=, 3=, 4=]
	{\EPolHier{#1}[#2][#3][#4]-\CComp}

\newcommandx{\UPolHier}[4][2=, 3=, 4=]
	{\ULHier{#1}[#2][\argb{\mathrm{P}}{#3}][#4]}
\newcommandx{\UPolHierH}[4][2=, 3=, 4=]
	{\UPolHier{#1}[#2][#3][#4]-\HComp}
\newcommandx{\UPolHierE}[4][2=, 3=, 4=]
	{\UPolHier{#1}[#2][#3][#4]-\EComp}
\newcommandx{\UPolHierC}[4][2=, 3=, 4=]
	{\UPolHier{#1}[#2][#3][#4]-\CComp}

\newcommandx{\DAriHier}[4][2=, 3=, 4=]
	{\DLHier{#1}[#2][\argb{0}{#3}][#4]}
\newcommandx{\DAriHierH}[4][2=, 3=, 4=]
	{\DAriHier{#1}[#2][#3][#4]-\HComp}
\newcommandx{\DAriHierE}[4][2=, 3=, 4=]
	{\DAriHier{#1}[#2][#3][#4]-\EComp}
\newcommandx{\DAriHierC}[4][2=, 3=, 4=]
	{\DAriHier{#1}[#2][#3][#4]-\CComp}

\newcommandx{\EAriHier}[4][2=, 3=, 4=]
	{\ELHier{#1}[#2][\argb{0}{#3}][#4]}
\newcommandx{\EAriHierH}[4][2=, 3=, 4=]
	{\EAriHier{#1}[#2][#3][#4]-\HComp}
\newcommandx{\EAriHierE}[4][2=, 3=, 4=]
	{\EAriHier{#1}[#2][#3][#4]-\EComp}
\newcommandx{\EAriHierC}[4][2=, 3=, 4=]
	{\EAriHier{#1}[#2][#3][#4]-\CComp}

\newcommandx{\UAriHier}[4][2=, 3=, 4=]
	{\ULHier{#1}[#2][\argb{0}{#3}][#4]}
\newcommandx{\UAriHierH}[4][2=, 3=, 4=]
	{\UAriHier{#1}[#2][#3][#4]-\HComp}
\newcommandx{\UAriHierE}[4][2=, 3=, 4=]
	{\UAriHier{#1}[#2][#3][#4]-\EComp}
\newcommandx{\UAriHierC}[4][2=, 3=, 4=]
	{\UAriHier{#1}[#2][#3][#4]-\CComp}

\newcommandx{\DAnaHier}[4][2=, 3=, 4=]
	{\DLHier{#1}[#2][\argb{1}{#3}][#4]}
\newcommandx{\DAnaHierH}[4][2=, 3=, 4=]
	{\DAnaHier{#1}[#2][#3][#4]-\HComp}
\newcommandx{\DAnaHierE}[4][2=, 3=, 4=]
	{\DAnaHier{#1}[#2][#3][#4]-\EComp}
\newcommandx{\DAnaHierC}[4][2=, 3=, 4=]
	{\DAnaHier{#1}[#2][#3][#4]-\CComp}

\newcommandx{\EAnaHier}[4][2=, 3=, 4=]
	{\ELHier{#1}[#2][\argb{1}{#3}][#4]}
\newcommandx{\EAnaHierH}[4][2=, 3=, 4=]
	{\EAnaHier{#1}[#2][#3][#4]-\HComp}
\newcommandx{\EAnaHierE}[4][2=, 3=, 4=]
	{\EAnaHier{#1}[#2][#3][#4]-\EComp}
\newcommandx{\EAnaHierC}[4][2=, 3=, 4=]
	{\EAnaHier{#1}[#2][#3][#4]-\CComp}

\newcommandx{\UAnaHier}[4][2=, 3=, 4=]
	{\ULHier{#1}[#2][\argb{1}{#3}][#4]}
\newcommandx{\UAnaHierH}[4][2=, 3=, 4=]
	{\UAnaHier{#1}[#2][#3][#4]-\HComp}
\newcommandx{\UAnaHierE}[4][2=, 3=, 4=]
	{\UAnaHier{#1}[#2][#3][#4]-\EComp}
\newcommandx{\UAnaHierC}[4][2=, 3=, 4=]
	{\UAnaHier{#1}[#2][#3][#4]-\CComp}

\newcommandx{\DBorHier}[4][2=, 3=, 4=]
	{\DBHier{#1}[#2][\argb{\mathrm{B}}{#3}][#4]}
\newcommandx{\DBorHierH}[4][2=, 3=, 4=]
	{\DBorHier{#1}[#2][#3][#4]-\HComp}
\newcommandx{\DBorHierE}[4][2=, 3=, 4=]
	{\DBorHier{#1}[#2][#3][#4]-\EComp}
\newcommandx{\DBorHierC}[4][2=, 3=, 4=]
	{\DBorHier{#1}[#2][#3][#4]-\CComp}

\newcommandx{\EBorHier}[4][2=, 3=, 4=]
	{\EBHier{#1}[#2][\argb{\mathrm{B}}{#3}][#4]}
\newcommandx{\EBorHierH}[4][2=, 3=, 4=]
	{\EBorHier{#1}[#2][#3][#4]-\HComp}
\newcommandx{\EBorHierE}[4][2=, 3=, 4=]
	{\EBorHier{#1}[#2][#3][#4]-\EComp}
\newcommandx{\EBorHierC}[4][2=, 3=, 4=]
	{\EBorHier{#1}[#2][#3][#4]-\CComp}

\newcommandx{\UBorHier}[4][2=, 3=, 4=]
	{\UBHier{#1}[#2][\argb{\mathrm{B}}{#3}][#4]}
\newcommandx{\UBorHierH}[4][2=, 3=, 4=]
	{\UBorHier{#1}[#2][#3][#4]-\HComp}
\newcommandx{\UBorHierE}[4][2=, 3=, 4=]
	{\UBorHier{#1}[#2][#3][#4]-\EComp}
\newcommandx{\UBorHierC}[4][2=, 3=, 4=]
	{\UBorHier{#1}[#2][#3][#4]-\CComp}

\newcommand{\HComp}
	{\txtname{hard}\xspace}

\newcommand{\EComp}
	{\txtname{easy}\xspace}

\newcommand{\CComp}
	{\txtname{complete}\xspace}

\newtheorem{definition}{Definition}[section]

\newtheorem{theorem}{Theorem}[section]
\newtheorem{corollary}{Corollary}[section]
\newtheorem{conjecture}{Conjecture}[section]

\newcounter{flushenumerate}
\newenvironment{flushenumerate}
	{%
	\begin{list}{\arabic{flushenumerate})}%
	{\setlength{\leftmargin}{1.50em}}%
	\setlength{\labelwidth}{1.00em}
	\setlength{\itemindent}{0.00em}
	\setlength{\labelsep}{0.50em}
	\usecounter{flushenumerate}%
	}%
	{\end{list}}

% ...

\newcommand{\prcrel}{<}
\newcommandx{\prcRel}[3][1=, 2=, 3=]
	{\mthrel{\prcrel#3}[#1][#2]}

\newcommand{\deprel}{\rightsquigarrow}
\newcommandx{\depRel}[3][1=, 2=, 3=]
	{\mthrel{\deprel#3}[#1][#2]}

\newcommand{\colrel}{\cong}
\newcommandx{\colRel}[3][1=, 2=, 3=]
	{\mthrel{\colrel#3}[#1][#2]}

\newcommand{\entrel}{\preceq}
\newcommandx{\entRel}[3][1=, 2=, 3=]
	{\mthrel{\entrel#3}[#1][#2]}

\newcommand{\bcset}{B}
\newcommandx{\BCSet}[3][1=, 2=, 3=]
	{\mthset{\bcset#3}[#1][#2]}

\newcommand{\witfun}{wit}
\newcommandx{\witFun}[4][1=, 2=, 3=, 4=]
	{\mthargfun{\witfun#4}[#1][#2]{#3}}

\newcommand{\satfun}{sat}
\newcommandx{\satFun}[4][1=, 2=, 3=, 4=]
	{\mthargfun{\satfun#4}[#1][#2]{#3}}

\newcommand{\tupfun}{t}
\newcommandx{\tupFun}[4][1=, 2=, 3=, 4=]
	{\mthargfun{\tupfun#4}[#1][#2]{#3}}

\newcommand{\forfun}{fr}
\newcommandx{\forFun}[4][1=, 2=, 3=, 4=]
	{\mthargfun{\forfun#4}[#1][#2]{#3}}

\newcommand{\forset}{Fr}
\newcommandx{\ForSet}[3][1=, 2=, 3=]
	{\mthset{\forset#3}[#1][#2]}

\newcommand{\bisrel}{\sim}
\newcommandx{\bisRel}[3][1=, 2=, 3=]
	{\mthrel{\bisrel#3}[#1][#2]}

% ...

\usepackage{tikz}
\usetikzlibrary{arrows,shapes}

\usepackage{wrapfig}
\usepackage[caption=false,font=footnotesize]{subfig}

\tikzstyle{every node} =
	[draw = none, fill = white, thin]
\tikzstyle{every edge} +=
	[black, thick]

\tikzstyle{noall} =
	[draw = none, fill = none]
\tikzstyle{nodraw} =
	[draw = none, fill = white]
\tikzstyle{nofill} =
	[draw = black, fill = none]

\tikzstyle{cnode} =
	[circle, draw = black]
\tikzstyle{snode} =
	[regular polygon, regular polygon sides = 4, draw = gray!50]
\tikzstyle{lnode} =
	[diamond, draw = gray!75]
\tikzstyle{pnode} =
	[regular polygon, regular polygon sides = 5, draw = gray]

\AfterEndPreamble
	{

	\newcommand{\figbis}
		{
		\begin{wrapfigure}[8]{i}{0.200\textwidth}
			\vspace{-1.75em}
			\begin{center}
				\footnotesize
				\scalebox{1.00}[1.00]
					{
					\begin{tikzpicture}[node distance = 2em, bend angle = 22.5]

						\node [noall]
									(I)
									[]
									{};

						\node [nodraw]
									(R1)
									[node distance = 2.5em, left of = I]
									{$\RelStrName[1]$};

						\node [nodraw]
									(R2)
									[node distance = 2.5em, right of = I]
									{$\RelStrName[2]$};

						\node [nodraw]
									(0L)
									[below of = R1]
									{$0$};
						\node [nodraw]
									(1L)
									[right of = 0L]
									{$1$};
						\node [nodraw]
									(2L)
									[node distance = 5em, below of = 1L]
									{$2$};

						\node [nodraw]
									(0R)
									[below of = R2]
									{$0$};
						\node [nodraw]
									(1R)
									[right of = 0R]
									{$1$};
						\node [nodraw]
									(2R)
									[node distance = 5em, below of = 1R]
									{$2$};
						\node [nodraw]
									(1PR)
									[right of = 1R]
									{$1'$};
						\node [nodraw]
									(2PR)
									[node distance = 5em, below of = 1PR]
									{$2'$};

						\path[-]
							(1L)	edge	[bend right]
													node [] {$\rSym$}
													node [font = \scriptsize, pos = 0.10, sloped, above]
													{$\aSym$}
													node [font = \scriptsize, pos = 0.90, sloped, below]
													{$\bSym$}
													(2L)
							;

						\path[-]
							(1R)	edge	[bend right]
													node [] {$\rSym$}
													node [font = \scriptsize, pos = 0.10, sloped, above]
													{$\aSym$}
													node [font = \scriptsize, pos = 0.90, sloped, below]
													{$\bSym$}
													(2R)
							(1PR)	edge	[bend left]
													node [] {$\rSym$}
													node [font = \scriptsize, pos = 0.10, sloped, above]
													{$\aSym$}
													node [font = \scriptsize, pos = 0.90, sloped, below]
													{$\bSym$}
													(2PR)
							;

					\end{tikzpicture}
					}
					\vspace{-1.25em}
				\caption{\label{fig:bis} \small Two one-binding bisimilar \RelStr{s}.}
			\end{center}
			\vspace{-1em}
		\end{wrapfigure}
		}

	\newcommand{\figcggnobind}
		{
		\begin{figure}[htbp]
			\vspace{-2.00em}
			\centering
			\parbox{0.225\textwidth}
				{%
				\begin{center}
					\footnotesize
					\scalebox{0.90}[0.90]
						{
						\begin{tikzpicture}[node distance = 3em, bend angle = 120]

							\node [noall]
										(I)
										[]
										{};

							\node [nodraw]
										(R3)
										[node distance = 2.25em, left of = I]
										{$\RelStrName[3]$};

							\node [nodraw]
										(R4)
										[node distance = 2.25em, right of = I]
										{$\RelStrName[4]$};

							\node [nodraw]
										(0L)
										[below of = R3]
										{$0$};

							\node [nodraw]
										(0R)
										[below of = R4]
										{$0$};
							\node [nodraw]
										(1R)
										[node distance = 5em, below of = 0R]
										{$1$};

							\path[-]
								(0L)	edge	[out = 135, in = 225, loop, distance = 2.5em]
														node [] {$\rSym$}
														node [font = \scriptsize, pos = 0.10, sloped, above]
														{$\aSym$}
														node [font = \scriptsize, pos = 0.90, sloped, below]
														{$\bSym$}
														(0L)
								;

							\path[-]
								(0R)	edge	[out = 45, in = 315, loop, distance = 2.5em]
														node [] {$\rSym$}
														node [font = \scriptsize, pos = 0.10, sloped, above]
														{$\aSym$}
														node [font = \scriptsize, pos = 0.90, sloped, below]
														{$\bSym$}
														(0R)
								(1R)	edge	[out = 45, in = 315, loop, distance = 2.5em]
														node [] {$\rSym$}
														node [font = \scriptsize, pos = 0.10, sloped, above]
														{$\aSym$}
														node [font = \scriptsize, pos = 0.90, sloped, below]
														{$\bSym$}
														(1R)
								;

							\path[->, >=stealth]
								(0L)	edge	[dashed]
														node [] {$\fFun$}
														(0R)
											edge	[dashed]
														node [] {$\gFun$}
														(1R)
								;

						\end{tikzpicture}
						}
				\end{center}
				\vspace{-1.50em}
				\caption{\label{fig:cggnind} \small \!\!\!\! \CGFOL and \GNFOL
				indistinguishable \RelStr{s}.}
				}
			\quad
			\parbox{0.225\textwidth}
				{%
				\begin{center}
					\footnotesize
					\scalebox{0.79}[0.85]
						{
						\begin{tikzpicture}[node distance = 4em, bend angle = 45]

							\node [noall]
										(I)
										[]
										{};

							\node [nodraw]
										(R5)
										[node distance = 4.75em, left of = I]
										{$\RelStrName[5]$};

							\node [nodraw]
										(R6)
										[node distance = 4.75em, right of = I]
										{$\RelStrName[6]$};

							\node [nodraw]
										(0L)
										[node distance = 2.5em, below of = R5]
										{$0$};
							\node [nodraw]
										(1L)
										[below left of = 0L]
										{$1$};
							\node [nodraw]
										(2L)
										[below right of = 0L]
										{$2$};
							\node [cnode]
										(3L)
										[node distance = 5.5em, below of = 1L]
										{$3$};
							\node [nodraw]
										(4L)
										[node distance = 5.5em, below of = 2L]
										{$4$};

							\node [nodraw]
										(0R)
										[node distance = 2.5em, below of = R6]
										{$0$};
							\node [nodraw]
										(1R)
										[below left of = 0R]
										{$1$};
							\node [nodraw]
										(2R)
										[below right of = 0R]
										{$2$};
							\node [cnode]
										(3R)
										[node distance = 5.5em, below of = 1R]
										{$3$};
							\node [nodraw]
										(4R)
										[node distance = 5.5em, below of = 2R]
										{$4$};

							\path[-]
								(0L)	edge	[bend right]
														node [] {$\rSym$}
														node [font = \scriptsize, pos = 0.10, sloped, above]
														{$\aSym$}
														node [font = \scriptsize, pos = 0.90, sloped, above]
														{$\bSym$}
														(1L)
											edge	[bend left]
														node [] {$\rSym$}
														node [font = \scriptsize, pos = 0.10, sloped, above]
														{$\aSym$}
														node [font = \scriptsize, pos = 0.90, sloped, above]
														{$\bSym$}
														(2L)
								(1L)	edge	[]
														node [] {$\rSym$}
														node [font = \scriptsize, pos = 0.10, sloped, below]
														{$\aSym$}
														node [font = \scriptsize, pos = 0.85, sloped, below]
														{$\bSym$}
														(3L)
											edge	[]
														node [] {$\rSym$}
														node [font = \scriptsize, pos = 0.10, sloped, above]
														{$\aSym$}
														node [font = \scriptsize, pos = 0.90, sloped, below]
														{$\bSym$}
														(4L)
								(2L)	edge	[]
														node [] {$\rSym$}
														node [font = \scriptsize, pos = 0.10, sloped, above]
														{$\aSym$}
														node [font = \scriptsize, pos = 0.85, sloped, below]
														{$\bSym$}
														(3L)
											edge	[]
														node [] {$\rSym$}
														node [font = \scriptsize, pos = 0.10, sloped, above]
														{$\aSym$}
														node [font = \scriptsize, pos = 0.90, sloped, above]
														{$\bSym$}
														(4L)
								;

							\path[-]
								(0R)	edge	[bend right]
														node [] {$\rSym$}
														node [font = \scriptsize, pos = 0.10, sloped, above]
														{$\aSym$}
														node [font = \scriptsize, pos = 0.90, sloped, above]
														{$\bSym$}
														(1R)
											edge	[bend left]
														node [] {$\rSym$}
														node [font = \scriptsize, pos = 0.10, sloped, above]
														{$\aSym$}
														node [font = \scriptsize, pos = 0.90, sloped, above]
														{$\bSym$}
														(2R)
								(1R)	edge	[]
														node [] {$\rSym$}
														node [font = \scriptsize, pos = 0.10, sloped, below]
														{$\aSym$}
														node [font = \scriptsize, pos = 0.85, sloped, below]
														{$\bSym$}
														(3R)
											edge	[]
														node [] {$\rSym$}
														node [font = \scriptsize, pos = 0.10, sloped, above]
														{$\aSym$}
														node [font = \scriptsize, pos = 0.90, sloped, below]
														{$\bSym$}
														(4R)
								(2R)	edge	[]
														node [] {$\rSym$}
														node [font = \scriptsize, pos = 0.10, sloped, above]
														{$\aSym$}
														node [font = \scriptsize, pos = 0.90, sloped, above]
														{$\bSym$}
														(4R)
								;

						\end{tikzpicture}
						}
				\end{center}
				\vspace{-1.25em}
				\caption{\label{fig:obind} \small \OBFOL indistinguishable \RelStr{s}.}
				}
		\end{figure}
		}

	\newcommand{\figcolgrp}
		{
		\begin{figure}[htbp]
			\vspace{-1.50em}
			\centering
			\parbox{0.225\textwidth}
				{%
				\begin{center}
					\footnotesize
					\scalebox{1.00}[1.00]
						{
						\begin{tikzpicture}[node distance = 2.5em, bend angle = 45]

							\node [noall]
										(I)
										[]
										{};

							\node []
										(A)
										[right of = I]
										{$\aSym$};
							\node []
										(B)
										[right of = A]
										{$\bSym$};
							\node []
										(C)
										[right of = B]
										{$\cSym$};

							\node []
										(1)
										[below of = I]
										{$\schSym[1]$};
							\node []
										(2)
										[below of = 1]
										{$\schSym[2]$};
							\node []
										(3)
										[below of = 2]
										{$\schSym[3]$};

							\node []
										(1A)
										[right of = 1]
										{$\bullet$};
							\node []
										(1B)
										[right of = 1A]
										{$\bullet$};
							\node []
										(1C)
										[right of = 1B]
										{$\bullet$};
							\node []
										(2A)
										[right of = 2]
										{$\bullet$};
							\node []
										(2B)
										[right of = 2A]
										{$\bullet$};
							\node []
										(2C)
										[right of = 2B]
										{$\bullet$};
							\node []
										(3A)
										[right of = 3]
										{$\bullet$};
							\node []
										(3B)
										[right of = 3A]
										{$\bullet$};
							\node []
										(3C)
										[right of = 3B]
										{$\bullet$};

							\path[-]
								(1A)	edge	[]
														(2A)
								(2A)	edge	[]
														(3A)
								(1B)	edge	[]
														(2B)
								(2B)	edge	[]
														(3B)
								(1C)	edge	[]
														(2C)
								(2C)	edge	[]
														(3C)
								;

						\end{tikzpicture}
						}
				\end{center}
				\vspace{-1.25em}
				\caption{\label{fig:colgrp1} \small Collapsing graph $\CName[1]$.}
				}
			\quad
			\parbox{0.225\textwidth}
				{%
				\begin{center}
					\footnotesize
					\scalebox{1.00}[1.00]
						{
						\begin{tikzpicture}[node distance = 2.5em, bend angle = 45]

							\node [noall]
										(I)
										[]
										{};

							\node []
										(A)
										[right of = I]
										{$\aSym$};
							\node []
										(B)
										[right of = A]
										{$\bSym$};
							\node []
										(C)
										[right of = B]
										{$\cSym$};

							\node []
										(1)
										[below of = I]
										{$\schSym[1]$};
							\node []
										(2)
										[below of = 1]
										{$\schSym[2]$};
							\node []
										(3)
										[below of = 2]
										{$\schSym[3]$};

							\node []
										(1A)
										[right of = 1]
										{$\bullet$};
							\node []
										(1B)
										[right of = 1A]
										{$\bullet$};
							\node []
										(1C)
										[right of = 1B]
										{$\bullet$};
							\node []
										(2A)
										[right of = 2]
										{$\bullet$};
							\node []
										(2B)
										[right of = 2A]
										{$\bullet$};
							\node []
										(2C)
										[right of = 2B]
										{$\bullet$};
							\node []
										(3A)
										[right of = 3]
										{$\bullet$};
							\node []
										(3B)
										[right of = 3A]
										{$\bullet$};
							\node []
										(3C)
										[right of = 3B]
										{$\bullet$};

							\path[-]
								(1A)	edge	[]
														(2A)
								(2A)	edge	[]
														(3A)
								(1B)	edge	[]
														(2B)
								(2B)	edge	[]
														(3B)
								(1C)	edge	[]
														(2C)
								(2C)	edge	[]
														(3C)
								(3B)	edge	[]
														(3C)
								;

						\end{tikzpicture}
						}
				\end{center}
				\vspace{-1.25em}
				\caption{\label{fig:colgrp2} \small Collapsing graph $\CName[2]$.}
				}
		\end{figure}
		}

	\newcommand{\figdepgrp}
		{
		\begin{figure}[htbp]
			\vspace{-1.40em}
			\centering
			\parbox{0.145\textwidth}
				{%
				\begin{center}
					\footnotesize
					\scalebox{1.00}[1.00]
						{
						\begin{tikzpicture}[node distance = 2.5em, bend angle = 45]

							\node [noall]
										(I)
										[]
										{};

							\node []
										(A)
										[node distance = 2em,right of = I]
										{$\aSym$};
							\node []
										(B)
										[right of = A]
										{$\bSym$};
							\node []
										(C)
										[right of = B]
										{$\cSym$};

							\node []
										(1)
										[node distance = 2em, below of = I]
										{$\schSym[1]$};
							\node []
										(2)
										[below of = 1]
										{$\schSym[2]$};
							\node []
										(3)
										[below of = 2]
										{$\schSym[3]$};

							\node []
										(1A)
										[node distance = 2em, right of = 1]
										{$\bullet$};
							\node []
										(1B)
										[right of = 1A]
										{$\bullet$};
							\node []
										(1C)
										[right of = 1B]
										{$\bullet$};
							\node []
										(2A)
										[below of = 1A]
										{$\bullet$};
							\node []
										(2B)
										[right of = 2A]
										{$\bullet$};
							\node []
										(2C)
										[right of = 2B]
										{$\bullet$};
							\node []
										(3A)
										[below of = 2A]
										{$\bullet$};
							\node []
										(3B)
										[right of = 3A]
										{$\bullet$};
							\node []
										(3C)
										[right of = 3B]
										{$\bullet$};

							\path[->, >=stealth]
								(1A)	edge	[]
														(1B)
								(2A)	edge	[gray]
														(1B)
								(3A)	edge	[gray]
														(1B)
								(2B)	edge	[]
														(2C)
								(3B)	edge	[gray]
														(2C)
								(1B)	edge	[gray]
														(2C)
								;

						\end{tikzpicture}
						}
				\end{center}
				\vspace{-1.25em}
				\caption{\label{fig:depgrp1} \small \!\!Dependence graph $\DName[1]$.}
				}
			\quad\!\!\!
			\parbox{0.145\textwidth}
				{%
				\begin{center}
					\footnotesize
					\scalebox{1.00}[1.00]
						{
						\begin{tikzpicture}[node distance = 2.5em, bend angle = 45]

							\node [noall]
										(I)
										[]
										{};

							\node []
										(A)
										[node distance = 2em,right of = I]
										{$\aSym$};
							\node []
										(B)
										[right of = A]
										{$\bSym$};
							\node []
										(C)
										[right of = B]
										{$\cSym$};

							\node []
										(1)
										[node distance = 2em, below of = I]
										{$\schSym[1]$};
							\node []
										(2)
										[below of = 1]
										{$\schSym[2]$};
							\node []
										(3)
										[below of = 2]
										{$\schSym[3]$};

							\node []
										(1A)
										[node distance = 2em, right of = 1]
										{$\bullet$};
							\node []
										(1B)
										[right of = 1A]
										{$\bullet$};
							\node []
										(1C)
										[right of = 1B]
										{$\bullet$};
							\node []
										(2A)
										[below of = 1A]
										{$\bullet$};
							\node []
										(2B)
										[right of = 2A]
										{$\bullet$};
							\node []
										(2C)
										[right of = 2B]
										{$\bullet$};
							\node []
										(3A)
										[below of = 2A]
										{$\bullet$};
							\node []
										(3B)
										[right of = 3A]
										{$\bullet$};
							\node []
										(3C)
										[right of = 3B]
										{$\bullet$};

							\path[->, >=stealth]
								(1A)	edge	[]
														(1B)
								(2A)	edge	[gray]
														(1B)
								(3A)	edge	[gray]
														(1B)
								(1B)	edge	[gray]
														(2C)
								(2B)	edge	[]
														(2C)
								(3B)	edge	[gray]
														(2C)
								(1C)	edge	[gray]
														(2C)
								(2C)	edge	[loop right, gray]
														()
								(3C)	edge	[gray]
														(2C)
								;

						\end{tikzpicture}
						}
				\end{center}
				\vspace{-1.25em}
				\caption{\label{fig:depgrp2} \small \!\!Dependence graph $\DName[2]$.}
				}
			\quad\!\!
			\parbox{0.145\textwidth}
				{%
				\begin{center}
					\footnotesize
					\scalebox{1.00}[1.00]
						{
						\begin{tikzpicture}[node distance = 2.5em, bend angle = 45]

							\node [noall]
										(I)
										[]
										{};

							\node []
										(A)
										[node distance = 2em,right of = I]
										{$\aSym$};
							\node []
										(B)
										[right of = A]
										{$\bSym$};
							\node []
										(C)
										[right of = B]
										{$\cSym$};

							\node []
										(1)
										[node distance = 2em, below of = I]
										{$\schSym[1]$};
							\node []
										(2)
										[below of = 1]
										{$\schSym[2]$};
							\node []
										(3)
										[below of = 2]
										{$\schSym[3]$};

							\node []
										(1A)
										[node distance = 2em, right of = 1]
										{$\bullet$};
							\node []
										(1B)
										[right of = 1A]
										{$\bullet$};
							\node []
										(1C)
										[right of = 1B]
										{$\bullet$};
							\node []
										(2A)
										[below of = 1A]
										{$\bullet$};
							\node []
										(2B)
										[right of = 2A]
										{$\bullet$};
							\node []
										(2C)
										[right of = 2B]
										{$\bullet$};
							\node []
										(3A)
										[below of = 2A]
										{$\bullet$};
							\node []
										(3B)
										[right of = 3A]
										{$\bullet$};
							\node []
										(3C)
										[right of = 3B]
										{$\bullet$};

							\path[->, >=stealth]
								(1A)	edge	[]
														(1B)
								(2A)	edge	[gray]
														(1B)
								(3A)	edge	[gray]
														(1B)
								(2B)	edge	[]
														(2C)
								(3B)	edge	[gray]
														(2C)
								(1B)	edge	[gray]
														(2C)
								(1C)	edge	[gray]
														(3A)
								(2C)	edge	[gray]
														(3A)
								(3C)	edge	[bend left]
														(3A)
								;

						\end{tikzpicture}
						}
				\end{center}
				\vspace{-1.25em}
				\caption{\label{fig:depgrp3} \small \!\!Dependence graph $\DName[3]$.}
				}
		\end{figure}
		}

	}

\usepackage[ruled,vlined]{algorithm2e}

\DontPrintSemicolon

\SetKw{Signature}{signature}
\SetKwFor{Function}{function}{}{}

\SetKw{With}{with}{}{}

\AfterEndPreamble
	{

	\newcommand{\algsat}
		{
		\begin{algorithm}
			\caption{\label{alg:sat} \OBFOL Satisfiability Checker.}
			\Signature{$\satFun : \LanSigName(\VarSet)\text{-}\OBFOL \to \{ \Tt, \Ff
			\}$} \;
			\Function{$\satFun(\varphi)$}
				{
				\nl \ForEach{$\FSet \in \witFun(\varphi)$}
					{
					\nl $\forFun \gets \set{ (\qntElm, \bndElm) \!\in\! \SchSet \mapsto
					\der{\relElm} \!\in\! \der{\RelSet} }{ \qntelm \bndElm \der{\relElm}
					\!\in\! \FSet }$ \;
					\nl $\iElm \gets \Ff$ \;
					\nl \ForEach{$\SSet \subseteq \dom{\forFun} \cap \SchSet(\ASet)$ \With
					$\ASet \subseteq \ArgSet$}
						{
						\nl \If{$\SSet$ \emph{is-overlapping-over} $\ASet$}
							{
							\nl \If{$\witFun(\bigwedge_{\schElm \in \SSet} \forFun(\schElm)) =
							\emptyset$}
								{
								\nl $\iElm \gets \Tt$ \;
								}
							}
						}
					\nl \If{$\iElm = \Ff$}
						{
						\nl \Return $\Tt$ \;
						}
					}
				\nl \Return $\Ff$ \;
				}
		\end{algorithm}
		}

	}

\begin{document}

	\title{\Large\textbf{On the Remarkable Features of Binding Forms}}

	\author
		{%
		\IEEEauthorblockN{Fabio Mogavero and Giuseppe Perelli}%
		\IEEEauthorblockA{Universit\`a degli Studi di Napoli Federico II}%
		}

	\maketitle



\begin{abstract}


	Hilbert's ``\emph{Entscheidungsproblem}'' has given rise to a broad and
	productive line of research in mathematical logic, where the
	\emph{classification process} of decidable classes of first-order sentences
	represent only one of the remarkable results.
	According to the criteria used to identify the particular classes of interest,
	this process was declined into several research programs, of which some of the
	most deeply investigated are the ones classifying sentences in \emph{prenex
	normal form} in base of their \emph{prefix vocabulary}.

	Unfortunately, almost all of these approaches did not shed any light on the
	reasons why \emph{modal logic} is so robustly decidable.
	Trying to answer to this question, Andr\'eka, van Benthem, and N\'emeti
	introduced the \emph{guarded fragment} of first-order logic, which generalizes
	the modal framework by essentially retaining all its fundamental properties.
	They started, so, a completely new research program based on the way
	quantifications can be \emph{relativized}.
	Although this approach succeeded in its original task, we cannot consider it
	satisfactory in spotting the reasons why some complex extensions of modal
	logic are well-behaved.
	In particular, by just using the related results, we are not able to derive
	the decidability of multi-agent logics for strategic abilities.

	In this paper, aiming to lay the foundation for a more thorough understanding
	of some of these decidability questions, we introduce a new kind of
	classification based on the \emph{binding forms} that are admitted in a
	sentence, \ie, on the way the arguments of a relation can be bound to a
	variable.
	We describe a hierarchy of first-order fragments based on the Boolean
	combinations of these forms, showing that the less expressive one is already
	incomparable with the guarded logic and related extensions.
	We also prove, via a new model-theoretic technique, that it enjoys the
	finite-model property and a \PSpace satisfiability problem.

\end{abstract}




\begin{section}{Introduction}


	Since from the publication of the revolutionary negative solutions owed to
	Church~\cite{Chu36a,Chu36b,Chu36c} and Turing~\cite{Tur37a,Tur37b}, Hilbert's
	original ``\emph{Entscheidungsproblem}''~\cite{HA28} turned into a vast
	\emph{classification process} looking for all those classes of first-order
	sentences having a \emph{decidable satisfiability}~\cite{Bor84}.
	Depending upon the syntactic criteria used to identify the particular classes
	of interest~\cite{Gra03}, this process was declined into several research
	programs, among which we can mention, on one side, those limiting
	\emph{relation arities}~\cite{Low15} or \emph{free
	variables}~\cite{Mor75,GKV97} and, on the other one, those classifying
	sentences in \emph{prenex normal form} in base of their \emph{prefix
	vocabulary}~\cite{BGG97}.
	\\ \indent
	The research in this field has had a considerable impact, from both a
	theoretical and practical point of view, in numerous areas on the edge between
	mathematics and computer science, \eg, \emph{reverse
	mathematics}~\cite{Sim88}, \emph{descriptive complexity}, \emph{database
	theory}, and \emph{formal verification}, just to cite a few.
	However, almost all of the classic approaches did not shed any satisfactory
	light on a fundamental question: why are \emph{modal logic} and derived
	frameworks, like those ones featuring fixpoint constructs, so robustly
	decidable~\cite{Var96,Gra01}?
	Trying to find a plausible answer, Andr\'eka, van Benthem, and N\'emeti
	introduced the \emph{guarded fragment} of first-order logic~\cite{ABN98},
	which generalizes the modal framework by essentially retaining all its
	model-theoretic and algorithmic properties.
	They started, so, a completely new research program based on the way
	quantifications can be \emph{relativized} to some particular facts, avoiding
	the usual syntactic restrictions on quantifier patterns, relation arities, and
	number of variables.
	Pushing forward the idea that robust fragments of first-order logic owe their
	nice properties to some kind of guarded quantification, several extensions
	along this line of research were introduced in the literature, such as the
	\emph{clique guarded}~\cite{Gra99a}, \emph{loosely guarded}~\cite{Ben97}, the
	\emph{clique guarded}~\cite{Gra99a,Gra02}, the \emph{action
	guarded}~\cite{Ben00,GG00}, and the \emph{guarded fixpoint logic}~\cite{GW99}.
	This classification program has also important applications in database theory
	and description logic, where it is relevant to evaluate a query against
	guarded first-order theories~\cite{BCO10}.
	\\ \indent
	Only recently, ten Cate and Segoufin observed that the first-order translation
	of modal logic, besides the guarded nature of quantifications, presents
	another important peculiarity: negation is only applied to sentences or
	monadic formulas, \ie, formulas with a single free variable.
	Exploiting this observation, they introduced a new robust fragment of
	first-order logic, called \emph{unary negation}~\cite{CS11,CS13}, which
	extends modal logic, as well as other formalisms, like Boolean conjunctive
	queries, that cannot be expressed in terms of guarded quantifications.
	Since this new logic is incomparable to the guarded fragments, right after the
	original work,another logic was proposed, called \emph{guarded
	negation}~\cite{BCS11}, unifying both the approaches of classification.
	Syntactically, there is no primary universal quantification and the use of
	negation is only allowed if guarded by an atomic relation.
	In terms of expressive power, this fragment forms a strict extension of both
	the logics on which it is based, while preserving the same desirable
	properties.
	However, it has to be noted that it is still incomparable to more complex
	extensions of the guarded fragment, such as the clique guarded one.
	This way of analyzing formulas founded on the guarded nature of negation has
	also remarkable applications to database theory, where it is well-known that
	complementation makes queries hard to evaluate~\cite{BCO12}.
	\\ \indent
	Although these two innovative classification programs really succeeded in the
	original task to explain the nice properties of modal logic, we cannot
	consider them completely satisfactory with respect to the more general intent
	to identify the reasons why some of its complex extensions are so
	well-behaved.
	In particular, by just using the related model-theoretic and algorithmic
	results, we are not able to answer to the question about the decidability of
	several multi-agent logics for strategic abilities, such as the
	\emph{Alternating Temporal Logics} \ATL~\cite{WLWW06,SF06} and
	\ATLS~\cite{Sch08} and the one-goal fragment of \emph{Strategy Logic}
	\OGSL~\cite{MMV10b,MMPV11,MMPV12}, which do not intrinsically embed such kinds
	of relativization.
	For example, consider the \ATL formula $\varphi = \AAll{ \aSym, \bSym, \cSym }
	\neg \psi$ over a game structure with the four agents $\aSym$, $\bSym$,
	$\cSym$, and $\dSym$.
	Intuitively, it asserts that agent $\dSym$ has a strategy, which depends upon
	those chosen by the other three ones, ensuring that the property $\psi$ does
	not hold.
	Now, it is easy to observe that the underlying strategy quantifications can be
	represented as a prefix of the form $\forall^{3} \exists$ coupled with the
	quaternary relation described by $\psi$.
	Therefore, $\varphi$ is interpretable neither as a decidable prefix-vocabulary
	class nor as a two-variable formula.
	Moreover, quantifications are not guarded and negation is applied to the
	property $\psi$ that cannot be considered as monadic.
	Another explicit example is given by the \OGSL sentence $\AAll{\xSym}
	\EExs{\ySym} \AAll{\zSym} (\aSym, \xSym) (\bSym, \xSym) (\dSym, \ySym) (\cSym,
	\zSym) \psi$ asserting that, once $\aSym$ and $\bSym$ chose the common
	strategy $\xSym$, agent $\dSym$ can select its better response $\ySym$ to
	ensure $\psi$, in a way that is independent of the behavior $\zSym$ of
	$\cSym$.
	Here, we have a more complex quantification prefix of the form $\forall
	\exists \forall$ coupled with the quaternary relation $\psi$, in which two of
	its arguments are bound to the same variable.
	Also in this case, we are not able to cast this sentence in any of the
	decidable cases described by some of the classification programs already
	introduced in the literature.
	In particular, it is not guarded negation, since universal quantifications are
	used as primary construct, which is prohibited in that fragment.
	\\ \indent
	In this paper, trying to lay the foundation for a more thorough understanding
	of these decidability questions, we introduce a new classification program
	based on the \emph{binding forms} that are admitted in a sentence, \ie, on the
	way the arguments of a relation can be bound to a variable.
	To do this, similarly to the treatment of the attributes of a table in a
	database setting, we define a generalization of standard notions of language
	signature and relational structure, which makes syntactically explicit the
	arguments of interest.
	With more detail, every relation $\relElm$ is associated with a set of
	arguments $\{ \argElm[1], \ldots, \argElm[n] \}$, which are bound to the
	variables via a binding form $(\argElm[1], \varElm[1]) \cdots (\argElm[n],
	\varElm[n]) \relElm$ that replaces the usual writing $\relElm (\varElm[1],
	\ldots, \varElm[n])$.
	Our notation, although perfectly equivalent to the classic one, allows to
	introduce and analyze, in an easier and natural way, a hierarchy of four
	fragments of first-order logic based on the Boolean combinations of these
	forms.
	In particular, we study the simplest fragment, called \emph{one binding},
	proving that it is already incomparable to clique guarded and guarded
	negation.
	Simple examples of one-binding sentences are given by the translation of the
	game properties described above: $\forall \xSym \forall \ySym \forall \zSym
	\exists \wSym (\aSym, \xSym) (\bSym, \ySym) (\cSym, \zSym) (\dSym, \wSym) \neg
	\rSym$ and $\forall \xSym \exists \ySym \forall \zSym (\aSym, \xSym) (\bSym,
	\xSym) (\dSym, \ySym) (\cSym, \zSym) \rSym$, where in place of the temporal
	goal $\psi$ we have the relation $\rSym$ whose arguments $\{ \aSym, \bSym,
	\cSym, \dSym \}$ stand for the agents of the game.
	As main results, via a novel model-theoretic technique, we prove that our
	logic enjoys the finite-model property, both Craig's interpolation and Beth's
	definability, and a \PSpace satisfiability problem.

\end{section}




\begin{section}{Preliminaries}
	\label{sec:prl}

	Since from Codd's pioneering work on the structure of \emph{relational
	databases}~\cite{Cod70}, several kinds of first-order languages have been used
	to describe \emph{databases queries}~\cite{Cod72}.
	In particular, \emph{first-order logic} (\FOL, for short) has been established
	as the main theoretical framework in which to prove results about properties
	of query languages~\cite{Imm82,Var82,Imm86}.
	In this context, a table is usually represented as a mathematical relation
	between elements of a given domain, where its attributes are mapped to the
	indexes of that relation in a predetermined fixed way.
	Consequently, the attributes do not have any explicit matching element in the
	syntax of the language.
	In this paper, in order to introduce the \emph{binding-form fragments} of
	\FOL, we need to reformulate both the syntax and the semantics of the logic in
	a way that is much closer to the database setting.
	In particular, we explicitly associate a finite non-empty set of arguments to
	each relation.
	Those arguments are handled in the syntax via corresponding symbols.
	To this aim, we introduce an alternative version of classic \emph{language
	signatures} and \emph{relational structures}.

	\begin{subsection}{Language Signatures}
		\label{sec:prl;sub:lansig}

		A \emph{language signature} is a mathematical object that simply describes
		the structure of all non-logical symbols composing a first-order formula.
		The typology we introduce here is purely relational, since we do not make
		use of constant or function symbols.
		Also, in our reasonings, we do not explicitly consider distinguished
		relations as equality, equivalences, or orders.


		\begin{definition}[Language Signature]
			\label{def:lansig}
			A \emph{language signature} (\LanSig, for short) is a tuple $\LanSigName
			\defeq \LanSigStr$, where $\ArgSet$ and $\RelSet$ are the finite non-empty
			sets of \emph{argument} and \emph{relation} names and $\argFun : \RelSet
			\to \pow{\ArgSet} \setminus \{ \emptyset \}$ is the \emph{argument
			function} mapping every relation $\relSym \in \RelSet$ to its non-empty
			set of arguments $\argFun(\relSym) \subseteq \ArgSet$.
		\end{definition}

		In the following, as running example, we use the simple \LanSig $\LanSigName
		= \LanSigStr[][][\{ \aSym, \bSym \}][\{ \qSym, \rSym, \sSym \}]$ having
		$\qSym$ and $\rSym$ as two binary relations over the arguments
		$\argFun(\qSym) = \argFun(\rSym) = \{ \aSym, \bSym \}$ and $\sSym$ as a
		monadic relation over the argument $\argFun(\sSym) = \{ \aSym \}$.

	\end{subsection}

	\begin{subsection}{Relational Structures}
		\label{sec:prl;sub:relstr}

		Given a language signature, we define the interpretation of all symbols
		by means of a \emph{relational structure}, \ie, a carrier domain together
		with an association of each relation with an appropriate set of tuples on
		the elements of that domain.
		Since, in our framework, relations with the same arity may have different
		arguments, it is not sufficient for us to manage usual tuples as components
		of their interpretation.
		For this reason, we map a given relation $\relSym$ to a set of \emph{tuple
		functions} $\tupFun$ having $\argFun(\relSym)$ as support and elements of
		the carrier domain as values.


		\begin{definition}[Relational Structure]
			\label{def:relstr}
			A \emph{relational structure} over an \LanSig $\LanSigName = \LanSigStr$
			($\LanSigName$-\RelStr, for short) is a tuple $\RelStrName \defeq
			\RelStrStr$, where $\DomSet$ is the non-empty set of arbitrary objects
			called \emph{domain} and $\relFun : \RelSet \cto[\relSym]
			\pow{\argFun(\relSym) \to \DomSet}$ is the \emph{relation function}
			mapping every relation $\relSym \in \RelSet$ to a set
			$\relSym[][\RelStrName] \defeq \relFun(\relSym) \subseteq \argFun(\relSym)
			\to \DomSet$ of \emph{tuple functions} $\tupFun \in
			\relSym[][\RelStrName]$ from arguments of $\relSym$ to elements of the
			domain.
		\end{definition}


		The \emph{order} of an \RelStr $\RelStrName$ is given by the size
		$\ordFun[][][\RelStrName] \defeq \card{\DomSet}$ of its domain.
		A relational structure is \emph{finite} iff it has finite order.

		An example of a finite \RelStr over the previous \LanSig $\LanSigName$ is
		given by $\RelStrName \!=\! \RelStrStr[][][\{ 0, 1 \}]$, where
		$\qSym[][\RelStrName] \!=\! \{ (\aSym \!\mapsto\! 0, \bSym \!\mapsto\! 0),
		(\aSym \!\mapsto\! 0, \bSym \!\mapsto\! 1) \}$, $\rSym[][\RelStrName] \!=\!
		\{ (\aSym \!\mapsto\! 0, \bSym \!\mapsto\! 0), (\aSym \!\mapsto\! 1, \bSym
		\!\mapsto\! 1) \}$, and $\sSym[][\RelStrName] \!=\! \{ (\aSym \!\mapsto\! 1)
		\}$ are the interpretations of the three relations.

	\end{subsection}

\end{section}




\begin{section}{First-Order Logic}
	\label{sec:firordlog}

	We start by describing a slightly different but completely equivalent
	formalization of both the syntax and the semantics of \FOL according to the
	explained alternatives of language signature and relational structure.
	We also introduce a new family of fragments based on the kinds of binding
	forms that are admitted in a formula, \ie, on the ways an argument can be
	bound to a variable.

	From now on, if not differently stated, we use $\LanSigName = \LanSigStr$ to
	denote an \apriori fixed \LanSig.
	Also, $\VarSet$ represents an enu\-merable non-empty set of \emph{variables}.
	For sake of succinctness, to indicate the extension of $\LanSigName$ with
	$\VarSet$, we adopt the composed symbol $\LanSigName(\VarSet)$ in place of the
	tuple $\tupleb{\LanSigName}{\VarSet}$.

	\begin{subsection}{Syntax}
		\label{sec:firordlog;sub:syn}

		As far as the syntax of \FOL concerns, the novelty of our setting resides in
		the explicit presence of arguments as atomic components of a formula.
		In particular, a variables $\varElm$ cannot be directly applied to the index
		associated with an argument $\argElm$ of a relation $\relElm$ as in the
		usual writing $\relElm(\ldots, \varElm, \ldots)$, but a construct
		$(\argElm, \varElm)$, called \emph{binding}, is required to link $\varElm$
		to $\argElm$ in $\relElm$.

		\begin{definition}[\FOL Syntax]
			\label{def:folsyn}
			\FOL \emph{formulas} over $\LanSigName(\VarSet)$ are built by means of the
			following context-free grammar, where $\argElm \in \ArgSet$, $\relElm \in
			\RelSet$, and $\varElm \in \VarSet$:
			\begin{center}
				$\varphi \seteq \relElm \mid \neg \varphi \mid (\varphi \wedge \varphi)
				\mid (\varphi \vee \varphi) \mid \exists \varElm. \varphi \mid  \forall
				\varElm. \varphi \mid (\argElm, \varElm) \varphi$.
			\end{center}
			$\LanSigName(\VarSet)$-\FOL denotes the enumerable set of all formulas
			over $\LanSigName(\VarSet)$ generated by the above grammar.
		\end{definition}


		Consider the \LanSig described in Section~\ref{sec:prl;sub:lansig}.
		Then, $\varphi_{1} = \exists \xSym \forall \ySym \exists \zSym \allowbreak
		((\aSym, \xSym) (\bSym, \ySym) (\qSym \vee \neg \sSym) \wedge (\aSym, \ySym)
		(\bSym, \zSym) \rSym)$, $\varphi_{2} = \exists \xSym (\aSym, \xSym) \rSym$,
		and $\varphi_{3} = (\aSym, \xSym) \forall \ySym (\bSym, \ySym) \qSym$ are
		three formulas over $\LanSigName(\{ \xSym, \ySym, \zSym \})$.
		Note that, by fixing an ordering on the arguments, it is possible to rewrite
		our \FOL formulas in the classic syntax.
		For instance, $\varphi_{1}$ can be expressed by $\exists \xSym \forall \ySym
		\exists \zSym ((\qSym(\xSym, \ySym) \vee \neg \sSym(\xSym)) \wedge
		\rSym(\ySym, \zSym))$, once it is assumed that $\aSym < \bSym$.
		On the other hand, every classic \FOL formula can be easily translated in
		our syntax by means of numeric arguments that represent the positions in the
		relations.
		As an example, the transitivity property $\forall \xSym \forall \ySym
		\forall \zSym ((\neg \rSym(\xSym, \ySym) \vee \neg \rSym(\ySym, \zSym)) \vee
		\rSym(\xSym, \zSym))$ can be rewritten as follows: $\forall \xSym \forall
		\ySym \forall \zSym (((1, \xSym) (2, \ySym) \neg \rSym \vee (1, \ySym) (2,
		\zSym) \neg \rSym) \vee (1, \xSym) (2, \zSym) \rSym)$.
		Observe that the parentheses surrounding the binary Boolean connectives are
		employed to enforce the property of unique parsing of a formula.
		However, in the following, we simplify the notation by avoiding their use in
		the unambiguous cases.

		Usually, predicative logics, \ie, languages having explicit quantifiers,
		need a concept of free or bound \emph{placeholder} to formally evaluate the
		meaning of their formulas.
		Placeholders are used, indeed, to enlighten particular positions in a
		syntactic expression that are crucial to the definition of its semantics.
		Classic formalizations of \FOL just require one kind of placeholder
		represented by the variables on which the formulas are built.
		In our new setting, instead, both variables and arguments have this
		fundamental role.
		In particular, arguments are used to decouple variables from their
		association with a relation.
		Consequently, we need a way to check whether a variable is quantified or an
		argument is bound.
		To do this, we use the concept of \emph{free arguments/variables}, \ie, the
		subset of $\ArgSet \cup \VarSet$ containing all arguments that are free from
		binding and all variables occurring in some binding that are not quantified.

		\begin{definition}[Free Placeholders]
			\label{def:freeplc}
			The set of \emph{free arguments/variables} of an
			$\LanSigName(\VarSet)$-\FOL formula can be computed via the function
			$\freeFun{} : \LanSigName(\VarSet)\text{-\FOL} \!\to\! \pow{\ArgSet \cup
			\VarSet}$ defined as follows:
			\begin{enumerate}
				\item\label{def:freeplc:rel}
					$\freeFun{\relElm} \defeq \argFun(\relElm)$, where $\relElm \in
					\RelSet$;
				\item\label{def:freeplc:neg}
					$\freeFun{\neg \varphi} \defeq \freeFun{\varphi}$;
				\item\label{def:freeplc:conjdisj}
					$\freeFun{\varphi_{1} \Opr\! \varphi_{2}} \!\defeq\!
					\freeFun{\varphi_{1}} \cup \freeFun{\varphi_{2}}$, where $\Opr\!
					\!\in\! \{ \wedge, \vee \}$;
				\item\label{def:freeplc:qnt}
					$\freeFun{\Qnt\! \varElm. \varphi} \defeq \freeFun{\varphi} \setminus
					\{ \varElm \}$, where $\Qnt\! \!\in\! \{ \exists, \forall \}$;
				\item\label{def:freeplc:bnd}
					$\freeFun{(\argElm, \varElm) \varphi} \!\defeq\!
					\begin{cases}
						(\freeFun{\varphi} \!\setminus\! \{ \argElm \}) \!\cup\! \{ \varElm
						\}, &
						\text{if } \argElm \in \freeFun{\varphi};
						\\
						\freeFun{\varphi}, &
						\text{otherwise}.
					\end{cases}$
			\end{enumerate}
		\end{definition}
		Observe that, free arguments are introduced in Item~\ref{def:freeplc:rel}
		and removed in Item~\ref{def:freeplc:bnd}.
		Moreover, free variables are introduced in Item~\ref{def:freeplc:bnd} and
		removed in Item~\ref{def:freeplc:qnt}.

		A formula $\varphi$ without free arguments (resp., variables), \ie,
		$\argFun(\varphi) \defeq \freeFun{\varphi} \cap \ArgSet = \emptyset$ (resp.,
		$\varFun(\varphi) \defeq \freeFun{\varphi} \cap \VarSet = \emptyset$), is
		named \emph{argument-closed} (resp., \emph{variable-closed}).
		If $\varphi$ is both argument- and variable-closed, it is referred to as a
		\emph{sentence}.

		Consider the formulas $\varphi_{1}$, $\varphi_{2}$, and $\varphi_{3}$ given
		above.
		It is easy to verify that $\freeFun{\varphi_{1}} = \emptyset$,
		$\freeFun{\varphi_{2}} = \argFun(\varphi_{2}) = \{ \bSym \}$, and
		$\freeFun{\varphi_{3}} = \varFun(\varphi_{3}) = \{ \xSym \}$.
		Thus, $\varphi_{1}$ is a sentence, $\varphi_{2}$ is variable-closed, and
		$\varphi_{3}$ is argument-closed.

	\end{subsection}

	\begin{subsection}{Semantics}
		\label{sec:firordlog;sub:sem}

		The semantics of \FOL we introduce here is defined, as usual, \wrt an
		\RelStr.
		In fact, the peculiarities of our setting just concern the interpretation of
		binding constructs and the non-standard evaluation of relations.

		In order to formalize the meaning of a formula, we first need to describe
		the concept of \emph{assignment}, \ie, a partial function $\asgFun \in
		\AsgSet[\DSet] \defeq (\ArgSet \cup \VarSet) \pto \DSet$ mapping each
		placeholder in its domain to a value of an arbitrary set $\DSet$, which is
		used to set a valuation of all the free arguments and variables.
		For a given placeholder $\pElm \in \ArgSet \cup \VarSet$ and a value $\dElm
		\in \DSet$, the notation ${\asgFun}[\pElm \mapsto \dElm]$ represents the new
		assignment defined on $\dom{{\asgFun}[\pElm \mapsto \dElm]} \!\defeq\!
		\dom{\asgFun} \cup \{ \pElm \}$ that returns $\dElm$ on $\pElm$ and is equal
		to $\asgFun$ on the remaining part of its domain, \ie, ${\asgFun}[\pElm
		\mapsto \dElm](\pElm) \!\defeq\! \dElm$ and ${\asgFun}[\pElm \mapsto
		\dElm](\pElm') \!\defeq\! \asgFun(\pElm')$, for all $\pElm' \!\in\!
		\dom{\asgFun} \!\setminus\! \{ \pElm \}$.

		\begin{definition}[\FOL Semantics]
			\label{def:folsem}
			Let $\RelStrName$ be an $\LanSigName$-\RelStr and $\varphi$ an
			$\LanSigName(\VarSet)$-\FOL formula.
			Then, for all assignments $\asgFun \in \AsgSet[\DomSet]$ with
			$\freeFun{\varphi} \subseteq \dom{\asgFun}$, it holds that the relation
			$\RelStrName, \asgFun \!\models\! \varphi$ is inductively defined on the
			structure of $\varphi$ as follows.
			\begin{enumerate}
				\item\label{def:folsem:rel}
					For every relation $\relElm \in \RelSet$, it holds that $\RelStrName,
					\asgFun \models \relElm$ iff $\asgFun_{\rst \argFun(\relElm)} \in
					\relFun(\relElm)$.
				\item\label{def:folsem:bln}
					For all formulas $\varphi$, $\varphi_{1}$, and $\varphi_{2}$, it holds
					that:
					\begin{enumerate}
						\item\label{def:folsem:bln:neg}
							$\RelStrName, \asgFun \models \neg \varphi$ if not $\RelStrName,
							\asgFun \models \varphi$, that is $\RelStrName, \asgFun
							\not\models \varphi$;
						\item\label{def:folsem:bln:con}
							$\RelStrName, \asgFun \models \varphi_{1} \wedge \varphi_{2}$ iff
							$\RelStrName, \asgFun \models \varphi_{1}$ and $\RelStrName,
							\asgFun \models \varphi_{2}$;
						\item\label{def:folsem:bln:dis}
							$\RelStrName, \asgFun \models \varphi_{1} \vee \varphi_{2}$ iff
							$\RelStrName, \asgFun \models \varphi_{1}$ or $\RelStrName,
							\asgFun \models \varphi_{2}$.
					\end{enumerate}
				\item\label{def:folsem:qnt}
					For a variable $\varElm \in \VarSet$ and a formula $\varphi$, it holds
					that:
					\begin{enumerate}
						\item\label{def:folsem:qnt:exs}
							$\RelStrName, \asgFun \models \exists \varElm. \varphi$ iff there
							exists a value $\dElm \in \DomSet$ such that $\RelStrName,
							{\asgFun}[\varElm \mapsto \dElm] \models \varphi$;
						\item\label{def:folsem:qnt:unv}
							$\RelStrName, \asgFun \models \forall \varElm. \varphi$ iff, for
							all values $\dElm \in \DomSet$, it holds that $\RelStrName,
							{\asgFun}[\varElm \mapsto \dElm] \models \varphi$.
					\end{enumerate}
				\item\label{def:folsem:bin}
					For an argument $\argElm \in \ArgSet$, a variable $\varElm \in
					\VarSet$, and a formula $\varphi$, it holds that $\RelStrName, \asgFun
					\models (\argElm, \varElm) \varphi$ iff $\RelStrName,
					{\asgFun}[\argElm \mapsto \asgFun(\varElm)] \models \varphi$.
			\end{enumerate}
		\end{definition}
		Intuitively, Items~\ref{def:folsem:bln} and~\ref{def:folsem:qnt} define the
		classic semantics of Boolean connectives and first-order quantifiers,
		respectively.
		Item~\ref{def:folsem:rel}, instead, verifies an atomic relation $\relElm$ on
		a given assignment $\asgFun$ by checking whether the tuple function
		${\asgFun}_{\rst \argFun(\relElm)}$ obtained by restricting the assignment
		to the arguments of $\relElm$ is included in the interpretation
		$\relFun(\relElm)$.
		Finally, Item~\ref{def:folsem:bin} evaluates the binding construct
		$(\argElm, \varElm)$ by associating the argument $\argElm$ with the value
		of the variable $\varElm$ contained into the assignment.

		In order to complete the description of the semantics, we now give the
		notions of \emph{model} and \emph{satisfiability}.
		For an $\LanSigName$-\RelStr $\RelStrName$ and an
		$\LanSigName(\VarSet)$-\FOL sentence $\varphi$, we say that $\RelStrName$ is
		a \emph{model} of $\varphi$, in symbols $\RelStrName \models \varphi$, iff
		$\RelStrName, \emptyfun \models \varphi$, where $\emptyfun \in
		\AsgSet[\DomSet]$ simply denotes the empty assignment.
		We also say that $\varphi$ is \emph{satisfiable} iff there exists a model
		for it.
		Given two $\LanSigName(\VarSet)$-\FOL formulas $\varphi_{1}$ and
		$\varphi_{2}$, we say that $\varphi_{1}$ \emph{implies} $\varphi_{2}$, in
		symbols $\varphi_{1} \implies \varphi_{2}$, iff $\RelStrName, \asgFun
		\models \varphi_{1}$ implies $\RelStrName, \asgFun \models \varphi_{2}$, for
		each $\LanSigName$-\RelStr $\RelStrName$ and assignment $\asgFun \in
		\AsgSet[\DomSet]$ with $\freeFun{\varphi_{1}}, \freeFun{\varphi_{2}}
		\subseteq \dom{\asgFun}$.
		Moreover, we say that $\varphi_{1}$ is \emph{equivalent} to $\varphi_{2}$,
		in symbols $\varphi_{1} \equiv \varphi_{2}$, iff both $\varphi_{1} \implies
		\varphi_{2}$ and $\varphi_{2} \implies \varphi_{1}$ hold.

		Consider again the formulas $\varphi_{1}$, $\varphi_{2}$, and $\varphi_{3}$
		and the $\LanSigName$-\RelStr $\RelStrName$ described in
		Section~\ref{sec:prl;sub:relstr}.
		We have that $\RelStrName, \emptyfun \models \varphi_{1}$.
		Moreover, $\RelStrName, \asgFun \models \varphi_{2}$ and $\RelStrName,
		\asgFun \not\models \varphi_{3}$, where $\asgFun(\xSym) = 0$ and
		$\asgFun(\bSym) = 1$.

	\end{subsection}

	\begin{subsection}{Fragments}
		\label{sec:firordlog;sub:frg}

		We now describe a family of four syntactic fragments of \FOL by means of a
		normal form where relations over the same set of arguments may be clustered
		together by a unique sequence of bindings.
		With more detail, we consider Boolean combinations of sentences in prenex
		normal form in which quantification prefixes are coupled with Boolean
		combinations of these relation clusters called \emph{binding forms}.
		Each fragment is then characterized by a specific constraint on the possible
		combinations of binding forms.

		A \emph{quantification prefix} $\qntElm \in \QntSet \subseteq \set{ \exists
		\varElm, \forall \varElm }{ \varElm \in \VarSet }^{*}$ is a finite word on
		the set of quantifications $\set{ \exists \varElm, \forall \varElm }{
		\varElm \in \VarSet }$ over $\VarSet$, in which each variable occurs at most
		once.
		It naturally induces an injective partial function $\qntElm : \VarSet \pto
		\numco{0}{\card{\qntElm}}$ assigning to each variable $\varElm \in
		\varFun(\qntElm) \defeq \dom{\qntElm}$ in its domain the position
		$\qntElm(\varElm) \in \numco{0}{\card{\qntElm}}$ where it occurs in
		$\qntElm$.
		So, either $(\qntElm)_{\qntElm(\varElm)} = \exists \varElm$ or
		$(\qntElm)_{\qntElm(\varElm)} = \forall \varElm$.
		Similarly to a quantification prefix, a \emph{binding prefix} $\bndElm \in
		\BndSet \subseteq \set{ (\argElm, \varElm) }{ \argElm \in \ArgSet \land
		\varElm \in \VarSet }^{*}$ is a finite word on the set of bindings $\set{
		(\argElm, \varElm) }{ \argElm \in \ArgSet \land \varElm \in \VarSet }$ over
		$\ArgSet$ and $\VarSet$, in which each argument occurs at most once.
		By abuse of notation, $\bndElm$ can be also interpreted as a partial
		function $\bndElm : \ArgSet \pto \VarSet$ assigning to each argument
		$\argElm \in \argFun(\bndElm) \defeq \dom{\bndElm}$ in its domain the
		variable $\bndElm(\argElm) \in \varFun(\bndElm) \defeq \rng{\bndElm}$ it is
		bound to in $\bndElm$.
		As an example, the strings $\bndSym[1] = (\aSym, \xSym) (\bSym, \ySym)$,
		$\bndSym[2] = (\aSym, \ySym) (\bSym, \zSym)$, and $\qntSym = \exists \xSym
		\forall \ySym \exists \zSym$ are the binding and quantification prefixes
		occurring in the formula $\varphi_{1} = \qntSym ( \bndSym[1] (\qSym \vee
		\neg \sSym) \wedge \bndSym[2] \rSym )$ previously given, where
		$\bndSym[1](\aSym) \!=\! \xSym$, $\bndSym[1](\bSym) \!=\! \bndSym[2](\aSym)
		\!=\! \ySym$, $\bndSym[2](\bSym) \!=\! \zSym$, and $\qntSym(\xSym) <
		\qntSym(\ySym) < \qntSym(\zSym)$.

		In the following, by Boolean combination over a given set of elements
		$\ESet$ we simply mean a syntactic expression obtained by the grammar $\phi
		\seteq \eElm \mid \neg \phi \mid \phi \wedge \phi \mid \phi \vee \phi$,
		where $\eElm \in \ESet$.
		The set of all these combinations is denoted by $\BCSet(\ESet)$.
		Moreover, $\witFun : \BCSet(\ESet) \to \pow{\pow{\ESet}}$ is the function
		assigning to each combination $\phi \in \BCSet(\ESet)$ the family
		$\witFun(\phi) \subseteq \pow{\ESet}$ of subsets of $\ESet$ that are witness
		for $\phi$, \ie, $\ESet' \!\in\! \witFun(\phi)$ iff $\ESet' \!\models\!
		\phi$, for all $\ESet' \!\subseteq\! \ESet$, where the relation $\models$ is
		classically interpreted.
		For instance, the Boolean combination $\phi = (\eSym[1] \vee \neg \eSym[2])
		\wedge (\neg \eSym[1] \vee \eSym[2])$ over $\ESet = \{ \eSym[1], \eSym[2]
		\}$ has just $\emptyset$ and $\ESet$ itself as possible witnesses.

		Finally, a \emph{derived relation} $\der{\relElm} \in \der{\RelSet} \defeq
		\bigcup_{\ASet \subseteq \ArgSet} \BCSet(\set{ \relElm \in \RelSet }{
		\argFun(\relElm) = \ASet })$ is a Boolean combination over the set of
		relations in $\RelSet$ having all the same arguments.
		An example for the \LanSig $\LanSigName$ of Section~\ref{sec:prl;sub:lansig}
		is given by $\qSym \wedge \rSym$.
		On the contrary, $\qSym \vee \neg \sSym$ is not a derived relation, since
		$\argFun(\rSym) \neq \argFun(\sSym)$.

		\begin{definition}[Binding Fragments]
			\label{def:binfrg}
			\emph{Boolean binding formulas} over $\LanSigName(\VarSet)$ are built by
			means of the following context-free grammar, where $\qntElm \in \QntSet$,
			$\bndElm \in \BndSet$, and $\der{\relElm} \in \der{\RelSet}$:
			\begin{center}
				$\varphi \seteq \qntElm \psi \mid (\varphi \wedge \varphi) \mid (\varphi
				\vee \varphi)$;
				\\
				$\psi \seteq \bndElm \der{\relElm} \mid (\psi \wedge \psi) \mid(\psi
				\vee \psi)$.
			\end{center}
			$\LanSigName(\VarSet)$-\BBFOL denotes the enumerable set of all formulas
			over $\LanSigName(\VarSet)$ generated by the principal rule $\varphi$.
			Moreover, the \emph{conjunctive}, \emph{disjunctive}, and \emph{one
			binding} fragments of \FOL (\CBFOL, \DBFOL, and \OBFOL, for short) are
			obtained, respectively, by weaken the rule $\psi$ as follows: $\psi \seteq
			\bndElm \der{\relElm} \mid (\psi \wedge \psi)$, $\psi \seteq \bndElm
			\der{\relElm} \mid (\psi \vee \psi)$, and $\psi \seteq \bndElm
			\der{\relElm}$.
		\end{definition}

		As an example, consider the \BBFOL sentence $\exists \xSym \forall \ySym
		\exists \zSym \allowbreak(( (\aSym, \xSym) (\bSym, \ySym) \qSym \vee
		(\aSym, \xSym) \neg \sSym ) \wedge (\aSym, \ySym) (\bSym, \zSym) \qSym)$.
		It is not hard to see that it is equivalent to the formula $\varphi_{1}$
		given above.
		In general, by applying a simple generalization of the classic translation
		procedure used to obtain a prenex normal form, it is always possible to
		transform a \FOL formula in a \BBFOL equivalent one, with only a linear
		blow-up in its length.


		\begin{theorem}[Binding Normal Form]
			\label{thr:binnorfor}
			For each $\LanSigName(\VarSet)$-\FOL formula, there is an equivalent
			$\LanSigName(\VarSet)$-\BBFOL one.
		\end{theorem}

		An immediate consequence of the previous theorem is that \BBFOL inherits all
		computational and model-theoretic properties of \FOL.
		Consequently, the following holds.

		\begin{corollary}[\BBFOL Negative Properties]
			\label{cor:bmfolnegprp}
			\BBFOL does not enjoy the finite-model property and has an undecidable
			satisfiability problem.
		\end{corollary}

		This result spurred us to investigate the simplest fragment \OBFOL that
		turns out to be well-behaved both from a model-theoretic and algorithmic
		point of view.
		Actually, we conjecture that \CBFOL enjoys the same nice properties of
		\OBFOL.

		\begin{conjecture}[\CBFOL Positive Properties]
			\label{con:cmfolposprp}
			\CBFOL enjoys the finite-model property and a decidable satisfiability
			problem.
		\end{conjecture}

		On the other hand, it is easy to prove that \DBFOL is not
		model-theoretically as well-behaved as \OBFOL.

		\begin{theorem}[\DBFOL Negative Property]
			\label{thr:dmfolnegprp}
			\DBFOL does not enjoy the finite-model property.
		\end{theorem}
		\begin{proof}
			Consider the conjunction $\varphi = \varphi_{irr} \wedge \varphi_{unb}
			\wedge \varphi_{trn}$ of the following three \DBFOL sentences, asserting
			that the binary relation $\relElm$ on arguments $\{ \aElm, \bElm \}$  is
			irreflexive, unbounded, and transitive: $\varphi_{irr} \!=\! \forall \xElm
			(\aElm, \xElm) (\bElm, \xElm) \neg \relElm$, $\varphi_{unb}  \!=\! \forall
			\xElm \exists \yElm (\aElm, \xElm) (\bElm, \yElm) \relElm$, and
			$\varphi_{trn} \!=\! \forall \xElm \forall \yElm \forall (\zElm (\aElm,
			\xElm) (\bElm, \yElm) \neg \relElm \!\vee\! (\aElm, \yElm) (\bElm, \zElm)
			\neg \relElm \!\vee\! (\aElm, \xElm) (\bElm, \zElm)  \relElm)$.
			It is easy to see that $\varphi$ has only infinite models, since it forces
			$\relElm$ to be a strict partial order without upper bound~\cite{EF95}.
		\end{proof}

		Despite this negative result, we conjecture that the satisfiability problem
		for \DBFOL is decidable.
		Actually, we ground our statement on the observation that classic
		undecidability proofs, as the reduction from the domino
		problem~\cite{Wan61}, seem to require both conjunctions and disjunctions of
		binding forms.

		\begin{conjecture}[\DBFOL Positive Property]
			\label{con:dmfolposprp}
				\DBFOL enjoys a decidable satisfiability problem.
		\end{conjecture}

	\end{subsection}

	\begin{subsection}{Skolemization}
		\label{sec:firordlog;sub:skl}

		Skolemization is a standard tool in model theory that, by replacing each
		occurrence of an existential variable $\varElm$ with a functional symbols
		ranging over the universal variables from which $\varElm$ depends,
		syntactically transform a formula while preserving its satisfiability.
		Following the same principle, we give a semantic notion of this concept that
		is at the base of our novel model-theoretic technique for \OBFOL.
		With more detail, we define a machinery called Skolem Map that, taken an
		assignment over the universal variables of a quantification prefix, returns
		a new assignment over all variables, by complying with the functional
		dependences.

		Let $\qntElm \in \QntSet$ be a quantification prefix.
		Then, $\exists(\qntElm) \defeq \set{ \varElm \in \varFun(\qntElm) }{
		(\qntElm)_{\qntElm(\varElm)} = \exists \varElm }$ and $\forall(\qntElm)
		\defeq \varFun(\qntElm) \setminus \exists(\qntElm)$ denote the sets of
		existential and universal variables quantified in $\qntElm$.
		Moreover, $\prcRel[\qntElm] \subseteq \varFun(\qntElm) \times
		\varFun(\qntElm)$ and $\depRel[\qntElm] \subseteq \forall(\qntElm) \times
		\exists(\qntElm)$ represent, respectively, the \emph{precedence ordering} on
		the variables in $\varFun(\qntElm)$ and the corresponding relation of
		\emph{functional dependence} induced by $\qntElm$, \ie, for all $\varElm[1],
		\varElm[2] \in \varFun(\qntElm)$, it holds that $\varElm[1] \prcRel[\qntElm]
		\varElm[2]$ iff $\qntElm(\varElm[1]) < \qntElm(\varElm[2])$ and
		$\depRel[\qntElm] \defeq \prcRel[\qntElm] \cap (\forall(\qntElm) \times
		\exists(\qntElm))$.
		For instance, if $\qntElm = \exists \xSym \forall \ySym \exists \zSym$ then
		$\exists(\qntElm) = \{ \xSym, \zSym \}$, $\forall(\qntElm) = \{ \ySym \}$,
		$\xSym \prcRel[\qntElm] \ySym \prcRel[\qntElm] \zSym$, and $\ySym
		\depRel[\qntElm] \zSym$.
		In the following, we often indicate by $\AsgSet[\DSet](\PSet) \defeq \set{
		\asgFun \in \AsgSet[\DSet] }{ \dom{\asgFun} = \PSet }$ the set of all
		assignments defined on the subset of placeholders $\PSet \subseteq \ArgSet
		\cup \VarSet$.

		\begin{definition}[Skolem Map]
			\label{def:sklmap}
			Let $\qntElm \in \QntSet$ be a quantification prefix and $\DSet$ an
			arbitrary set.
			Then, a \emph{Skolem map} for $\qntElm$ over $\DSet$ is a function $\smFun
			\in \SMSet[\DSet](\qntElm) \subseteq \AsgSet[\DSet](\forall(\qntElm)) \to
			\AsgSet[\DSet](\varFun(\qntElm))$ satisfying the following properties:
			\begin{enumerate}
				\item
					\label{def:sklmap:unv}
					$\smFun(\asgFun)(\varElm) = \asgFun(\varElm)$, for all $\asgFun \in
					\AsgSet[\DSet](\forall(\qntElm))$ and $\varElm \in \forall(\qntElm)$;
				\item
					\label{def:sklmap:exs}
					$\smFun(\asgFun[1])(\varElm) = \smFun(\asgFun[2])(\varElm)$, for all
					$\asgFun[1], \asgFun[2] \in \AsgSet[\DSet](\forall(\qntElm))$ and
					$\varElm \in \exists(\qntElm)$ such that $\asgFun[1]_{\rst
					\DepSet[\qntElm](\varElm)} = \asgFun[2]_{\rst
					\DepSet[\qntElm](\varElm)}$, where $\DepSet[\qntElm](\varElm) \defeq
					\set{ \varElm' \in \forall(\qntElm) }{ \varElm'
					\depRel[\qntElm] \varElm }$.
			\end{enumerate}
		\end{definition}
		Intuitively, Item~\ref{def:sklmap:unv} ensures that the Skolem map $\smFun$
		behaves as the identity function over all universal variables, while
		Item~\ref{def:sklmap:exs} forces the value $\smFun(\asgFun)(\varElm)$
		associated with an existential variable $\varElm$ to be function only of the
		values $\asgFun(\varElm')$ associated with the universal variables $\varElm'
		\in \DepSet[\qntElm](\varElm)$ from which $\varElm$ depends.
		%

		Consider the set $\DSet = \{ 0, 1 \}$ and the quantification prefix $\qntSym
		= \exists \xSym \forall \ySym \exists \zSym$.
		Then, a possible Skolem map for $\qntSym$ over $\DSet$ is the function
		$\smFun$ such that $\smFun(\asgFun)(\xSym) \!=\! 0$ and
		$\smFun(\asgFun)(\ySym) \!=\! \smFun(\asgFun)(\zSym) \!=\! \asgFun(\ySym)$,
		for all $\asgFun \in \AsgSet[\DSet](\{ \ySym \})$.

		In the remaining part of this section, we \apriori fix an
		$\LanSigName$-\RelStr $\RelStrName = \RelStrStr$, an
		$\LanSigName(\VarSet)$-\FOL formula $\psi$, a quantification prefix $\qntElm
		\in \QntSet$ with $\varFun(\qntElm) \subseteq \varFun(\psi)$, and an
		assignment $\asgFun \in \AsgSet[\DomSet]$ with $\freeFun{\qntElm \psi}
		\subseteq \dom{\asgFun}$.

		A Skolem map can be seen as a tool to remove a quantification prefix
		$\qntElm$ from a formula $\qntElm \psi$ providing a suitable evaluation for
		some of the free variables in $\psi$.
		In fact, one can note a strict connection with the notion of satisfiability,
		as the following definition suggests.

		\begin{definition}[Skolem Satisfiability]
			\label{def:sklsat}
			Let $\smFun \in \SMSet[\DomSet](\qntElm)$ be a Skolem map for $\qntElm$
			over $\DomSet$.
			Then, \emph{$\RelStrName$ satisfies $\psi$ given $\asgFun$ and $\smFun$},
			in symbols $\RelStrName, \asgFun \cmodels[\smFun] \psi$, iff $\RelStrName,
			\asgFun' \models \psi$, for all $\asgFun' \in
			\AsgSet[\DomSet](\dom{\asgFun} \cup \varFun(\qntElm))$ such that $\asgFun
			\!\subseteq\! \asgFun'\!$ and $\asgFun'_{\rst \varFun(\qntElm)} \!\!\in\!
			\rng{\smFun}$.
		\end{definition}

		To better understand the meaning of this definition, consider again the
		$\LanSigName$-\RelStr $\RelStrName$ described in
		Section~\ref{sec:prl;sub:relstr}, the previous formula $\varphi_{1} \!=\!
		\qntSym \psi$ with $\qntSym \!=\! \exists \xSym \forall \ySym \exists \zSym$
		and $\psi \!=\! \bndSym[1] (\qSym \wedge \neg \sSym) \wedge \bndSym[2]
		\rSym$, and the Skolem map $\smFun$ for $\qntSym$ of the example above.
		Then, one can verify that $\RelStrName, \emptyfun \cmodels[\smFun] \psi$.

		Driven by the intuition behind Skolem maps, it is not hard to prove by
		structural induction that the standard notion of satisfiability is actually
		equivalent to the Skolem one.

		\begin{theorem}[Skolem Satisfiability]
			\label{thr:sklsat}
			$\RelStrName, \asgFun \!\models\! \qntElm \psi$ iff there is a Skolem map
			$\smFun \!\in\! \SMSet[\DomSet](\qntElm)$ for $\qntElm$ over $\DomSet$
			such that $\RelStrName, \asgFun \cmodels[\smFun] \psi$.
		\end{theorem}

	\end{subsection}

\end{section}




\begin{section}{Expressiveness Comparison}
	\label{sec:expcom}

	In order to compare the expressive power of our family of logics, in
	particular \OBFOL, to the well-known \emph{clique guarded} (\CGFOL, for
	short)~\cite{Gra99a,Gra02} and \emph{guarded negation} (\GNFOL, for
	short)~\cite{BCS11} fragments of \FOL, we shell study a suitable concept of
	bisimulation between \RelStr{s}, for which the satisfiability of \OBFOL turns
	out to be invariant.

	\begin{subsection}{Bisimulation}
		\label{sec:expcom;sub:bis}

		We now introduce a novel definition of bisimulation between \RelStr{s} that
		result to be bisimilar iff every mapping between partial assignments can be
		extended to a single tuple functions in a way that preserves the
		interpretation of the relations.

		In the following, by $\asgFun{[\PSet \mapsto \dElm]}$ we denote the
		assignment that agrees with $\asgFun$ on $\dom{\fFun} \setminus \PSet$ and
		is equal to $\dElm$ on all placeholders in $\PSet$.

		\begin{definition}[One-Binding Bisimulation]
			\label{def:onebndbis}
			Let $\RelStrName[1] = \RelStrStr[1]$ and $\RelStrName[2] = \RelStrStr[2]$
			be two $\LanSigName$-\RelStr{s}.
			Then, $\RelStrName[1]$ is \emph{one-binding bisimilar} to $\RelStrName[2]$
			iff there exists a total relation $\bisRel \subseteq \allowbreak
			\bigcup_{\ASet \subseteq \ArgSet} \AsgSet[ {\DomSet[1]} ]\!(\ASet)
			\!\times\! \AsgSet[ {\DomSet[2]} ]\!(\ASet)$ for which the following hold.
			\begin{flushenumerate}
				\item\label{def:onebndbis:forbck}
					For all $\ASet \subseteq \ArgSet$, $\ASet' \subseteq \ArgSet \setminus
					\ASet$, and pairs of assignments $\asgFun[1] \!\in\! \AsgSet[
					{\DomSet[1]} ]\!(\ASet)$ and $\asgFun[2] \!\in\! \AsgSet[ {\DomSet[2]}
					]\!(\ASet)$, if $\asgFun[1] \bisRel \asgFun[2]$ then:
					\begin{description}
						\item[\textbf{forth}]
							for all $\dElm[1] \in \DomSet[1]$, there is $\dElm[2] \in
							\DomSet[2]$ such that $\asgFun[1]{[\ASet' \mapsto \dElm[1]]}
							\bisRel \asgFun[2]{[\ASet' \mapsto \dElm[2]]}$;
						\item[\textbf{back}]
							for all $\dElm[2] \in \DomSet[2]$, there is $\dElm[1] \in
							\DomSet[1]$ such that $\asgFun[1]{[\ASet' \mapsto \dElm[1]]}
							\bisRel \asgFun[2]{[\ASet' \mapsto \dElm[2]]}$.
					\end{description}
				\item\label{def:onebndbis:rel}
					For all $\relElm \!\in\! \RelSet$, $\asgFun[1] \!\!\in\!\! \AsgSet[
					{\DomSet[1]} ]\!(\argFun(\relElm))$, and $\asgFun[2] \!\!\in\!\!
					\AsgSet[ {\DomSet[2]} ]\!(\argFun(\relElm))$, with $\asgFun[1] \bisRel
					\asgFun[2]$, it holds that $\asgFun[1] \in \relElm[][ {\RelStrName[1]}
					]$ iff $\asgFun[2] \in \relElm[][ {\RelStrName[2]} ]$.
			\end{flushenumerate}
		\end{definition}

		\figbis
		Consider the \LanSig $\LanSigName = \LanSigStr[][][\{ \aSym, \bSym \}][\{
		\rSym \}]$, where $\argFun(\rSym) = \{ \aSym, \bSym \}$, and the two
		$\LanSigName$-\RelStr{s} $\RelStrName[1] = \RelStrStr[1]$ and
		$\RelStrName[2] = \RelStrStr[2]$ given in Figure~\ref{fig:bis}, where
		$\DomSet[1] = \{ 0, 1, 2 \}$, $\rSym[][ {\RelStrName[1]} ] = \{ (\aSym
		\!\mapsto\! 1, \bSym \!\mapsto\! 2) \}$, $\DomSet[2] = \DomSet[1] \cup \{
		1', 2' \}$, and $\rSym[][ {\RelStrName[2]} ] = \rSym[][ {\RelStrName[1]} ]
		\cup \{ (\aSym \!\mapsto\! 1', \bSym \!\mapsto\! 2') \}$.
		It is not hard to see that they are one-binding bisimilar.
		Indeed, the forth condition is trivially satisfied, since the first
		structure is contained into the second one.
		For the back condition, it is enough to observe that each pair of elements
		in relation (resp., not in relation) in $\RelStrName[2]$ has a corresponding
		pair in $\RelStrName[1]$.

		Finally, by structural induction, it is easy to prove that the existence of
		a one-binding bisimulation between two \RelStr{s} implies their
		indistinguishability \wrt \OBFOL sentences.
		In other words, \OBFOL is invariant under one-binding bisimulation.

		\begin{theorem}[One-Binding Invariance]
			\label{thr:onebndinv}
			Let $\RelStrName[1]$ and $\RelStrName[2]$ be two one-binding bisimilar
			$\LanSigName$-\RelStr{s}.
			Then, for all $\LanSigName(\VarSet)$-\OBFOL sentences $\varphi$, it holds
			that $\RelStrName[1] \models \varphi$ iff $\RelStrName[2] \models
			\varphi$.
		\end{theorem}

	\end{subsection}

	\begin{subsection}{Expressiveness}
		\label{sec:expcom;sub:exp}

		We now have the tool to compare the expressive power of \OBFOL first \wrt
		its extensions \CBFOL and \DBFOL and, successively, to the other fragments
		\CGFOL and \GNFOL.

		\begin{theorem}[\OBFOL \vs \CBFOL and \DBFOL]
			\label{thr:obfol-vs-cbfoldbfol}
			\OBFOL is strictly less expressive than \CBFOL and \DBFOL.
		\end{theorem}
		\begin{proof}
			Observe preliminary that, since \OBFOL is a syntactic fragment of both
			\CBFOL and \DBFOL, they are at least as expressive as the former.
			Now, consider the one-binding bisimilar \RelStr{s} $\RelStrName[1]$ and
			$\RelStrName[2]$ of Figure~\ref{fig:bis}.
			By Theorem~\ref{thr:onebndinv} (One-Binding Invariance), there is no
			\OBFOL sentence that distinguishes them.
			On the contrary, for the \CBFOL sentence $\varphi = \exists \xSym \exists
			\ySym \exists \zSym \exists \wSym ((\aSym, \xSym) (\bSym, \ySym) \rSym
			\wedge (\aSym, \zSym) (\bSym, \wSym) \rSym \wedge (\aSym, \xSym) (\bSym,
			\wSym) \neg \rSym)$, it holds that $\RelStrName[1] \not\models \varphi$
			and $\RelStrName[2] \models \varphi$.
			Thus, $\varphi$ has no \OBFOL equivalent.
			To verify the thesis for \DBFOL too, it is enough to observe that there is
			a \DBFOL sentence equivalent to $\neg \varphi$.
		\end{proof}

		\figcggnobind

		We can now focus on \CGFOL and \GNFOL.
		To evaluate the comparisons, it is useful to do some preliminary
		observations.
		Consider the \LanSig $\LanSigName$ of the previous example and the two
		simple $\LanSigName$-\RelStr{s} $\RelStrName[3] = \RelStrStr[3]$ and
		$\RelStrName[4] = \RelStrStr[4]$ depicted in Figure~\ref{fig:cggnind}, where
		$\DomSet[3] = \{ 0 \}$, $\rSym[][ {\RelStrName[3]} ] = \{ (\aSym \!\mapsto\!
		0, \bSym \!\mapsto\! 0) \}$, $\DomSet[4] = \{ 0, 1 \}$, and $\rSym[][
		{\RelStrName[4]} ] = \rSym[][ {\RelStrName[3]} ] \cup \{ (\aSym \!\mapsto\!
		1, \bSym \!\mapsto\! 1) \}$.
		It is easy to observe that they are both clique-guarded and guarded-negation
		bisimilar, via the the morphisms $\fFun$ and $\gFun$ represented by the
		dashed arrows~\cite{Gra99a,Gra02,BCS11}.
		Now, let $\LanSigName'$ be a new \LanSig extending the previous one with a
		monadic relation $\pSym$ having $\argFun(\pSym) = \{ \bSym \}$.
		Moreover, consider the two $\LanSigName'$-\RelStr{s} $\RelStrName[5] =
		\RelStrStr[5]$ and $\RelStrName[6] = \RelStrStr[6]$, depicted in
		Figure~\ref{fig:obind}, where $\pSym[][ {\RelStrName[5]} ] = \pSym[][
		{\RelStrName[6]} ] = \{ (\bSym \!\mapsto\! 3) \}$.
		It is not hard to see that they are one-binding bisimilar, since the element
		$2$, when associated to the argument $\aSym$ in $\RelStrName[6]$, can be
		mapped to the element $0$ in $\RelStrName[5]$ in order to be sure that it
		has just an ``$\rSym$-successor'' that does not satisfy $\pSym$.

		\begin{theorem}[\OBFOL \vs \CGFOL and \GNFOL]
			\label{thr:obfol-vs-cgfolgnfol}
			\OBFOL is incomparable with \CGFOL and \GNFOL.
		\end{theorem}
		\begin{proof}
			To show that \CGFOL and \GNFOL are not as expressive as \OBFOL, consider
			the clique-guarded and guarded-negation bisimilar \RelStr{s}
			$\RelStrName[3]$ and $\RelStrName[4]$ described above.
			By known invariance theorems~\cite{Gra99a,Gra02,BCS11}, there are neither
			\CGFOL nor \GNFOL sentences that distinguish them.
			On the contrary, for the \OBFOL sentence $\varphi_{1} = \forall \xSym
			\forall \ySym (\aSym, \xSym) (\bSym, \ySym) \rSym$, it holds that
			$\RelStrName[3] \models \varphi_{1}$ and $\RelStrName[4] \not\models
			\varphi_{1}$.
			Thus, $\varphi_{1}$ has no \CGFOL or \GNFOL equivalent.
			For the converse, consider the one-binding bisimilar \RelStr{s}
			$\RelStrName[5]$ and $\RelStrName[6]$ described above.
			By Theorem~\ref{thr:onebndinv}, there is no \OBFOL sentence able to
			distinguish them.
			On the contrary, for the sentence $\varphi_{2} = \exists \xSym \forall
			\ySym ( (\aSym, \xSym) (\bSym, \ySym) \rSym \rightarrow ( \exists \zSym
			(\aSym, \ySym) (\bSym, \zSym) \rSym \wedge (\bSym, \zSym) \pSym ) )$, it
			holds that $\RelStrName[5] \models \varphi_{2}$ and $\RelStrName[6]
			\not\models \varphi_{2}$.
			Thus, $\varphi_{2}$ has no \OBFOL equivalent.
			At this point, it remains just to observe that $\varphi_{2}$ is a guarded
			sentence.
			So, it belongs to both \CGFOL and \GNFOL.
		\end{proof}

	\end{subsection}

\end{section}




\begin{section}{Model-Theoretic Analysis}
	\label{sec:modthranl}

	We now come to the more technical part of this paper, in which we describe a
	novel model-theoretic tool for \OBFOL that allows to obtain the results of
	\emph{finite-model property}, \emph{Craig interpolation}, \emph{and Beth
	definability}.
	Moreover, it represents a key point in the decidability procedure given in the
	last section.

	The tool is based on two characterization theorems that tightly correlate the
	concept of \emph{entanglement}, a semantic property on the sets of Skolem maps
	$\smFun[i]$ associated with the subsentences $\qntElm[i] \bndElm[i]
	\der{\relElm}[i]$ of a given sentence of interest $\varphi$, to that of
	\emph{overlapping}, a syntactic property on the structure of pairs of
	quantification and binding prefixes $(\qntElm[i], \bndElm[i])$.
	Intuitively, on one hand, a set $\{ \smFun[i] \}$ is entangled if the
	underlying quantifications force the corresponding derived relations $\{
	\der{\relElm}[i] \}$ to be verified on a same tuple function $\tupFun$, \ie,
	in other words, if there exists a set of assignments $\{ \asgFun[i] \!\in\!
	\rng{\smFun[i]} \}$ such that $\tupFun \!=\! \asgFun[i] \cmp \bndElm[i]$, for
	all indexes $i$.
	On the other hand, a set $\{ (\qntElm[i], \bndElm[i]) \}$ is overlapping if
	each argument is bound to an existential variable at most once and there is a
	strict total order $\depRel$ between them that agrees, via $\{ \bndElm[i] \}$,
	with all functional dependences of the prefixes $\{ \qntElm[i] \}$.
	These concepts can be seen as two sides of the same coin, which allow to
	define a measure of the degree of inconsistency that one can find among the
	derived relations of $\varphi$.
	So, less entangled is the set of Skolem maps, \ie, less overlapping are the
	prefixes, higher is the possibility for $\varphi$ to be satisfiable.
	In particular, a sentence having a set of non overlapping prefixes can be
	always associated with a set of untangled Skolem maps.

	To have a deeper intuition behind the idea just discussed, we now describe
	four examples, built on the \LanSig $\LanSigName \!=\! \LanSigStr[][][\{
	\aSym, \bSym, \cSym \}][\{ \qSym, \allowbreak \rSym \}]$ with $\argFun(\qSym)
	\!=\! \argFun(\rSym) \!=\! \{ \aSym, \bSym, \cSym \}$, which covers some
	interesting cases of correlation among the concepts.
	Consider the $\LanSigName$-\OBFOL sentence $\varphi_{1} = \qntSym[1]
	\bndElm[1] \qSym \wedge \qntSym[2] \bndElm[2] \neg \rSym \wedge \qntSym[3]
	\bndElm[3] (\qSym \leftrightarrow \rSym)$ having $\qntSym[1] = \forall \xSym
	\exists \ySym \forall \zSym$, $\qntSym[2] = \forall \ySym \exists \zSym
	\forall \xSym$, $\qntSym[3] = \forall \xSym \forall \ySym \forall \zSym$, and
	$\bndSym[1] \!=\! \bndSym[2] \!=\! \bndSym[3] \!=\! (\aSym, \xSym) (\bSym,
	\ySym) (\cSym, \zSym)$.
	It is not hard to see that $\varphi_{1}$ is unsatisfiable, since it requires
	the inconsistent derived relations $\qSym$, $\neg \rSym$, and $\qSym
	\leftrightarrow \rSym$ to hold on a same tuple function $\tupFun$.
	Indeed, independently of the Skolem maps $\smFun[1]$, $\smFun[2]$, and
	$\smFun[3]$ that one may associate with the quantification prefixes
	$\qntSym[1]$, $\qntSym[2]$, and $\qntSym[3]$, there are always three
	assignments $\asgFun[1] \!\in\! \rng{\smFun[1]}$, $\asgFun[2] \!\in\!
	\rng{\smFun[2]}$, and $\asgFun[3] \!\in\! \rng{\smFun[3]}$ such that $\tupFun
	\!=\! \asgFun[1] \cmp \bndSym[1] \!=\! \asgFun[2] \cmp \bndSym[2] \!=\!
	\asgFun[3] \cmp \bndSym[3]$.
	So, $\smFun[1]$, $\smFun[2]$, and $\smFun[3]$ are necessarily entangled.
	This is because the pairs $(\qntElm[1], \bndElm[1])$, $(\qntElm[2],
	\bndElm[2])$, and $(\qntElm[3], \bndElm[3])$ are overlapping, due to the fact
	that we can order the arguments as follows: $\aSym \depRel \bSym \depRel
	\cSym$.
	Given this order, it is simple to find the tuple function $\tupFun$.
	First choose an arbitrary value for $\aSym$.
	Then, compute the corresponding value for $\bSym$ via $\smFun[1]$.
	Finally, extract from  $\smFun[2]$ the value for $\cSym$.
	Now, let $\varphi_{2}$ be the sentence obtained from $\varphi_{1}$ by
	replacing the prefixes $\qntSym[3]$ and $\bndElm[3]$ with the following ones:
	$\qntSym[3] \!=\! \forall \xSym \forall \ySym$ and $\bndSym[3] \!=\! (\aSym,
	\xSym) (\bSym, \ySym) (\cSym, \ySym)$.
	We have that $\varphi_{2}$ is satisfiable on a simple $\LanSigName$-\RelStr of
	order $2$, where the Skolem map $\smFun[2]$ is such that
	$\smFun[2](\asgFun)(\zSym) \!\neq\! \asgFun(\ySym)$.
	Indeed, there are no assignments $\asgFun[2] \!\in\! \rng{\smFun[2]}$ and
	$\asgFun[3] \!\in\! \rng{\smFun[3]}$ for which $\asgFun[2] \cmp \bndSym[2]
	\!=\! \asgFun[3] \cmp \bndSym[3]$, since $(\asgFun[2] \cmp \bndSym[2])(\bSym)
	\!\neq\! (\asgFun[2] \cmp \bndSym[2])(\cSym)$ but $(\asgFun[3] \cmp
	\bndSym[3])(\bSym) \!=\! (\asgFun[3] \cmp \bndSym[3])(\cSym)$.
	Hence, $\smFun[2]$ and $\smFun[3]$ are untangled.
	Therefore, there is no tuple function $\tupFun$ on which $\varphi_{2}$ may
	require the three derived relations all together.
	The syntactic reason here is that $(\qntElm[2], \bndElm[2])$ and $(\qntElm[3],
	\bndElm[3])$ does not allow to find the required order between the argument,
	since from them it follows that $\cSym \depRel \cSym$.
	Similarly to the previous example, consider the sentence $\varphi_{3}$ drawn
	from $\varphi_{1}$ by substituting $\forall \zSym \exists \xSym \forall \ySym$
	for the prefix $\qntSym[3]$.
	Also in this case, $\varphi_{3}$ is satisfiable on a simple
	$\LanSigName$-\RelStr of order $2$, where the three Skolem maps are chosen,
	for instance, as follows: $\smFun[1](\asgFun)(\ySym) \!=\! \asgFun(\xSym)$,
	$\smFun[2](\asgFun)(\zSym) \!=\! \asgFun(\ySym)$, and
	$\smFun[3](\asgFun)(\xSym) \!\neq\! \asgFun(\zSym)$.
	As a matter of fact, there are no assignments $\asgFun[1] \!\in\!
	\rng{\smFun[1]}$, $\asgFun[2] \!\in\! \rng{\smFun[2]}$, and $\asgFun[3]
	\!\in\! \rng{\smFun[3]}$ such that $\tupFun \!=\! \asgFun[1] \cmp \bndSym[1]
	\!=\! \asgFun[2] \cmp \bndSym[2] \!=\! \asgFun[3] \cmp \bndSym[3]$, since the
	definitions of $\smFun[1]$ and $\smFun[2]$ require $\asgFun(\aSym) \!=\!
	\asgFun(\bSym) \!=\! \asgFun(\cSym)$ while that of $\smFun[3]$ implies
	$\asgFun(\aSym) \!\neq\! \asgFun(\cSym)$.
	Here, the pairs $(\qntElm[1], \bndElm[1])$, $(\qntElm[2], \bndElm[2])$, and
	$(\qntElm[3], \bndElm[3])$ are not overlapping, since we can obtain a cycle
	$\aSym \depRel \bSym \depRel \cSym \depRel \aSym$.
	Finally, derive the sentence $\varphi_{4}$ from $\varphi_{1}$, where the
	prefix $\qntSym[3]$ is set to $\exists \zSym \forall \xSym \forall \ySym$.
	It is easy to see that $\varphi_{4}$ is again satisfiable, since one can
	choose the two Skolem maps $\smFun[2]$ and $\smFun[3]$ in such a way that
	$\smFun[2](\asgFun[2])(\zSym) \!\neq\! \smFun[3](\asgFun[3])(\zSym)$,
	independently of the assignments $\asgFun[2]$ and $\asgFun[3]$.
	Consequently, there is no common tuple function $\tupFun$.
	Indeed, the argument $\cSym$ is existential in both $(\qntElm[2], \bndElm[2])$
	and $(\qntElm[3], \bndElm[3])$.

	\begin{subsection}{Entangled Quantifications}
		\label{sec:modthranl;sub:entqnt}

		We start by dealing with the semantic part of our tool.
		First, we introduce the notions of \emph{schemas} and \emph{coupling maps}
		that formalize, in a suitable way, the corresponding concepts of pairs of
		prefixes and set of Skolem maps explained in the informal comment above.
		Then, we describe a new notion of satisfiability of a sentence, given a
		coupling map, that generalizes the one for Skolem maps.
		Finally, we define the concept of entanglement and show how to use it to
		prove the finite-model property.

		A \emph{schema} $\schElm \defeq (\qntElm, \bndElm) \in \SchSet \defeq \set{
		(\qntElm, \bndElm) \!\in\! \QntSet \!\times\! \BndSet }{ \varFun(\qntElm)
		\!=\! \varFun(\bndElm)}$ is a pair of two prefixes, one of quantification
		$\qntElm(\schElm) \!\defeq\! \qntElm$ and the other of binding
		$\bndElm(\schElm) \!\defeq\! \bndElm$, on the sets of arguments
		$\argFun(\schElm) \!\defeq\! \argFun(\bndElm)$ and same variables
		$\varFun(\schElm) \!\defeq\! \varFun(\qntElm) \!=\! \varFun(\bndElm)$.
		By $\SchSet(\ASet) \!\defeq\! \set{ \schElm \!\in\! \SchSet \!}{\!
		\argFun(\schElm) \!=\! \ASet }$ we denote the subset of schemas having
		argument set $\ASet \!\subseteq\! \ArgSet$.
		In what follows, we \apriori fix a finite set of schemas $\SSet \subseteq
		\SchSet$, its size $n \defeq \card{\SSet}$, and both the maximal numbers of
		existential variables $h \defeq \max_{\schElm \in \SSet}
		\card{\exists(\qntElm(\schElm))}$ and arguments $k \defeq \max_{\schElm \in
		\SSet} \card{\argFun(\schElm)}$.

		\begin{definition}[Coupling Map]
			\label{def:cplmap}
			Let $\DSet$ be an arbitrary set.
			Then, a \emph{coupling map} for $\SSet$ over $\DSet$ is a function $\cmFun
			\in \CMSet[\DSet](\SSet) \defeq \SSet \cto[\schElm]
			\SMSet[\DSet](\qntElm(\schElm))$ assigning to every schema $\schElm \in
			\SSet$ a Sk\"{o}lem map $\cmFun(\schElm) \in
			\SMSet[\DSet](\qntElm(\schElm))$ for $\qntElm(\schElm)$ over $\DSet$.
		\end{definition}

		Similarly to the Skolem satisfiability, we now introduce the \emph{coupling
		satisfiability}.
		As a matter of technical convenience, we do not treat formulas explicitly,
		but use a suitable function $\forFun$ representing a set of possible
		sentences of the form $\qntElm \bndElm \der{\relElm}$.

		\begin{definition}[Coupling Satisfiability]
			\label{def:cplsat}
			Let $\RelStrName = \RelStrStr$ be an $\LanSigName$-\RelStr and $\cmFun \in
			\CMSet[\DomSet](\SSet)$ a coupling map for $\SSet$ over $\DomSet$.
			Moreover, let $\forFun \in \ForSet(\SSet) \defeq \SSet \cto[\schElm] \set{
			\der{\relElm} \in \der{\RelSet} }{ \argFun(\der{\relElm}) =
			\argFun(\relElm) }$ be a \emph{formula function} for $\SSet$ assigning to
			each of its schemas $\schElm$ a derived relation $\forFun(\schElm)$ on the
			arguments of $\schElm$ itself.
			Then, \emph{$\RelStrName$ satisfies $\forFun$ given $\cmFun$}, in symbols
			$\RelStrName \cmodels[\cmFun] \forFun$, iff $\RelStrName
			\cmodels[\cmFun(\schElm)] \forFun(\schElm)$, for all $\schElm \in \SSet$.
		\end{definition}

		Due to the marked similarity between Skolem and coupling satisfiability, the
		next result follows as an immediate corollary of Theorem~\ref{thr:sklsat}
		(Skolem Satisfiability).

		\begin{corollary}[Coupling Satisfiability]
			\label{cor:cplsat}
			Let $\RelStrName = \RelStrStr$ be an $\LanSigName$-\RelStr and $\forFun
			\in \ForSet(\SSet)$ a formula function for $\SSet$.
			Then, $\RelStrName \models \qntElm \bndElm \der{\relElm}$, for all
			$\schElm = (\qntElm, \bndElm) \in \SSet$ with $\der{\relElm} =
			\forFun(\schElm)$, iff there exists a coupling map $\cmFun \in
			\CMSet[\DomSet](\SSet)$ for $\SSet$ over $\DomSet$ such that $\RelStrName
			\cmodels[\cmFun] \forFun$.
		\end{corollary}

		We give here the fundamental definition of \emph{coupling entanglement} that
		allows to verify the existence of a tuple function shared by the Skolem maps
		in the range of a coupling map.

		\begin{definition}[Coupling Entanglement]
			\label{def:cplent}
			Let $\cmFun \in \CMSet[\DSet](\SSet)$ be a coupling map for $\SSet$ over a
			set $\DSet$.
			Then, the \emph{entanglement} of $\cmFun$ \wrt $\SSet' \subseteq \SSet
			\cap \SchSet(\ASet)$ over $\ASet' \subseteq \ASet \subseteq \ArgSet$ is
			the set of functions from $\ASet'$ to $\DSet$ defined as follows:
			$\EntSet[\cmFun](\SSet', \ASet') \defeq \set{ \tupFun \in \ASet' \to \DSet
			}{ \forall \schElm \in \SSet' \:.\: \exists \asgFun \in
			\rng{\cmFun(\schElm)} \:.\: \tupFun = (\asgFun \cmp
			\bndElm(\schElm))_{\rst \ASet'} }$.
			If $\EntSet[\cmFun](\SSet', \ASet') \neq \emptyset$ then $\cmFun$ is said
			to be \emph{entangled} \wrt $\SSet'$ over $\ASet'$, otherwise, it is
			\emph{untangled}.
		\end{definition}

		As already mentioned in the informal description above, the concept of
		entanglement indirectly induces a preorder among coupling maps.
		Such a preorder plays a key role in all proofs of model-theoretic and
		algorithmic properties for \OBFOL.

		\begin{definition}[Coupling Preorder]
			\label{def:cplprd}
			Let $\cmFun[1] \in \CMSet[ {\DSet[1]} ](\SSet)$ and $\cmFun[2] \in \CMSet[
			{\DSet[2]} ](\SSet)$ be two coupling maps for $\SSet$ over the sets
			$\DSet[1]$ and $\DSet[2]$.
			Then, $\cmFun[1]$ is \emph{at least as entangled as} $\cmFun[2]$, in
			symbols $\cmFun[2] \entRel \cmFun[1]$, iff, whenever $\cmFun[2]$ is
			entangled \wrt $\SSet' \subseteq \SSet \cap \SchSet(\ASet)$ over $\ASet'
			\subseteq \ASet \subseteq \ArgSet$, so is $\cmFun[1]$, \ie, $\EntSet[
			{\cmFun[2]} ](\SSet', \ASet') \neq \emptyset$ implies $\EntSet[
			{\cmFun[1]} ](\SSet', \ASet') \neq \emptyset$.
		\end{definition}

		As stated in the next theorem, the main feature for such a preorder is that
		it is downward-bounded, namely it admits minimal elements.
		In particular, among those elements it is always possible to find a finite
		coupling map, \ie, a map whose range is composed only by Skolem maps defined
		over a finite domain.
		The proof of this property, reported in the next section, is one of the most
		important result we derive from the announced characterization theorems.

		\begin{theorem}[Finite Minimal Coupling]
			\label{thr:finmincpl}
			There exists a coupling map $\cmFun[][\star] \in \CMSet[ {\DSet[][\star]}
			](\SSet)$ for $\SSet$ over a finite set $\DSet[][\star]$ with
			$\card{\DSet[][\star]} \leq n \!\cdot\! h \!\cdot\! 2^{(n \cdot k)!}$ that
			is minimal \wrt $\entRel$.
		\end{theorem}

		Informally, the property of entanglement describes the degree of possible
		inconsistencies among the derived relations of a given sentence.
		Therefore, the lesser entangled a coupler map is, the easier a model for
		that sentence can be found.
		Next theorem actually formalize this intuitive idea, by showing that, once a
		formula function $\forFun$ is satisfied by a given coupling map $\cmFun[1]$,
		it is satisfied by all coupling maps $\cmFun[2]$ that are less entangled
		than $\cmFun[1]$.

		\begin{theorem}[Satisfiability Preservation]
			\label{thr:satprs}
			Let $\cmFun[1] \!\in\! \CMSet[ {\DSet[1]} ](\SSet)$ and $\cmFun[2] \in
			\CMSet[ {\DSet[2]} ](\SSet)$ be two coupling maps for $\SSet$ over the
			sets $\DSet[1]$ and $\DSet[2]$ such that $\cmFun[2] \entRel \cmFun[1]$.
			Moreover, let $\forFun \in \ForSet(\SSet)$ be a formula function for
			$\SSet$.
			If there is an $\LanSigName$-\RelStr $\RelStrName[1] = \RelStrStr[1][][
			{\DSet[1]} ]$ such that $\RelStrName[1] \cmodels[ {\cmFun[1]} ] \forFun$
			then there exists an $\LanSigName$-\RelStr $\RelStrName[2] =
			\RelStrStr[2][][ {\DSet[2]} ]$ such that $\RelStrName[2] \cmodels[
			{\cmFun[2]} ] \forFun$ as well.
		\end{theorem}

		By suitably exploiting the results stated in the previous two theorems, we
		can finally prove that \OBFOL has the finite-model property.

		\begin{theorem}[Finite Model Property]
			\label{thr:finmodprp}
			Every $\LanSigName(\VarSet)$-\OBFOL satisfiable sentence is finite
			satisfiable.
		\end{theorem}
		\begin{proof}
			Since $\varphi$ is satisfiable, there is an $\LanSigName$-\RelStr
			$\RelStrName = \RelStrStr$ such that $\RelStrName \models \varphi$.
			This implies the existence of a witness for $\varphi$ satisfied by
			$\RelStrName$, \ie, a set of subsentences $\FSet \in \witFun(\varphi)$
			such that $\RelStrName \models \phi$, for all $\phi \in \FSet$.
			Now, let $\SSet \defeq \set{ (\qntElm, \bndElm) \in \SchSet }{ \exists
			\der{\relElm} \in \der{\RelSet} \:.\: \qntElm \bndElm \der{\relElm} \in
			\FSet }$ be the set of schemas for $\FSet$ and $\forFun \in
			\ForSet(\SSet)$ the formula function for $\SSet$ assigning to each of its
			schemas $\sigma \in \SSet$ a derived relation $\der{\relElm} =
			\forFun(\schElm)$ such that $\qntElm \bndElm \der{\relElm} \in \FSet$.
			\Wlogx, we can assume that $\forFun$ covers $\FSet$, \ie, for each
			sentence $\phi \in \FSet$, there is a schema $\schElm = (\qntElm, \bndElm)
			\in \SSet$ such that $\phi = \qntElm \bndElm \der{\relElm}$ with
			$\der{\relElm} = \forFun(\schElm)$.
			Indeed, we can prevent two sentences from sharing the same schema by
			simply renaming their variables.
			At this point, by Corollary~\ref{cor:cplsat} (Coupling Satisfiability), we
			derive the existence of a coupling map $\cmFun \in \CMSet[\DomSet](\SSet)$
			such that $\RelStrName \cmodels[\cmFun] \forFun$.
			Moreover, by Theorem~\ref{thr:finmincpl} (Finite Minimal Coupling), there
			is a coupling map $\cmFun[][\star] \in \CMSet[ {\DSet[][\star]} ](\SSet)$
			over a finite set $\DSet[][\star]$ with $\card{\DSet[][\star]} \leq n
			\!\cdot\! h \!\cdot\! 2^{(n \cdot k)!}$ that is minimal \wrt $\entRel$.
			Consequently, $\cmFun[][\star] \entRel \cmFun$.
			Now, by Theorem~\ref{thr:satprs} (Satisfiability Preservation), there is
			an $\LanSigName$-\RelStr $\RelStrName[][\star] = \RelStrStr[][\star][
			{\DSet[][\star]} ]$ such that $\RelStrName[][\star] \cmodels[
			{\cmFun[][\star]} ] \forFun$.
			Thus, again by Corollary~\ref{cor:cplsat}, we have that
			$\RelStrName[][\star] \models \phi$, for all $\phi \in \FSet$.
			Hence, $\RelStrName[][\star] \models \varphi$.
			To conclude the proof, it is enough to observe that $\RelStrName[][\star]$
			has order $\AOmicron{n \!\cdot\! h \!\cdot\! 2^{(n \cdot k)!}}$.
		\end{proof}

	\end{subsection}

	\begin{subsection}{Entanglement Characterization}
		\label{sec:modthranl;sub:entchr}

		We can now deal with the syntactic part of our tool.
		To introduce the concept of overlapping used to state the characterization
		theorems, we make use of two suitable graphs defined on the structural
		properties of a given sets of schemas.

		Let $\schElm \in \SchSet$ be a schema.
		Then, $\exists(\schElm) \defeq \set{ \argElm \in \argFun(\schElm) }{
		\bndElm(\schElm)(\argElm) \in \exists(\qntElm(\schElm)) }$ and
		$\forall(\schElm) \defeq \argFun(\schElm) \setminus \exists(\schElm)$ denote
		the sets of existential and universal arguments in $\schElm$, \ie, the
		arguments $\argElm \in \argFun(\schElm)$ that are associated with an
		existential or universal variable $\bndElm(\schElm)(\argElm)$ in the
		quantification prefix $\qntElm(\schElm)$ via the binding prefix
		$\bndElm(\schElm)$.
		Moreover, $\colRel[\schElm] \subseteq \argFun(\schElm) \times
		\argFun(\schElm)$ and $\depRel[\schElm] \subseteq \forall(\schElm) \times
		\exists(\schElm)$ represent, respectively, the \emph{collapsing equivalence}
		and the \emph{functional dependence} induced by $\schElm$, \ie, for all
		$\argElm[1], \argElm[2] \in \argFun(\schElm)$, it holds that $\argElm[1]
		\colRel[\schElm] \argElm[2]$ iff $\bndElm(\schElm)(\argElm[1]) =
		\bndElm(\schElm)(\argElm[2])$ and, for all $\argElm[1] \in \forall(\schElm)$
		and $\argElm[2] \in \exists(\schElm)$, it holds that $\argElm[1]
		\depRel[\schElm] \argElm[2]$ iff $\bndElm(\schElm)(\argElm[1])
		\depRel[\qntElm(\schElm)] \bndElm(\schElm)(\argElm[2])$.
		As an example, for the schema $\schElm \!=\! (\forall \xSym \exists \ySym,
		(\aSym, \xSym) (\bSym, \ySym) (\cSym, \xSym))$, it holds that
		$\exists(\schElm) \!=\! \{ \bSym \}$, $\forall(\schElm) \!=\! \{ \aSym,
		\cSym \}$, $\aSym \depRel[\schElm] \bSym$, $\cSym \depRel[\schElm] \bSym$,
		and $\aSym \colRel[\schElm] \cSym$.

		The vertexes of the above mentioned graphs are arguments coupled with the
		schemas in which they occur, as the following description reports.
		Let $\ASet' \!\!\subseteq\!\! \ASet \!\!\subseteq\!\! \ArgSet$ be two sets
		of arguments and $\SSet' \!\subseteq\! \SSet \cap \SchSet(\ASet)$ a set of
		schemas.
		Then, $\ArgSet[\SSet'][\ASet'] \!\defeq\! \set{ (\schElm, \argElm) \in
		\SSet' \times \ASet' }{ \ASet' \!\subseteq\! \argFun(\schElm) }$ denotes the
		set of \emph{extended arguments} for $\SSet'$ over $\ASet'$, \ie, the pairs
		$\eElm \defeq (\schElm, \argElm) \in \ArgSet[\SSet'][\ASet']$ of a schema
		$\schElm(\eElm) \defeq \schElm \in \SSet'$ and one of its arguments
		$\argElm(\eElm) \defeq \argElm \in \argFun(\schElm) \cap \ASet'$.
		Also, $\exists_{\SSet'}^{\ASet'} \!\defeq\! \set{ (\schElm, \argElm) \!\in\!
		\ArgSet[\SSet'][\ASet'] }{ \argElm \!\in\! \exists(\schElm) }$ and
		$\forall_{\SSet'}^{\ASet'} \!\defeq\! \ArgSet[\SSet'][\ASet'] \setminus
		\exists_{\SSet'}^{\ASet'}$ represent the set of existential and universal
		extended arguments.

		We now introduce the first graph, which is used to represent the relation
		among the arguments that are somehow collapsed by the structure of the
		schemas, \ie, intuitively, they are forced to assume the same value.
		The \emph{collapsing graph} for $\SSet'$ over $\ASet'$ is the symmetric
		directed graph $\CName[\SSet'][\ASet'] \defeq \GrpStr[][][
		{\ArgSet[\SSet'][\ASet']} ][ {\colRel[\SSet'][\ASet']} ]$ with the extended
		arguments as vertexes and the edge relation defined as follows:
		$\colRel[\SSet'][\ASet'] \!\defeq\! \set{ (\eElm[1], \eElm[2]) \!\in\!
		\ArgSet[\SSet'][\ASet'] \!\times\! \ArgSet[\SSet'][\ASet'] }{
		(\schElm(\eElm[1]) = \schElm(\eElm[2]) \implies \argElm(\eElm[1]) \colRel[
		{\schElm(\eElm[1])} ] \argElm(\eElm[2])) \land (\schElm(\eElm[1]) \neq
		\schElm(\eElm[2]) \implies \argElm(\eElm[1]) = \argElm(\eElm[2])) }^{+}$.

		\figcolgrp

		In Figure~\ref{fig:colgrp1}, we depict the collapsing graph $\CName[1]
		\!=\! \CName[\SSet][\ASet]$, where $\SSet \!=\! \{ \schElm[1] \!=\!
		(\qntSym[1], \bndElm[1]), \schElm[2] \!=\! (\qntSym[2], \bndElm[2]),
		\schElm[3] \!=\! (\qntSym[3], \bndElm[3]) \}$ and $\ASet \!=\! \{ \aSym,
		\bSym, \cSym \}$ are the set of schemas and arguments of the sentences
		$\varphi_{1}$, $\varphi_{3}$, and $\varphi_{4}$ given in the intuitive
		discussion above.
		Here, the dots represents the extended arguments obtained by the
		intersection of rows and columns.
		Moreover, we are omitting the transitive closure, since
		$\colRel[\SSet'][\ASet']$ is an equivalence relation.
		In Figure~\ref{fig:colgrp2}, we report the collapsing graph $\CName[2]$ for
		the sentence $\varphi_{2}$.
		Observe that the edge between the vertexes $(\schElm[3], \bSym)$ and
		$(\schElm[3], \cSym)$ is due to the binding $\bndElm[3]$.

		At this point, we define a property that precisely describe the case in
		which two existential arguments are forced to assume the same value.
		A collapsing graph $\CName[\SSet'][\ASet']$ is \emph{conflicting} iff there
		exist two existential extended arguments $\eElm[1], \eElm[2] \!\in\!
		\exists_{\SSet'}^{\ASet'}$ such that $\eElm[1] \!\colRel[\SSet'][\ASet']\!
		\eElm[2]$ and if $\schElm(\eElm[1]) \!=\! \schElm(\eElm[2])$ then
		$\argElm(\eElm[1]) \not\!\!\colRel[ {\schElm(\eElm[1])} ]
		\argElm(\eElm[2])$.

		As an example, the graph $\CName[1]$ is conflicting when it is referred to
		the sentence $\varphi_{4}$.
		This is due to the fact that the argument $\cSym$ is existentially
		quantified in both $\schElm[2]$ and $\schElm[3]$.

		The second graph we introduce keeps track of the chains of functional
		dependences between arguments that cross all the schemas.
		The \emph{dependence graph} for $\SSet'$ over $\ASet'$ is the directed graph
		$\DName[\SSet'][\ASet'] \!\defeq\! \GrpStr[][][ {\ArgSet[\SSet'][\ASet']} ][
		{\depRel[\SSet'][\ASet']} ]$ with the extended arguments as vertexes and the
		edge relation defined as follows: $\depRel[\SSet'][\ASet'] \defeq
		\colRel[\SSet'][\ASet'] \!\!\cmp \set{ (\eElm[1], \eElm[2]) \!\in
		\ArgSet[\SSet'][\ASet'] \!\times\! \ArgSet[\SSet'][\ASet'] }{
		\schElm(\eElm[1]) \!=\! \schElm(\eElm[2]) \land \argElm(\eElm[1]) \depRel[
		{\schElm(\eElm[1])} ] \argElm(\eElm[2]) }$.

		\figdepgrp

		In Figures~\ref{fig:depgrp1},~\ref{fig:depgrp2}, and~\ref{fig:depgrp3},
		we report the dependence graphs $\DName[i] \!=\! \DName[\SSet][\ASet]$
		corresponding to the sentences $\varphi_{1}$, $\varphi_{2}$, and
		$\varphi_{3}$, respectively.
		The black arrows represent the functional dependences inside a single
		schema, while the gray ones are obtained by the composition with the
		collapsing relation.
		Note that $\DName[1]$ is acyclic, so, we can be construct an order among
		the extended arguments that agrees with all functional dependences of the
		schemas.
		On the contrary, in $\DName[2]$, there is a loop on $(\schElm[2], \cSym)$
		due to the structures of $\schElm[2]$ and $\schElm[3]$.
		Finally, also $\DName[3]$ is not acyclic, due to cycle among $(\schElm[3],
		\aSym)$, $(\schElm[1], \bSym)$, and $(\schElm[2], \cSym)$.

		\begin{definition}[Overlapping Schemas]
			\label{def:ovrsch}
			A set of schemas $\SSet' \subseteq \SSet \cap \SchSet(\ASet)$ is
			\emph{overlapping} over a set of arguments $\ASet' \subseteq \ASet
			\subseteq \ArgSet$ iff the collapsing graph $\CName[\SSet'][\ASet']$ is
			not conflicting and the dependence graph $\DName[\SSet'][\ASet']$ is
			acyclic.
		\end{definition}

		We can now state the two characterization theorems at the base of all
		results we provide for \OBFOL.
		They essentially move along two opposite directions.
		Intuitively, the first shows that, if a set of schemas is overlapping, every
		coupling map is necessarily entangled over that set.
		The second, instead, shows that, if a set of schemas is not overlapping, it
		is always possible to find a finite untangled coupling map over such a set.

		\begin{theorem}[Entangled Coupling Map]
			\label{thr:entcplmap}
			For all coupling maps $\cmFun \in \CMSet[\DSet](\SSet)$ for $\SSet$ over
			an arbitrary set $\DSet$, if $\SSet' \subseteq \SSet \cap \SchSet(\ASet)$
			is overlapping over $\ASet' \subseteq \ASet \subseteq \ArgSet$ then
			$\cmFun$ is entangled \wrt $\SSet'$ over $\ASet'$.
		\end{theorem}

		\begin{theorem}[Untangled Coupling Map]
			\label{thr:untcplmap}
			There exists a coupling map $\cmFun[][\star] \in \CMSet[ {\DSet[][\star]}
			](\SSet)$ for $\SSet$ over a finite set $\DSet[][\star]$ with
			$\card{\DSet[][\star]} \!\leq\! n \cdot h \cdot 2^{(n \cdot k)!}$\!, such
			that if $\SSet' \!\!\subseteq\! \SSet \cap \SchSet(\ASet)$ is not
			overlapping over $\ASet' \subseteq \ASet \subseteq \ArgSet$ then
			$\cmFun[][\star]$ is untangled \wrt $\SSet'$ over $\ASet'$.
		\end{theorem}

		Both the theorems we have just stated form our characterization result on
		the correlation between the concepts of entanglement and that of
		overlapping, from which we can easily derive the existence of a finite
		minimal coupling.

		\begin{proof}[Proof of Theorem~\ref{thr:finmincpl}]
			By Theorem~\ref{thr:untcplmap} (Untangled Coupling Map), there is a
			coupling map $\cmFun[][\star] \in \CMSet[ {\DSet[][\star]} ](\SSet)$ for
			$\SSet$ over a finite set $\DSet[][\star]$ with $\card{\DSet[][\star]}
			\leq n \!\cdot\! h \!\cdot\! 2^{(n \cdot k)!}$ such that if
			$\cmFun[][\star]$ is entangled \wrt $\SSet' \subseteq \SSet \cap
			\SchSet(\ASet)$ over $\ASet' \subseteq \ASet \subseteq \ArgSet$ then
			$\SSet'$ is overlapping over $\ASet'$.
			Now, such a coupling map is necessarily minimal, \ie, $\cmFun[][\star]
			\entRel \cmFun$, for all coupling maps $\cmFun \in \CMSet[\DSet](\SSet)$
			for $\SSet$ over an arbitrary set $\DSet$.
			Indeed, by Theorem~\ref{thr:entcplmap} (Entangled Coupling Map), $\cmFun$
			is entangled \wrt $\SSet'$ over $\ASet'$ as well.
		\end{proof}

		As a further model-theoretic result, a constructive version of the Craig
		interpolation property for \OBFOL can be provided.
		Here, due to the lack of space, we just give a quick intuition behind the
		proof of the basic case in which the input formulas are of the form
		$\varphi_{1} = \qntElm[1] \bndElm[1] \der{\relElm}[1]$ and $\varphi_{2} =
		\qntElm[2] \bndElm[2] \der{\relElm}[2]$ with $\varphi_{1} \implies
		\varphi_{2}$.
		Indeed, it is not hard to see that the formula $\varphi = \qntElm[1]
		\bndElm[1] \der{\relElm}$, with $\der{\relElm}$ being the classic
		propositional interpolating between $\der{\relElm}[1]$ and
		$\der{\relElm}[2]$, is the interpolating formula for $\varphi_{1}$ and
		$\varphi_{2}$.
		A proof of the general case deeply relies on the characterization theorems.
		By means of classic techniques, the Beth definability property can be also
		obtained from the Craig interpolation one.

		\begin{theorem}[Craig Interpolation and Beth Definability]
			\label{thr:crgintbetdef}
			\OBFOL enjoys both Craig interpolation and Beth definability.
		\end{theorem}

	\end{subsection}

\end{section}




\begin{section}{Satisfiability}
	\label{sec:sat}

	We finally provide a \PSpace deterministic algorithm for the solution of the
	satisfiability problem for \OBFOL, which can be interpreted as a
	satisfiability-modulo-theory procedure.
	Indeed, by means of a syntactic preprocessing based on the concept of
	overlapping schemas, the search for a model of a \OBFOL sentence is reduced to
	that of a sequence of Boolean formulas over the set of derived relations.
	The correctness of such an approach is crucially based on the fundamental
	characterization of the entanglement property previously discussed.
	It is interesting to observe that the procedure is independent from the size
	of the model pointed out in Theorem~\ref{thr:finmodprp}.

	To understand the main idea behind the algorithm, it is useful to describe it
	through a simple one-round two-player turn-based game between the existential
	player, called Eloise, willing to show that a sentence $\varphi$ is
	satisfiable, and the universal player, called Abelard, trying to do exactly
	the opposite.
	First, Eloise choses a witness $\FSet \!\in\! \witFun(\varphi)$ for $\varphi$
	seen as a Boolean combination of simpler subsentences of the form $\qntElm
	\bndElm \der{\relElm}$.
	In this way, she identifies a formula function $\forFun \!=\! \set{ (\qntElm,
	\bndElm) \!\in\! \SchSet \mapsto \der{\relElm} \!\in\! \der{\RelSet} }{
	\qntelm \bndElm \der{\relElm} \!\in\! \FSet }$ that describes $\FSet$ by
	associating each derived relation with the corresponding schema.
	Then, Abelard choses a subset of overlapping schemas $\SSet \!\subseteq\!
	\dom{\forFun} \!\cap\! \SchSet(\ASet)$ over a set of arguments $\ASet
	\!\subseteq\! \ArgSet$.
	At this point, Eloise wins the play iff the Boolean formula $\psi \!=\!
	\bigwedge_{\schElm \in \SSet} \forFun(\schElm)$, obtained as the conjunction
	of all the derived relations associated with the schemas in $\SSet$, is
	satisfiable.
	If this is not the case, Abelard has spotted a subset $\set{ \qntElm \bndElm
	\forFun(\schElm) }{ (\qntElm, \bndElm) \!\in\! \SSet }$ of the witness $\FSet$
	that requires to verify the unsatisfiable property $\psi$ on a certain
	valuation of arguments in $\ASet$.
	Thus, $\FSet$ cannot have a model.
	Consequently, the sentence $\varphi$ is satisfiable iff Eloise has a winning
	strategy for this game.

	\algsat

	We can now describe the pseudo-code of the algorithm.
	The deterministic counterpart of Eloise's choice is the selection of a witness
	$\FSet \!\in\! \witFun(\varphi)$ in the loop at Line~1, which is followed by
	the computation of the corresponding formula function $\forFun$.
	At each iteration, a flag $\iElm$ is also set to $\Ff$, with the aim to
	indicate that $\FSet$ is not inconsistent (a witness is consistent until
	proven otherwise).
	After this, we find the deterministic counterpart of Abelard's choice,
	implemented by the combination of a loop and a conditional statement at
	Lines~4 and~5, where a subset of overlapping schemas $\SSet \!\subseteq\!
	\dom{\forFun} \!\cap\! \SchSet(\ASet)$ over a set of arguments $\ASet
	\!\subseteq\! \ArgSet$ is selected.
	This is done in order to verify the inconsistency of the conjunction
	$\bigwedge_{\schElm \in \SSet} \forFun(\schElm)$ at Line~6.
	If this check is positive then the flag $\iElm$ is switched to $\Tt$.
	Once all choices for Abelard are analyzed, the computation reaches Line~8,
	where it is verified whether an inconsistency was previously found.
	In the negative case, the algorithm terminates by returning $\Tt$, with the
	aim to indicate that a good guess for Eloise is possible.
	In the case all witnesses are analyzed, finding for each of them an
	inconsistency, Eloise has no winning strategy.
	Thus, the algorithm ends by returning $\Ff$.

	At this point, it remains to evaluate the complexity of the algorithm \wrt the
	length of the sentence $\varphi$.
	First, observe that it only requires a Boolean flag $\iElm$, three sets
	$\FSet$, $\SSet$, and $\ASet$ and a function $\forFun$, whose sizes are all
	linear in the input.
	Moreover, the membership at Line~1 is verifiable in \PTime, the emptiness of
	the witness set at Line~6 can be computed in \NPTime, and the check for the
	overlapping property at Line~5 can be easily executed in \PSpace.
	Consequently, the whole complexity is \PSpace.

	\begin{theorem}[Decidability of Satisfiability]
		\label{thr:decsat}
		The satisfiability problem for \OBFOL is \PSpace.
	\end{theorem}

	%
	%
	%
	%
	%

	%
	%
	%

\end{section}




\begin{section}{Discussion}

	Trying to understand the reasons why some power extensions of modal logic are
	decidable, we introduced and studied a new family of \FOL fragments based on
	the combinations of binding forms admitted in a sentence.
	In other words, we provided an innovative criterion to classify \FOL formulas
	focused on the associations between arguments and variables.
	One of the main features of this classification is the avoiding of usual
	syntactic restrictions on quantifier patterns, number of variables, relation
	arities, and relativization of quantifications.
	Therefore, it represents a completely new framework in which to study
	model-theoretic and algorithmic properties for particular extensions of modal
	logic, like \SL~\cite{MMV10b}.

	We analyzed the expressiveness of the introduced fragments, showing that the
	simplest one, called \emph{one binding} (\OBFOL), is already incomparable with
	other important sublogics of \FOL, such as the \emph{clique guarded} and the
	\emph{guarded negation}.
	Moreover, we proved that it enjoys the finite-model property, a \PSpace
	satisfiability problem, and a constructive version of Craig's interpolation
	and Beth's definability.
	To do this, exploiting the fine structure of binding forms, we developed a
	technical characterization of first-order quantifications, which can be
	considered of interest by its own.

	An important and immediate application of the nice properties of \OBFOL is the
	obtaining of similar results for the one-goal fragment of
	\SL~\cite{MMPV11,MMPV12}.
	Indeed, from an high-level point of view, every \OGSL sentence has a \OBFOL
	corresponding one, with same quantification and binding prefixes, in which the
	inner \LTL temporal properties are replaced by suitable derived relations
	having the agents as arguments.
	The elements of the domain of quantification represent, therefore, the
	strategies of  the game.
	Now, since we proved that a \OBFOL sentence just requires a finite model to be
	satisfied, it is enough to have a finite number of strategies for the
	corresponding \OGSL sentence.
	Consequently, the game only needs finitely many actions, \ie, the model is
	bounded.
	By means of a similar idea, the same property for automata over game
	structures~\cite{SF06} and, thus, for the alternating \MuCalculus can be also
	derived.
	Finally, observe that a proof of our conjecture about \CBFOL immediately
	results in a proof of a related open problem about the conjunctive-goal
	fragment of \SL~\cite{MMS13a}.

	As future works, besides a deeper study of the \emph{conjunctive} and
	\emph{disjunctive} fragments (\CBFOL and \DBFOL), there are several other
	directions that could be investigated.
	First, it would be interesting to analyze the model-checking complexity for
	\OBFOL and its connection with fragments of classic query languages.
	Then, we conjecture that some of its extensions, like those ones with
	equality, counting quantifiers, or fixpoint constructs, still preserve the
	same good model-theoretic and algorithmic properties.
	Finally, as a long term project, it would be nice to come up with a wider
	framework in which the incomparable fragments \CGFOL, \GNFOL, and \OBFOL can
	be simply seen as particular cases of a unique decidable logic.

\end{section}





\begin{section}*{Acknowledgments}

	We thank M. Benerecetti and M.Y. Vardi for useful comments and discussions on
	a first draft of this article.

\end{section}


	\footnotesize
	\bibliographystyle{IEEEtran}
	\bibliography{References}

\begin{thebibliography}{10}
\providecommand{\url}[1]{#1}
\csname url@samestyle\endcsname
\providecommand{\newblock}{\relax}
\providecommand{\bibinfo}[2]{#2}
\providecommand{\BIBentrySTDinterwordspacing}{\spaceskip=0pt\relax}
\providecommand{\BIBentryALTinterwordstretchfactor}{4}
\providecommand{\BIBentryALTinterwordspacing}{\spaceskip=\fontdimen2\font plus
\BIBentryALTinterwordstretchfactor\fontdimen3\font minus
  \fontdimen4\font\relax}
\providecommand{\BIBforeignlanguage}[2]{{%
\expandafter\ifx\csname l@#1\endcsname\relax
\typeout{** WARNING: IEEEtran.bst: No hyphenation pattern has been}%
\typeout{** loaded for the language `#1'. Using the pattern for}%
\typeout{** the default language instead.}%
\else
\language=\csname l@#1\endcsname
\fi
#2}}
\providecommand{\BIBdecl}{\relax}
\BIBdecl

\bibitem{Chu36a}
A.~Church, ``{An Unsolvable Problem of Elementary Number Theory.}'' \emph{AJM},
  vol.~58, pp. 345--363, 1936.

\bibitem{Chu36b}
------, ``{A Note on the Entscheidungsproblem.}'' \emph{JSL}, vol.~1, no.~1,
  pp. 40--41, 1936.

\bibitem{Chu36c}
------, ``{Correction to a Note on the Entscheidungsproblem.}'' \emph{JSL},
  vol.~1, no.~3, pp. 101--102, 1936.

\bibitem{Tur37a}
A.~Turing, ``{On Computable Numbers, with an Application to the
  Entscheidungsproblem.}'' \emph{PLMS}, vol.~42, pp. 230--265, 1937.

\bibitem{Tur37b}
------, ``{On Computable Numbers, with an Application to the
  Entscheidungsproblem. A correction.}'' \emph{PLMS}, vol.~43, pp. 544--546,
  1937.

\bibitem{HA28}
D.~Hilbert and W.~Ackermann, \emph{{Grundz\"uge der Theoretischen Logik.}},
  ser. Die Grundlehren der Mathematischen Wissenschaften.\hskip 1em plus 0.5em
  minus 0.4em\relax Springer, 1928.

\bibitem{Bor84}
E.~B{\"o}rger, ``{Decision Problems in Predicate Logic.}'' in
  \emph{LC'82}.\hskip 1em plus 0.5em minus 0.4em\relax Elsevier, 1984, vol.
  112, pp. 263--301.

\bibitem{Gra03}
E.~Gr{\"a}del, ``{Decidable Fragments of First-Order and Fixed-Point Logic.}''
  in \emph{KWLCS'03}, 2003.

\bibitem{Low15}
L.~L{\"o}wenheim, ``{\"Uber M\"oglichkeiten im Relativkalk\"ul.}'' \emph{MA},
  vol.~76, no.~4, pp. 447--470, 1915.

\bibitem{Mor75}
M.~Mortimer, ``{On Languages with Two Variables.}'' \emph{MLQ}, vol.~21, no.~1,
  pp. 135--140, 1975.

\bibitem{GKV97}
E.~Gr{\"a}del, P.~Kolaitis, and M.~Vardi, ``{On the Decision Problem for
  Two-Variable First-Order Logic.}'' \emph{BSL}, vol.~3, no.~1, pp. 53--69,
  1997.

\bibitem{BGG97}
E.~B{\"o}rger, E.~Gr{\"a}del, and Y.~Gurevich, \emph{{The Classical Decision
  Problem.}}, ser. {Perspectives in Mathematical Logic.}\hskip 1em plus 0.5em
  minus 0.4em\relax Springer, 1997.

\bibitem{Sim88}
S.~Simpson, ``{Partial Realizations of Hilbert's Program.}'' \emph{JSL},
  vol.~53, no.~2, pp. 349--363, 1988.

\bibitem{Var96}
M.~Vardi, ``{Why is Modal Logic So Robustly Decidable?}'' in
  \emph{DCFM'96}.\hskip 1em plus 0.5em minus 0.4em\relax American Mathematical
  Society, 1996, pp. 149--184.

\bibitem{Gra01}
E.~Gr{\"a}del, ``{Why are Modal Logics so Robustly Decidable?}'' in
  \emph{CTTCS'01}, 2001, pp. 393--408.

\bibitem{ABN98}
H.~Andr{\'e}ka, J.~van Benthem, and I.~N{\'e}meti, ``{Modal Languages And
  Bounded Fragments Of Predicate Logic.}'' \emph{JPL}, vol.~27, no.~3, pp.
  217--274, 1996.

\bibitem{Gra99a}
E.~Gr{\"a}del, ``{Decision Procedures for Guarded Logics.}'' in \emph{CADE'99},
  ser. LNCS 1632.\hskip 1em plus 0.5em minus 0.4em\relax Springer, 1999, pp.
  31--51.

\bibitem{Ben97}
J.~van Benthem, ``{Dynamic Bits And Pieces.}'' University of Amsterdam,
  Amsterdam, Netherlands, Tech. Rep., 1997.

\bibitem{Gra02}
E.~Gr{\"a}del, ``{Guarded Fixed Point Logics and the Monadic Theory of
  Countable Trees.}'' \emph{TCS}, vol. 288, no.~1, pp. 129--152, 2002.

\bibitem{Ben00}
J.~van Benthem, ``{Modal Logic In Two Gestalts.}'' in \emph{AIML'98}.\hskip 1em
  plus 0.5em minus 0.4em\relax CSLI Publications, 2000, pp. 91--118.

\bibitem{GG00}
E.~Goncalves and E.~Gr{\"a}del, ``{Decidability Issues for Action Guarded
  Logics.}'' in \emph{DL'00}, 2000, pp. 123--132.

\bibitem{GW99}
E.~Gr{\"a}del and I.~Walukiewicz, ``{Guarded Fixed Point Logic.}'' in
  \emph{LICS'99}.\hskip 1em plus 0.5em minus 0.4em\relax IEEE Computer Society,
  1999, pp. 45--54.

\bibitem{BCO10}
V.~B{\'a}r{\'a}ny, G.~Gottlob, and M.~Otto, ``{Querying the Guarded
  Fragment.}'' in \emph{LICS'10}.\hskip 1em plus 0.5em minus 0.4em\relax IEEE
  Computer Society, 2010, pp. 1--10.

\bibitem{CS11}
B.~ten Cate and L.~Segoufin, ``{Unary Negation.}'' in \emph{STACS'11}, ser.
  LIPIcs 9.\hskip 1em plus 0.5em minus 0.4em\relax Leibniz-Zentrum fuer
  Informatik, 2011, pp. 344--355.

\bibitem{CS13}
------, ``{Unary Negation.}'' \emph{LMCS}, vol.~9, no.~3, pp. 1--46, 2013.

\bibitem{BCS11}
V.~B{\'a}r{\'a}ny, B.~ten Cate, and L.~Segoufin, ``{Guarded Negation.}'' in
  \emph{ICALP'11}, ser. LNCS 6756.\hskip 1em plus 0.5em minus 0.4em\relax
  Springer, 2011, pp. 356--367.

\bibitem{BCO12}
V.~B{\'a}r{\'a}ny, B.~ten Cate, and M.~Otto, ``{Queries with Guarded
  Negation.}'' \emph{PVLDB}, vol.~5, no.~11, pp. 1328--1339, 2012.

\bibitem{WLWW06}
D.~Walther, C.~Lutz, F.~Wolter, and M.~Wooldridge, ``{\ATL Satisfiability is
  Indeed ExpTime-Complete.}'' \emph{JLC}, vol.~16, no.~6, pp. 765--787, 2006.

\bibitem{SF06}
S.~Schewe and B.~Finkbeiner, ``{Satisfiability and Finite Model Property for
  the Alternating-Time muCalculus.}'' in \emph{CSL'06}.\hskip 1em plus 0.5em
  minus 0.4em\relax Springer, 2006, pp. 591--605.

\bibitem{Sch08}
S.~Schewe, ``{\ATLS Satisfiability is 2ExpTime-Complete.}'' in \emph{ICALP'08},
  ser. LNCS 5126.\hskip 1em plus 0.5em minus 0.4em\relax Springer, 2008, pp.
  373--385.

\bibitem{MMV10b}
F.~Mogavero, A.~Murano, and M.~Vardi, ``{Reasoning About Strategies.}'' in
  \emph{FSTTCS'10}, ser. LIPIcs 8.\hskip 1em plus 0.5em minus 0.4em\relax
  Leibniz-Zentrum fuer Informatik, 2010, pp. 133--144.

\bibitem{MMPV11}
F.~Mogavero, A.~Murano, G.~Perelli, and M.~Vardi, ``{Reasoning About
  Strategies: On the Model-Checking Problem.}'' arXiv, Tech. Rep., 2011.

\bibitem{MMPV12}
------, ``{What Makes \ATLS Decidable? A Decidable Fragment of Strategy
  Logic.}'' in \emph{CONCUR'12}, ser. LNCS 7454.\hskip 1em plus 0.5em minus
  0.4em\relax Springer, 2012, pp. 193--208.

\bibitem{Cod70}
E.~Codd, ``{A Relational Model of Data for Large Shared Data Banks.}''
  \emph{CACM}, vol.~13, no.~6, pp. 377--387, 1970.

\bibitem{Cod72}
------, ``{Relational Completeness of Data Base Sublanguages.}'' \emph{DS}, pp.
  65--98, 1972.

\bibitem{Imm82}
N.~Immerman, ``{Relational Queries Computable in Polynomial Time (Extended
  Abstract).}'' in \emph{STOC'82}.\hskip 1em plus 0.5em minus 0.4em\relax
  Association for Computing Machinery, 1982, pp. 147--152.

\bibitem{Var82}
M.~Vardi, ``{The Complexity of Relational Query Languages (Extended
  Abstract).}'' in \emph{STOC'82}.\hskip 1em plus 0.5em minus 0.4em\relax
  Association for Computing Machinery, 1982, pp. 137--146.

\bibitem{Imm86}
N.~Immerman, ``{Relational Queries Computable in Polynomial Time.}'' \emph{IC},
  vol.~68, no. 1-3, pp. 86--104, 1986.

\bibitem{EF95}
{H.-D.~Ebbinghaus and J.~Flum}, \emph{{Finite Model Theory.}}, ser.
  {Perspectives in Mathematical Logic.}\hskip 1em plus 0.5em minus 0.4em\relax
  Springer, 1995.

\bibitem{Wan61}
H.~Wang, ``{Proving Theorems by Pattern Recognition II.}'' \emph{BSTJ},
  vol.~40, pp. 1--41, 1961.

\bibitem{MMS13a}
F.~Mogavero, A.~Murano, and L.~Sauro, ``{On the Boundary of Behavioral
  Strategies.}'' in \emph{LICS'13}.\hskip 1em plus 0.5em minus 0.4em\relax IEEE
  Computer Society, 2013, pp. 263--272.

\end{thebibliography}






\end{document}